# The Task Analysis Cell Assembly Perspective.

Dan Diaper                Chris Huyck

DDD SYSTEMS          Middlesex University

ddiaper@ntlworld.com      c.huyck@mdx.ac.uk

*Abstract*

An entirely novel synthesis combines the applied cognitive psychology of a task analytic approach with a neural cell assembly perspective that models both brain and mind function during task performance; similar cell assemblies could be implemented as an artificially intelligent neural network. A simplified cell assembly model is introduced and this leads to several new representational formats that, in combination, are demonstrated as suitable for analysing tasks. The advantages of using neural models are exposed and compared with previous research that has used symbolic artificial intelligence production systems, which make no attempt to model neurophysiology. For cognitive scientists, the approach provides an easy and practical introduction to thinking about brains, minds and artificial intelligence in terms of cell assemblies. In the future, subsequent developments have the potential to lead to a new, general theory of psychology and neurophysiology, supported by cell assembly based artificial intelligences.

## .1      Introduction

There already exists a strong relationship between a cognitive ergonomics Task Analysis (TA) method and Artificial Intelligence (AI) of the symbolic sort. These are, respectively, Goals, Operations, Methods and Selection rules (GOMS, e.g. Card, Moran and Newell, 1983; Kieras, 2004) and production systems such as ACT-R (e.g. Anderson and Lebiere, 1998, Anderson 2007) and EPIC (Meyer and Kieras, 1997). Anderson and Lebiere claim that such systems "*are the only modelling formalism capable of spanning a broad range of tasks, dealing with complex cognition ...*" (p3), and in their enthusiasm go so far as to claim for ATC-R "*a profound sense of psychological reality*" (p13); Anderson (2007) sees EPIC as a precursor to ACT-R 6.0, contributing "Perceptual-Motor" modules. EPIC's developers are rather more cautious in their claims (e.g. Kieras and Meyer, 1994; Meyer and Kieras, 1997).

A fundamental problem with these production system symbolic AI approaches involves "cognitive architecture" which Anderson (2007, p7) defines as "*a specification of the structure of the brain at a level of abstraction that explains how it achieves the function of the mind.*" There is a problem concerning his "level of abstraction" notion. At the level of program code, these symbolic AI systems make no attempt to mimic the human brain, other than as functional, i.e. psychological, modules, although Anderson (2007) attempts, *post hoc*, to relate some of these to brain areas. The theoretical issue concerns simulation fidelity, here how well one thing, a symbolic AI, can mimic another, the brain, when at the level of operation they are completely different types of thing. This paper proposes a solution by using a different sort of AI, one which does attempt simulation of how both the brain and the mind operates and which uses a single, common modelling representation for both.

There are hundreds of different TA methods and virtually all of them have a cognitive, psychological component, although the psychology generally is not that good. As Kieras (2004) rightly notes, "*a task analysis for system design must be rather more informal and primarily heuristic in flavour compared to scientific research.*" Based on the cognitive psychology of Card, Moran and Newell (1980), GOMS is one of the more psychologically sophisticated of TA methods yet is easy to criticise as scientifically inadequate. For example, when a task performer needs to access Long Term Memory (LTM), a GOMS analysis can identify this but is pretty well independent of alternative theories of human LTM architectures and processes, i.e. a GOMS analysis would hardly change whether one modelled human LTM like computer backing store, or as memory traces with different strengths, or as multiple traces.

The basic theoretical argument in GOMS, and generally in TA, is that some cognitive representations and processes similar to those identified during an analysis must occur. For example, at some point in a task it might be necessary to store information temporarily, which the TA might call using Short Term Memory (STM), but whether this is the STM of Miller (1956) is moot, never mind the Baddeley and Hitch (1974, Baddeley, 1976) alternative architecture of their Working Memory, which has been considerably developed subsequently, e.g. Oberauer *et al*. (2018), and there are a number of other temporary and buffer like stores that are hypothesised to be common in all human minds, although the precise theoretical specification of these remain controversial, e.g. Morey, *et al*. (2018). Similarly, most TAs will identify when decisions are made in tasks, but the cognitive decision making mechanisms are left unspecified.

Given the difficulty of predicting human performance, e.g. for its traditional application of training design, GOMS is really very good, although Kieras (2004, Kieras and Butler, 2014) are carefully cautious about this, and there are exceptions (e.g. Jorritsma *et al*., 2015). While no one has ever successfully developed a general task taxonomy, i.e. a specification of sub-tasks or other task components that, together, could be used to specify any task performed by people (e.g. Balbo *et al*., 2004), GOMS does produce a modular, reusable output that resembles program pseudo-code. Indeed, it is a short, obvious step to implement such generic GOMS modules as software tool support to facilitate predicting task performance and, on such coat-tails, to implement the GOMS model as a symbolic AI. Given the tight binding between GOMS and systems like ACT-R and EPIC, it is unsurprising that they share similar theoretical limitations.

This paper's proposal involves a modern take (Huyck & Passmore, 2013) on Hebbian Cell Assemblies (CAs). Hebb's (1949) theory is that concepts are represented in the brain by a collection of neurons firing, e.g. there is not a Grandmother Cell that represents one of one's grandmothers, but rather there is a Grandmother CA, a collection of neurons that can fire persistently, with or without external stimulus from the environment. Though Hebb's 1949 work predates work on cognitive architecture (Newell, 1990), Hebb's cognitive architecture is elegant and straightforward: each mental representation of a concept is represented in the brain as a unique CA., i.e. this is the identity thesis of Smart (2007 for a summary) and of his colleague Ullin (1956); also of the similar, independent work of Feigl (1958) who says of mental events, that they "*are identical with certain (presumably configurational) aspects of the neural processes*".

CAs are normally implemented as a simplified model of neurons to mimic how the human brain might operate. The main proposal in this paper is that it is possible to model the behavioural and cognitive psychology of task performance using a putative CA based brain model and, in theory, the same model could be implemented as a CA based AI. One problem for GOMS and ACT-R that a CA approach automatically deals with are memory representation issues; Hebb's theory is one of LTM, i.e. CAs represent the conceptual contents of LTM.

Attractive if not completely compelling evidence for the CA approach is that like nearly all Artificial Neural Networks (ANNs), CA based ones are self-organising, i.e. they can learn. This is the Achilles' Heel of nearly all symbolic AIs, they need human programmers first. Thus, if a GOMS model changes then its symbolic AI equivalent would have to be reprogrammed. Anderson (2007) discusses learning in some detail (e.g. Chapter 5), but it is hardly surprising that ACT-R can model human learning since, at least in theory, following a TA is should be able to model any task performed by people, including ones involving learning. There is, however, a critical difference between being able to model human learning and the basic, inherent, inevitable and unstoppable learning that is fundamental to ANNs, including CA-based ones.

The 'Cell Assembly roBots' (CABots) demonstrate in a virtual environment the learning of both aspects of the environment and new objects within it, and it has a problem solving capability, all without the intervention of human programmers (Huyck & Mitchell, 2018). While ACT-R, and other cognitive architectures like Soar (Laird *et al.,* 1987) can learn, these typically work by parameter setting or generating new rules using old rules. They are not capable of, for instance, symbol grounding (Harnad, 1990). CAs provide an ability, for instance, to ground symbols, suggested as early as Hebb (1949).

There is considerable evidence, summarised by (Huyck & Passmore, 2013) that much of the human brain does use a CA architecture. The Strong CA Hypothesis, that all brain function is by CAs, is almost certainly untrue, although specialised brain areas may develop during neonatal tuning from a general CA architecture, e.g. Blakemore and Cooper (1970); and that cortical plasticity allows some recovery of function after localised brain insult, also might be plausibly explained by general purpose CAs becoming tuned in adulthood. The Weak CA Hypothesis, that the brain's default architecture is CA based, remains plausible.

On a more cautious note, much of our current understanding of CAs comes from work on ANNs. There is a serious issue of the biological plausibility of such ANNs. For example, while it is now possible to simulate a billion neurons in real time in a system (Furber, *et al.*, 2013), these artificial neurons are really represented as a rather simple algebraic equation and, as such, are an extremely simplified model of the brain's physiology. While, for example, Huyck's Fatiguing Leaky Integrate and Fire (FLIF) neurons (Huyck and Parvizi, 2012) are a better simulation of brain neurons than early ANNs, e.g. perceptrons (Rumelhard and McClelland, 1986), or, earlier, compartmental models (Hodkin and Huxley, 1952), they fail to model fundamental neural physiological properties such as spike trains. Even FLIF neurons fail to model basic physiology such as different neurotransmitters, other temporal neuron properties, and much else.

An absolutely crucial, and it seems sometimes overlooked, property of even quite simple CAs is that Byrne & Huyck (2010) have proved that they can be Turing machines, i.e. that, given enough neurons, they can compute the result of any legal mathematical or logical expression. The critical consequence of this is that anything that can be written using traditional programming approaches, including symbolic AI code, can be done using simulated neuron based CAs. At the moment, run-time efficiency remains a major problem, but it is believable that performance will continue to improve in the relatively short term future. On the other hand, Huyck's CABot already runs in real time on a PC.

Hebb's original theory has been considerably developed, particularly in recent years. A simple but critical improvement is that Hebb's concepts have been extended to most mental content and, indeed, to representing processes. On the latter, CAs naturally represent processes as CAs change over time, e.g. a grandmother CA is updated during a visit to her, and this is akin to a run time process description of computer program code (Osterweil, 1987; Diaper and Kadoda, 1999). CAs can also represent processes by providing structure to CAs pre-ignition, for example, for doing mental arithmetic, Natural Language (NL) parsing, and for other sorts of common problem solving and planning.

As concepts, Hebb's CAs can fire persistently over time and this remains a fundamental property of newer CA models, although, more accurately, they have the capability of persistence because in some tasks this may not be required, e.g. in a self-terminating, visual, serial search task the target CA would not persist for long if the target is the first item, but may have to persist for minutes in other circumstances. Critically for the brain, CAs can be ignited for longer than it takes a neuron to fatigue. Therefore, for CA persistence on the order of several seconds and above, there must be a pool of non-firing neurons that can be swapped in to replace fatiguing neurons so as to maintain an ignited CA (see PotN, section 2). Furthermore, with very long term CA persistence a member neuron might fire, fatigue, recover and then re-fire. Indeed, it is essential that the particular neurons that are firing in an ignited CA change over time so that the CA can perform processes, for example, doing a calculation (Tetzlaff *et al*., 2015). Even when a CA functions as an LTM item, this will change over ignitions, even when general learning is slight (see the QPID model below).

The brain has around $10^{11}$ neurons (Smith, 2010) and the size range of ignited CAs has been suggested as $10^3$ to $10^7$ neurons (Huyck and Passmore, 2013), although the upper estimates probably refer to "*super-CAs composed of many sub-CAs*". Even with all these brain neurons, most neurons will, at different times, have membership of different CAs, although CA type may be restricted, e.g. a neuron in the visual cortex might always be involved with visual processing, but be in millions of different CAs during its existence.

In the absence of alternative theories and appropriate physiological evidence, a simple model is that CAs can exist in four states: Quiescent, Priming, Ignited, and Decaying. For simplicity, it is assumed that all four states are physiologically similar, i.e. that the Q, P and D states are but weaker versions of a CA in the I-state, with fewer neurons but these may still be shared, at different times, across numerous CAs. Functionally, however, the four states may differ significantly: Q-state CAs are structured for permanent storage. The role of CAs in their P-state is to prepare a CA for ignition and support processes such as attentional mechanisms involving competition between CAs. The reality in brains in undoubtedly very complicated and a P-state CA probably has a very different structure at the start of priming to just before ignition as it evolves into a form ready for ignition; it is also possible that CAs may exist in the P-state without on some occasions ever igniting. The D-state is involved with preparing a CA for its LTM storage and may be equally as complex in its structures and functions. The physiology and functionality of these transition states is not so much under researched as virtually unresearched.

In a typical QPID cycle the new Q-state is not quite the same as its precursor. When the notionally same CA is ignited on different occasions, not only will these differ as to the set of neurons involved in each ignition, but the CA itself will not be quite the same. Thus, the functional definition of a CA must be at a sufficiently high level of description that such differences usually can be ignored. From Scott-Phillips *et al*. (2011) in the context of their distinction between proximate and ultimate explanations in evolutionary theory, it may be that functional and physical descriptions are of different types: the physical, brain TACAP models being proximate and addressing "How?" questions and the mental, functional ones may be ultimate models and addressing a highly specialised epistemological type of "Why?" questions.

Some concept of levels of description, of detail, is common in many areas of human endeavour. The super- and sub-CA proposal and the QPID model fits neatly with the extensive use of the levels concept in TA, and with this paper's CA based approach. Emphasising that a TA model is an analysts' model and different from that of task participants and other involved parties, e.g. managers, (Diaper, 2004 and *ibid*.), one difficult "judgement call" (Kieras, 2004) is the level at which a particular TA is pitched. Most TA methods involve some form of task decomposition into subtasks, and sub-subtasks, down to the level analysts select (N.B. different levels may be chosen for different parts of the same task). Many methods do simply decompose tasks, but not all. For example, the old but still popular Hierarchical Task Analysis (HTA) method (Annett and Duncan, 1967; Annett, 2004) decomposes task goals rather than recorded task components. As such, HTA is an analysis technique that can be used after task data is collected and represented.

This last point about HTA is crucial to this paper, which similarly only discusses an analysis technique and not a full TA method. Traditionally, a TA method early on will involve multiple information sources and data collection techniques; observation of performance, interviews and questionnaires are common, but many other data sources and techniques have been used over the decades. In nearly all TA methods, whatever data is collected, it is combined to produce some sort of Activity List (AL), otherwise known as a 'task protocol' (N.B. this is different from a 'task transcript').

While varying greatly in style, generally an AL is a prose description of how a task is performed and the strong recommendation of Diaper (1989a, 2004, and *ibid*.) is that an AL should consist of a list of short sentences that each describe a task step, at the level chosen, and each line should identify a main agent and the action(s) performed towards one or more things (agents or objects), perhaps using other things (tools). It is some such AL representation that HTA and this paper's work uses as input to their analysis techniques. One word of caution, however, data collected with one TA method in mind may not be suitable if other analysis techniques are then used; missing data being the most obvious problem, but there are more subtle ones.

This paper is not proposing yet another TA method or, even, analysis technique, at least, not at the moment. This is one of the reasons why "Perspective" appears in its title. A perspective is "a point of view" and in the scheme of things as used here, is a very general theoretical formulation, perhaps a high level framework. Within a Popperian (e.g. Popper, 1979) scientific epistemology, the claim is that only well specified hypotheses can be experimentally tested and that disproof of one does not necessarily disprove the more general framework from which that false hypothesis was derived; metaphorically, pruning twigs from a knowledge tree may not damage its main branches.

Perspective, as used in this paper, is a "General Theory of Psychology" (section 5.3.3), perhaps akin to cognitive psychology's one that has the axiom, "The mind is an information processing device." The claim is that all psychological phenomena can be described and, ultimately, explained within the perspective defined by its primary axioms (Diaper and Stanton, 2004). As an extension, the CA equivalent would be something like, "The mind and brain are information processing devices that both use common, although differently described, cell assemblies."

Cognitive psychology's axiom is implementation independent, i.e. it has no constraints on how the brain works, its architecture, processes and so forth, because it is only concerned with mental models, of information processing. In contrast, the CA perspective provides for a firm cognitive architecture that relates and explains concomitant brain and cognitive function. This and similar issues are more properly and completely covered in the Discussion (section 5.2).

The version of TACAP that is used in this paper is described in section 2. There remains, however, one further major issue concerning the "Perspective" in the TACAP title.

TACAP, as used in this paper, deliberately exploits the limitations of TA to provide a *demonstration* of what may be possible and an example of potential utility. The emphasis is that it is only a demonstration and this leads to what at first might seem an odd claim: we do not care if everything in the demonstration is wrong.

It is very likely that none of the brain CAs identified in this paper will ever be found to exist, but using the TA defence (see above), something similar must occur, and it remains possible that in a training programme based on a TACAP approach, some CAs from the TA will cause similar CAs in trainees. Similarly, the mental, functional TA descriptions provided may also all be wrong, but this may also be a matter of poor TA, which is not at all a concern in a demonstration. As for AI, the proposals concerning similarities with brains cannot be worse than that for the GOMS to ACT-R/EPIC relationship, howsoever the CAs proposed are wrong in detail, since the symbolic systems have no claim to any hardware realism. What is provided in this paper is a demonstration of the potential of a CA based perspective within the practical, engineering limitations of TA.

Before returning to the topics introduced above in the Discussion section (section 5), the TACAP version developed (section 2) and its application and the results (sections 3 plus Appendix I, and 4, respectively) are reported.

.2    The Task Analysis Cell Assembly Model

An advantage of this first demonstration using TA and the CA notion is that it can exploit TA's heuristic approach (see Kieras, 2004, quoted above) and, as argued immediately above, as applied psychology the description of task performance needs only to be plausible.

The following subsections outline the models and representations finally used. In their development a considerable variety of things were tried and rejected, sometimes simply because they were just too awkward to use. Leaving such to historians of science, this paper tends to concentrate on what was found successful and relatively easy to use. One of the biggest determinates of the development programme was consistency. Most TAs are performed iteratively and our development work was an extreme example as we would not only return to initially analysed task steps and re-analyse them, but sometimes we would even change the graphics and notational style to what was found later to be a better approach. Indeed, particularly during the early analysis stages, decisions were made about the nature of the CAs and their relationships that quite fundamentally changed the earlier analyses, which had to be redone, some of them several times.

### .2.1 The Simplified Cell Assembly Model (SCAM)

The standard, simple graphical representation of a CA plots number of neurons firing against time (Kaplan *et al.*, 1991, Huyck and Passmore, 2013). Unsurprisingly, this graph resembles that of an individual neuron's firing and, indeed, most negative feedback systems.

The lifecycle of a simple CA is: (a) there is a background level of neuron firing (quite a lot in the brain, but it is not organised, section 4.2.3); (b) a CA starts to develop, usually due to "priming" from already ignited CAs, and the number of neurons firing in the CA starts to increase, probably in an exponential manner (N.B. competition between a number of alternative CAs at this stage may be a critical part of autonomous cognitive decision making and attentional mechanisms); (c) sufficient neurons fire such that the CA's "threshold" is reached; (d) at which point a large number of neurons rapidly join the CA which then "ignites"; (e) as with most negative feedback systems, there is an overshoot as firing neuron CA membership climbs to its ignition state; (f) after the overshoot the function stabilises at a level which may be several times higher than threshold; (g) the CA then persists and there is a slow decay in the number of neurons firing to support the ignited CA, due to neuron fatigue, if nothing else; (h) at some point the CA will extinguish, either because there are insufficient neurons firing to maintain ignition, or because the CA becomes inhibited by the firing of other CAs; (i) the CA's neuron activity drops below threshold and the CA decays, although what it decays to may depend on the type and context of a CA, i.e. it may decay to background levels, or have a refractory period like neurons and be harder to re-ignite, or it may remain above background so that it is primed for re-ignition (sections 4.2.3 and 5.3.1).

Many CAs will be more complicated than this simple model, particularly ones that persist for long periods, minutes if not hours, as fatigued neurons are replaced. All sorts of things might change during a CAs persistence phase (g) due to CA competition, cooperation and, even, combination or division. Thus, this part of the model might present a saw tooth profile rather than a relatively smooth decline in the number of neurons firing in an ignited CA; for example, see Appendix I: CA 06 MSHWA – Motor Stride to Hot Water Area.

The Simplified Cell Assembly Model (SCAM) is shown in Figure 1 and each CA is represented as a single dimension array consisting of a unique identifier (ID) and eight parameters, four relating to number of neurons and four to elapsed time. We have not modelled the overshoot (f) because we have no idea as to its function, if it has one. Also, because so little is known or even theorised about background levels, priming and decay, this part of the SCAM is simplified to two parameters (P50% and D50%).

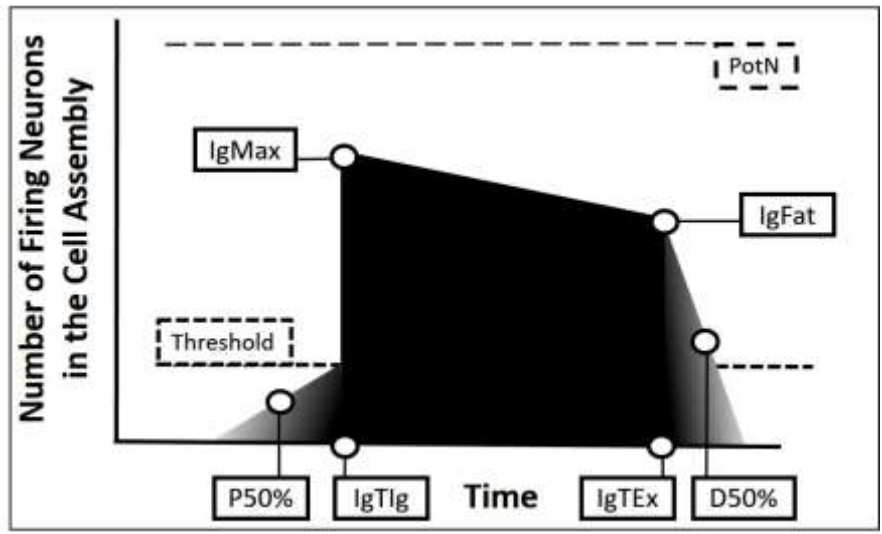

*Figure 1 The SCAM diagram. The four lower parameters are measures of time and the four floating ones are measures of neuron numbers.*

The four SCAM parameters associated with number of neurons are:

**PotN** – the **potential** number of **neurons** that could have membership of the CA;

**Thresh** – the **threshold** at which there are sufficient neurons firing to cause CA ignition;

**IgMax** – the **maximum** number of neurons that fire at CA **ignition**;

**IgFat** – the number of firing neurons after neuron **fatigue** at the end of CA **ignition**, i.e. at CA extinction.

N.B. In some cases IgFat may equal Thresh, in which case the CA will then decay, but in other cases the CA may be supressed so that at CA extinction IgFat > Thresh, as shown in the SCAM diagram.

The four SCAM parameters associated with time are:

**P50% -** the time at which a CA is **primed** to **50%** of the neurons firing that are required to reach its ignition threshold;

**IgTIg** – an **ignited** CA's **time** of **ignition**;

**IgTEx** – an **ignited** CA's **time** of **extinction**;

**D50%** – the time at which a CA **decays** to **50%** of the neurons that were firing at CA extinction (IgFat).

Even within the limited demonstration analysis, across CAs there is a considerable range of shapes to the SCAM diagrams and each of the eight parameters have some variation. This is desirable and if it were not so then a parameter could be treated as a constant.

For each CA identified, values for each parameter have to be estimated and while this is relatively straightforward from observational data for the four time parameters, those associated with the number of neurons may be wild guestimates. Far too little is known about brain CAs and the guestimates may be in error by an order of magnitude or two. On the other hand, generally the expectation is that errors will be consistent, so subsequent corrections based on new research might fix such errors by multiplying by a simple equation, or even just a constant. Explanations for choosing parameters for individual CAs are included in the main analysis (Appendix I).

While a crucial analysis component, with practice the SCAM diagrams became quite easy to visualise and their main representation during analysis was in the SCAM table.

### .2.2 The SCAM Table

Each identified CA is represented as a line in the SCAM table using the CA's unique identifier and the eight SCAM parameters. Table 1 shows the first few lines of the main analysis.

| No. | ID | PotN | Thresh | IgMax | IgFat | P50% | IgTIg | IgTEx | D50% | ID Acronym |
|---|---|---|---|---|---|---|---|---|---|---|
| 01 | CKEC | 10 | 2 | 7 | 6 | -1.0 | 0.0 | 0.4 | 0.5 | COG Kitchen Entrance Check |
| 02 | VKEG | 20 | 10 | 15 | 14 | -0.8 | 0.1 | 0.3 | 0.4 | VIS Kitchen Entrance General |
| 03 | CMC | 5 | 1 | 2 | 1.5 | -1.0 | 0.4 | 2.5 | 4.0 | COG Make Coffee |

*Table 1 Example lines from the main analysis' SCAM table (Table 2).*

While perhaps not ergonomically optimised, once one can visualise SCAM diagrams then the SCAM table becomes one of the three critical analysis tools. For example, the task timeline as represented by the ignition and extinction of each CA (IgTIg and IgTEx, respectively) can be seen by simply running down the table's two columns for these parameters.

The SCAM table had many uses during analysis and was crucial for iteration during analysis and for maintaining consistency and for error checking. Such roles are particularly important because of the complexity of another main analysis representation, the Cell Assembly Architecture Relationship (CAAR) diagram.

### .2.3 The Cell Assembly Architecture Relationship (CAAR) Diagram

The tidiness of the CAAR diagram shown in the results of the main analysis (Figure 9) belies its origins, which were pages of handwritten scrappy notes and diagrams. The basic procedure was to identify the next potential, small set of CAs that together would represent a cognitive task step. The possible inputs would have been identified during analysis of CAs earlier in the task and then the relationship between the CAs being analysed would be worked out; finally, the possible outputs would be identified.

In the CAAR diagram each CA identified is represented by a box and the relationships between CAs, i.e. how one CA ignites, maintains or supresses another, and what information is passed between CAs during their ignition, are represented by arrows. CA priming and decay also need to be considered.

The CAAR diagram has elapsed task time, approximately within graphical constraints, increasing vertically downward. Horizontally, CAs are arranged by type, from left to right: Perceptual; Cognitive; and Motor. The perceptual CAs are further subdivided as being Visual, Touch and Kinaesthetic ones  These CA types always represent the first character of each CA's ID, i.e. V/T/K, C, or M.

Figure 2 shows a generic CAAR diagram. It is the template for the pattern that was the most commonly found in the main analysis (section 4.3).

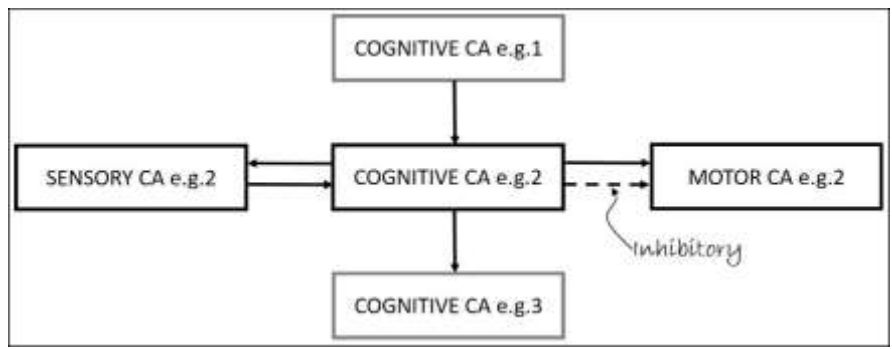

*Figure 2 Generic CAAR Diagram.*

In its present form the CAAR diagram shows only a limited amount of the information that it could contain. Iteration between CAs, e.g. where each of a pair is helping to maintain the other, is a critical property that is shown in neither diagrammatic or tabular representation; in the rough, hand drawn diagrams multiple arrow heads were used to show such iteration. Further possible refinements are left to the Discussion (section 5.3.1), although the reason why the CAAR acronym includes "Architecture" is that it is all the considered but currently unrepresented aspects of each CA, and how it relates to others that is architectural and potentially puts it beyond just a set of 1960s cognitive psychology style boxes and arrows. Description and explanation of many of these factors is included in the text associated with each CA in the full main analysis (Appendix I). Further description of the TACAP analysis techniques' method is in section 3.2.

## .3    The 'First Steps to Making Coffee' Example

It all started as a very quick investigation after the inspiration to join the TA and CA concepts. After a few days it became clear that the whole TACAP analysis enterprise would require considerable, long term, effort. There were weeks of trial and error as everything from the basic concepts, the notations, and the graphics had to be worked out. For example, at least half a dozen diagram styles were tried before developing the SCAM diagrams used in this paper; there were similar graphical problems with the CAAR diagram; and the SCAM table had to be reformatted a number of times as the eight SCAM parameters were themselves developed. In the end, just nine seconds of expert task behaviour was analysed, and it takes over sixty CAs to do so.

## .3.1  Task Selection and Data Collection Method

We appreciate the view of those who think that never again should there be a TA paper that uses the making_a_hot_beverage example. In our defence, the CA perspective is novel, so, on practical grounds, it is reasonable to choose an extremely familiar task, indeed, probably the one most commonly chosen to introduce students to TA. Furthermore, the demonstration analysis, in time, is rather short, so there is not much coffee making to worry the cognoscenti.

Data was collected using a repeated trials, self-observation, post sub-task recording, heuristic approach, i.e. the first author, who is a TA expert (hence "heuristic"), watched himself, many times (more than thirty) doing the first part of his making coffee routine and then making written notes after each trial. During some of the trials timing data (to the nearest 0.1 seconds) was recorded at two points during the task using a small mobile phone's digital stopwatch held in the left hand (section 4.1). The initial observations took place over several days and additional trials were done over the following month during the first stages of the main analysis.

Against any objections to this heuristic method, there are a number of advantages for what, we keep stressing, is only a demonstration of a possible analysis technique and not a new TA method. First, the task is a very highly practiced one, with a history of over 20 years in the current house and unchanged after about half a dozen years since the kitchen was remodelled. Second, its nigh invariant repeatability allowed access to renewed observations when they were needed, and they were. For example, the subject was unaware and failed to initially record what happened to the left hand while the kettle, grasped in the right, was moved to the sink for filling. Third, the subject was already expert at such self-observation because, using his TA expertise, he continuously works at prosaic task optimisation, ideally with an end result that he can continue to think of other things while performing the common and mundane. Thus, the data collection approach adopted provided high quality data, indeed, much higher quality than from most TAs that involve analysts recording the performance of other people.

As a further defence, the subject-analyst discovered new details of how he performed the task of which he was previously unaware, for example, the pattern of steps taken outside and across the kitchen (section 3.3.2), as well as the example of the empty left hand's actions mentioned above. At the least, this demonstrates that a TA was done and that the demonstration is not based merely on a desktop, thought-experiment exercise.

The three other residents in the house were also observed doing the same kettle filling task (see Appendix I: CA 06 MSHWA – Motor Stride to Hot Water Area) which, at least, demonstrated that a more traditional TA with subject and analyst separate was feasible.

## .3.2  Analysis Method

The AL resulting from the data collected was very simple and along the lines: enter kitchen; go to hot drinks preparation area; grip kettle in right hand; move kettle over sink; remove kettle lid with left hand; invert kettle to empty it; replace lid with left hand; move kettle under water spout; fill kettle. As soon as the analysis started, each such AL line was rapidly elaborated, often supported by further task observations, and soon the ordered list of identified CAs effectively became the AL used for further analysis.

At this early stage of research it is not feasible to provide a detailed method for the TACAP analysis technique, and it is undesirable to do so. Method specification in TA is extremely difficult, to the extent that Diaper (2001) suggests that it is necessary to develop analyst support software to support method specification. While such tools' superficial, primary function is to make analysis easier and less error prone, to teach and guide a neophyte analyst requires supporting software to have an explicit and detailed model of the method. The discipline required of programming means ready identification of missing and, much more frequently, underspecified parts of a method, which expert analysts bridge using their craft skills, often without being aware they are doing so. Indeed, HTA is often described as a "methodology" and its massive under-specification is seen as an advantage, for the experts who have served their apprenticeship.

Once the TACAP analysis technique settled down during the latter two thirds of the analysis, it was all done online, indeed, as if there was software tool support and with the analyst having the role for the desirable but missing program code. Of course there was a lot of printing for off-line checking and editing, but during analysis the only paper was a couple of very scrappy sheets with a hand drawn version of the CAAR diagram, and a lot of annotations, crossings out, etc. On-screen, centred was the main analysis document (Word); to the left was the SCAM table (Word) and to the right the CAAR diagram (PowerPoint). The acronym glossary (Appendix II) was also always available. Usually, a small set of CAs would be analysed as a group, the most common pattern being that shown in Figure 2 (Section 2.3).

The first step was to create an entry for each CA in the main document and to copy and paste (to minimise typographic errors) the ID and spelled out acronym into the glossary, and the ID into the SCAM table. As an example of cognitive architecture, the default is that at least one, already analysed, input will go to the new cognitive CA. There may be more than one known, analysed input, and during analysis, occasionally, there is a floating output, where an earlier CA must have this, but the analyst was not sure of its still unanalysed destination CA. Note, it is only a default, but with the advantage that exceptions, and there are some, are bought to the analyst's attention for especial consideration. It is an example of architecture in that other defaults could have been chosen, for example, making perceptual CAs the default and have some sort of Perception – Cognition – Motor cycle or left–then–right scan, i.e. P_C_M_P_C… or P_C_M_C_P_C…, respectively. The TACAP default model is more of a tree with C usually mediating between P and M, i.e. C_P_C_P… & C_M_C_M… . There are positive and negative arguments for any of these architectures, but they are all only defaults and the analysis allows alternatives, for example when perceptual and motor systems become tightly bound in some expert behaviours (section 4.3).

Once the new CA set's inputs have been cut and pasted to the tabular entries in the main document, then each CA is described as text (Appendix I) and the relationships between the CAs are added during writing the text, i.e. when a CA has an output to another member in the set being analysed, then the output is copy and pasted as input to the appropriate CA. This is just the sort of thing an analyst's support tool would do automatically. Also, while writing the text, the SCAM table is gradually filled in. In most cases the values assigned to entries in the SCAM table are explained in the Appendix I text, while attempting to avoid too much repetition. The order in which data was entered to the SCAM table was driven by the linear sequencing of the natural language text. After the first few CAs were analysed the SCAM diagrams were not drawn simply because the analyst could visualise them from the SCAM table and each diagram took quite some time to produce, which would have interfered with the main analysis processes; a trivial software tool is needed to draw the SCAM diagrams automatically from the table.

At the end of a CA's writing process, the outputs to yet unanalysed CAs will be entered. This text will be what is copy and pasted when it is the turn of these CAs to be analysed. This led to inserting some new IDs in the SCAM table ahead of their analysis.

A further feature of the Input/Output tabular specifications in the main analysis (Appendix I) is their punctuation. No punctuation between lines means that the two inputs or outputs occur in parallel and increasing punctuation strength, i.e. comma, and though rarely used in the main analysis (Appendix I), semicolon and colon, show increasing separation in time; a full stop indicates the termination of one input or output before the start of another, although both are within the main analysis' description of a particular CA. Checking the punctuation at the end of analysing a set of CAs was an important part of the error checking routines.

Unlike the SCAM diagrams, it was found important to regularly update the CAAR diagram during analysis. This was no simple transposition from its very rough paper representation to its accurate computer version. The CAAR diagram is a triumph of graphic design in that it shows over sixty CAs and their relationships in way that can be printed on a single sheet of A4 paper, without sacrificing readability. Many designs were tried and some of the earliest would have needed a dozen or so pages rather than just one. Furthermore, because it was prepared in PowerPoint, the analyst's default graphical editor for decades, it is actually quite easy to animate the diagram (Appendix III). This is returned to in the Discussion (sections 5.1 and 5.3.1).

On the other hand, using PowerPoint was a bit of a pig, even for a real expert, as the small scale pushed PowerPoint's resolution when drawing the arrows. It was essential to keep the CAAR diagram up to date, no matter that it was time consuming to do. When there was iteration in the analysis, returning and modifying CAs already analysed, then the CAAR diagram, the SCAM table and the main text's tabular specifications were always changed together. Usually, changing one analysed CA resulted in changing other ones as well.

The method adopted was designed to minimise error and facilitate error checking, e.g. every output must have its input, in the architecture, to another CA, which shows one of the chosen simplifications, not modelling the internal processes of a CA (section 2.1). It is necessary to check that every CA is correctly represented in each of the four main representations: the main analysis document, the CAAR diagram, the SCAM table and the SCAM diagrams. Especial care needs taking where previously analysed CAs have been changed by dividing or combining them as this will likely to have changed their IDs, which is the key identifier in all the main representations. The acronym glossary (Appendix II) was only updated occasionally once the analyst had learned his own acronymic CA IDs, and he used them all the time when reasoning about relationships between CAs.

*.3.3 Analysis Introduction*

Subsections 3.3.1 and 3.3.2 are intended to provide an introduction to the task and a flavour of the style used in the full main analysis in Appendix I. Subsection 3.3.3 contains a strong recommendation to readers that, before they read the results in section 4, that they familiarise themselves with some of main analysis in Appendix I and with the main representations used.

*.3.3.1 The Coffee Making Decision*

Prior to the start of the analysis in the kitchen, the subject has made the decision to make a small mug of coffee. This decision could be based on many things, from habit or time since last coffee, or thirst or other dehydration indicators, or just the need for a break, and so forth. Numerous CAs will have been involved in making this decision, but a critical issue is what one or more cognitive CAs are primed or already ignited at the kitchen's entrance. There may be intervening activities so that the time from making the decision to arriving at the kitchen entrance might be five or more minutes.

One possible model would involve the decision making CAs igniting a coffee making one that would persist until task completion. One could even suggest that this CA would contain a plan of what is involved in making a small mug of coffee. There is some evidence that this model is not that plausible. First, with intervening tasks then such a CA would have to persist, ignited, while many other CAs are deployed. Furthermore, the make coffee CA might just be part of a list of tasks to complete and such a dynamic task list CA would have complex behaviours as tasks are completed and, sometimes, the list order might be shuffled, some tasks deleted or postponed, and so forth. Note, arguments involving consciousness are weak to irrelevant, e.g. that people do not perform loads of intervening tasks while thinking "must make a coffee, must make a coffee, must …".

At a minimum, when the coffee making decision is made then a 'Make Coffee' CA must be ignited as a record of the decision. This CA can be of modest size as the decision record and, if one chooses, one could call it a "goal". There is evidence that this CA does not remain ignited in the widely reported phenomenon of one going to a room and then realising one cannot remember why one went there, i.e. the CA fails to reignite in its now appropriate context.

In the analysis that follows, the assumption is that the CA 'Make Coffee' has been previously ignited and remains sufficiently primed that it will reignite with suitable environmental input, e.g. from vision. The analysis starts at the kitchen entrance and the evidence suggests that the host of go-to-the-kitchen CAs that brought the subject to this spot all close down. This is suggested by the final kitchen entrance approach behaviour described in the next subsection.

*.3.3.2 Before the Kitchen's Entrance*

Before the kitchen entrance there is a shuffle zone. The following observations are a direct consequence of the research reported in this paper. The kitchen entrance has no door and there are four routes to arrive at the entrance, from North West to South East withershins respectively: corridor, stairs, lean-to, and lounge (Figure 3). Whichever route the subject takes to the kitchen entrance, he always arrives with his right foot planted in the centre of the kitchen entrance, that foot may be over the entrance's low, wooden floor bar, or the whole foot, up to a couple of centimetres clear, in front of or behind the bar (Figure 4B), but the right foot is always aligned at a right angle to the entrance's bar and at the centre of the entrance. Indeed, experiments requiring the left foot to be the kitchen entering step result in noticeably clumsy initial steps within the kitchen and the body, moving at a reasonable domestic speed, is unbalanced (e.g. balancing arm movements, hip and upper body twists and similar ergonomic inefficiencies). The right foot entrance is achieved by a shuffle in the area outside the kitchen, particularly easy to observe as, when necessary, a half step will be taken when coming down the corridor, and also, after descending the stairs, where although either foot may have started at the top, steps are adjusted in the shuffle zone. The shuffle zone is less clear from the lean-to because usually its door is closed before taking steps towards the kitchen entrance, but rationally a shuffle must exist because the right foot is inevitably correctly placed, as it is from the lounge, which requires a complex, short curved route of about 130 degrees so shuffling is again less easily observed.

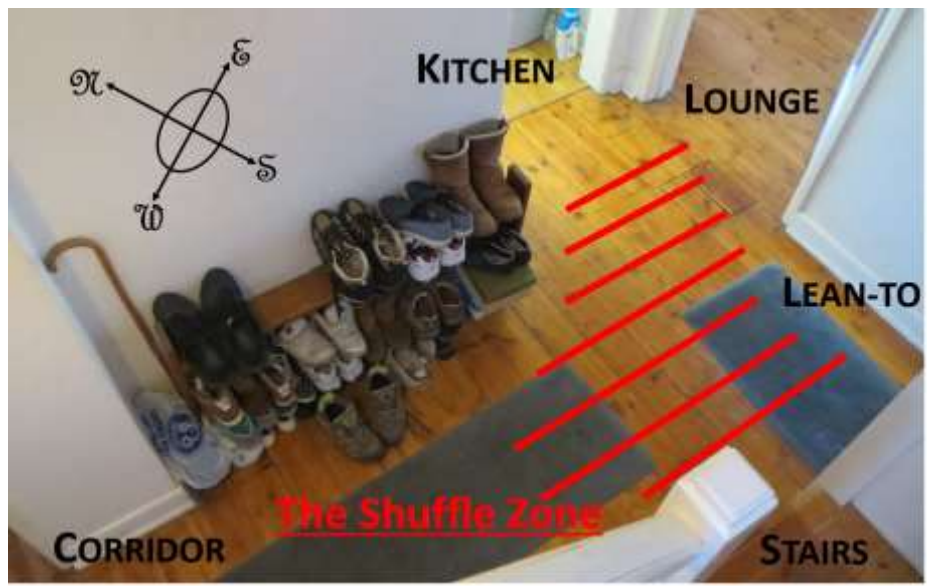

*Figure 3– The "Shuffle Zone" outside the kitchen entrance.*

### 3.3.3 Further Context

Rarely is "a picture worth a thousand words", which is, say, well into three typed sheets of A4. To cater for a divers readership, however, what is offered a quick, photographic story, hopefully, to help both task visualisation and comprehension. Just a bit from the first few seconds …

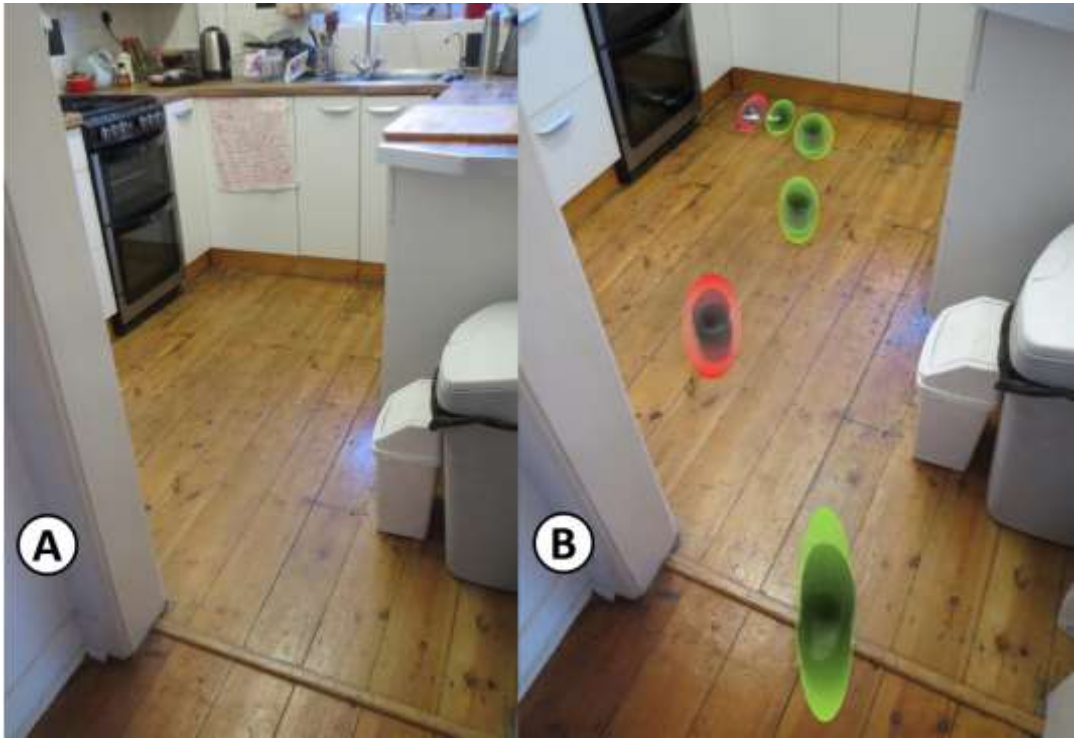

*Figure 4 (A) The kitchen entrance; (B) The "strides" across the kitchen: right foot in green; left in red.*

These photographs were taken opportunistically and the kitchen is "as found", without any prior preparation, or any tidying. Figure 4A shows the kitchen's ground geography, for illustration, but note the top of the photograph and the important context and focus of visual attention, already getting ready for kettle identification.

Figure 4B shows the "invariant" strides from the entrance to the hot water preparation area (Appendix I, CA 04 CAHWA to CA 06 MSHWA). The left foot, in red, takes the first and third strides and on the photograph the precision of foot placement is roughly represented by the shading. The right foot (shown in green) launches the strides and, from the shuffle zone (section 3.3.2), the foot may be before or over the bar on the entrance's floor. The next right foot stride is fairly precisely placed but with the left foot very accurately and correctly located, the right foot then makes a forward and then curving motion to locate the feet closely adjacent and, concomitantly, the whole body, well balanced in a tight corner space, where it is expertly placed. The visual and cognitive systems, however, are primarily concerned with the hot drinks preparation area, and how to pick up the kettle.

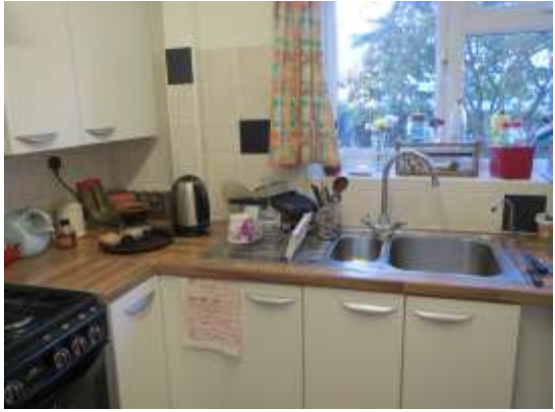 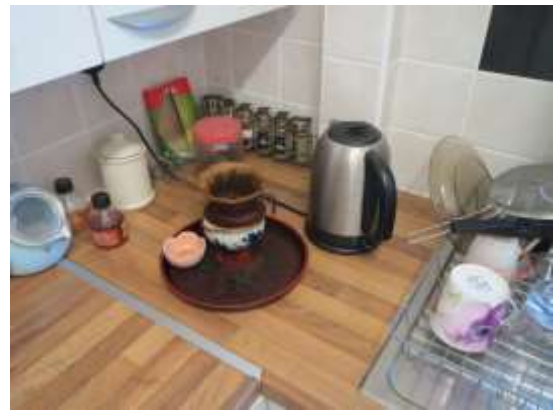

*Figure 5 General view of the kitchen.*     *Figure 6 View of the hot water preparation area.*

Figure 5 shows the general view of the kitchen, say about midstride on the right foot (see above).  The target is the kettle, but there are potential obstacles to its left and right.  The coffee filter cone to the left is where it usually is, but the draining board to the right often presents novel problems when not empty.

Figure 6 shows the view once at the hot water preparation area.  Binocular vision is an asset here, for detecting that the steel sieve handle to the right of the kettle is in front of it; and there is a lot of leftward lean on the translucent plate.

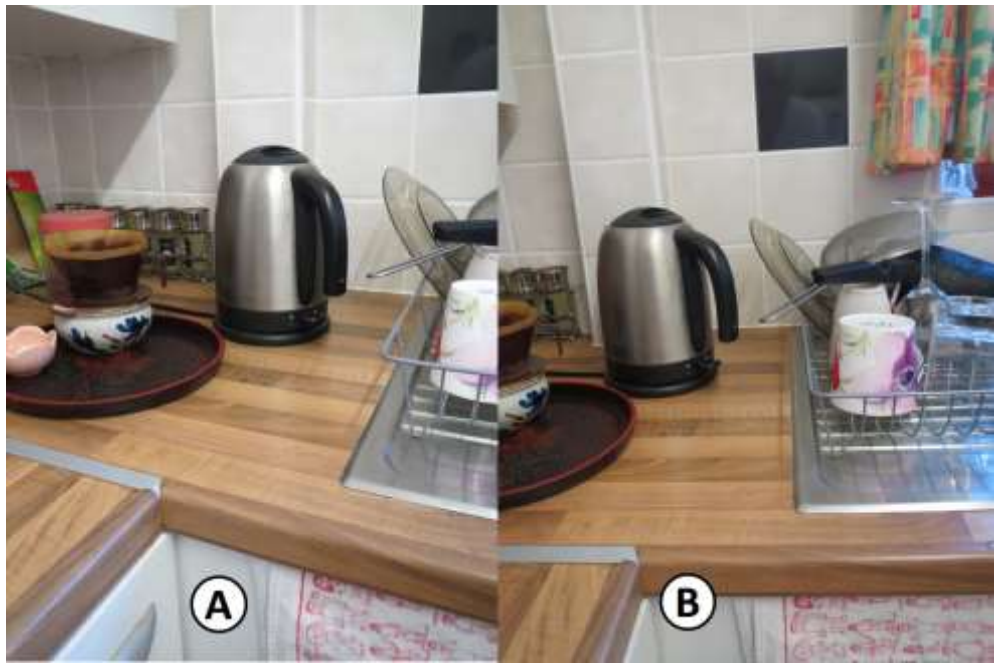

*Figure 7  The views from the hands' locations approaching the hot water preparation area: (A) right hand; (B) left hand.*

Fifty centimetres or so below the eyes, the view from the hands is rather different, and Figure 7 presents the start of the "flight path" views: 7A shows about where the right hand starts its final approach to the kettle and what it has to navigate (obviously some climb is essential); 7B shows the left hand's "view" and its target will later be somewhere around the black tile, catching up with the top of the kettle after it has been lifted (Appendix I: CA 33 MLHTKL).

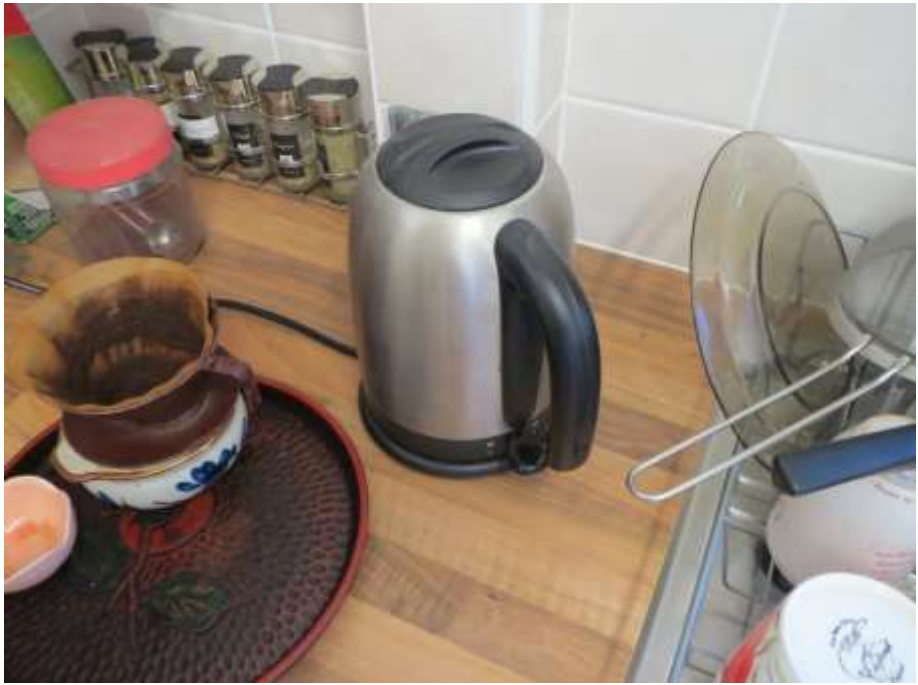

*Figure 8  View of the target kettle in the hot water preparation area.*

Moving the right hand to the, exactly identified, kettle handle without error, i.e. with no contact with any other objects, and, also, smoothly, curvaceously, etcetera, is a behavioural triumph. At this range the angle between the point of view in Figure 8 and the right hand's flight path (Figure 7A) is, in computational terms, impressive, massive, etc. On the other hand, it is just what CAs are so neat at describing, explaining and, even, is expected of them because they are flexible and capable, by themselves, of learning. The right hand is under visual negative feedback control, but it is typical of expert performance that only little control compensation is required from the planned motor output (N.B. this "planned" output in CA terms is just the initially ignited CA that, while ignited, evolves with sensory feedback, and other relevant inputs, and, perhaps its own temporal structure, i.e. as a process – see Introduction).

*3.3.4  The Main Analysis*

It is only for reasons of space that the main analysis is Appendix I and none of it is here in the main body of the paper. Section 4's "Results" are a high level description of the analysis, but in one sense the real results reported in the paper is the main analysis itself.

A completely new analysis technique has been developed and to understand the paper it is necessary for readers to have some understanding of the technique in application and the issues that were considered when assigning parameters to the SCAM table and relationships in the CAAR diagram. The issues considered include various psychological aspects and some basic neuroscience because the SCAM table, and the whole analysis, like other TAs, is performer centred and so the estimates of CA properties, size and so forth, relate to the human brain and not to possible ANN CA implementations (section 5.3.2). Although, as stated in the Introduction, if the estimates are in error by even a couple of orders of magnitude, then at this stage we are not at all concerned; it could be easily corrected by further research.

The main results in Appendix I contains graphical, tabular and textual descriptions of over sixty CAs. First time readers are strongly recommended to examine the first few CA descriptions (the fourth 'Cognitive Approach Hot Water Area (CAHWA)', is where the analysis starts to settle down after the first few analysed task steps). The initial descriptions tend to be longer and more descriptive and later ones rather briefer; and somethings are not repeatedly mentioned.

It is essential to consider the main analysis in Appendix I in conjunction with the SCAM table and the CAAR diagram, which are produced below in Table 2 and Figure 9.

| No. | ID | PotN | Thresh | IgMax | IgFat | P50% | IgTIg | IgTEx | D50% | ID Acronym |
|---|---|---|---|---|---|---|---|---|---|---|
| 01 | CKEC | 10 | 2 | 7 | 6 | -1.0 | 0.0 | 0.4 | 0.5 | COG Kitchen Entrance Check |
| 02 | VKEG | 20 | 10 | 15 | 14 | -0.8 | 0.1 | 0.3 | 0.4 | VIS Kitchen Entrance General |
| 03 | CMC | 5 | 1 | 2 | 1.5 | -1.0 | 0.4 | 2.5 | 4.0 | COG Make Coffee |
| 04 | CAHWA | 10 | 2 | 5 | 3 | 0.5 | 0.6 | 3.1 | 3.2 | COG Approaching Hot Water Area |
| 05 | VAHWA | 20 | 2 | 10 | 6 | 0.6 | 0.7 | 2.5 | 2.6 | VIS Approaching Hot Water Area |
| 06 | MSHWA | 10 | 2 | 7 | 6 | 0.6 | 0.7 | 3.0 | 3.1 | MOT Stride to Hot Water Area |
| 07 | CKHWA | 10 | 3 | 7 | 6 | 0.8 | 1.0 | 2.1 | 2.2 | COG Kettle Hot Water Area |
| 08 | VKHWA | 20 | 5 | 10 | 9 | 1.2 | 1.3 | 2.0 | 2.1 | VIS Kettle Hot Water Area |
| 09 | CKH | 5 | 1 | 3 | 2 | 1.5 | 1.6 | 3.5 | 3.6 | COG Kettle Handle |
| 10 | VKH | 10 | 3 | 7 | 6 | 1.6 | 1.8 | 3.3 | 3.4 | VIS Kettle Handle |
| 11 | MRAB | 5 | 1 | 2 | 2 | 1.9 | 2.0 | 2.1 | 2.2 | MOT Right Arm Ballistic |
| 12 | VRH | 15 | 2 | 5 | 4 | 2.0 | 2.1 | 3.2 | 3.3 | VIS Right hand |
| 13 | CRH | 12 | 3 | 7 | 6 | 2.1 | 2.2 | 3.4 | 3.5 | COG Right hand |
| 14 | CHWA | 15 | 5 | 10 | 8 | 2.2 | 2.4 | 3.5 | 3.7 | COG Hot water Area |
| 15 | CRHA | 25 | 5 | 15 | 12 | 2.3 | 2.5 | 3.6 | 3.7 | COG Right Hand Approach |
| 16 | VRHA | 25 | 10 | 15 | 14 | 2.3 | 2.6 | 3.3 | 3.4 | VIS Right Hand Approach |
| 17 | MRHA | 10 | 2 | 7 | 6 | 2.4 | 2.7 | 3.7 | 3.8 | MOT Right Hand Approach |
| 18 | TRHKH | 5 | 2 | 3 | 2 | 3.0 | 3.5 | 3.8 | 3.9 | TOU Right Hand to Kettle Handle |
| 19 | CRHG | 5 | 2 | 3 | 2 | 3.2 | 3.7 | 3.8 | 4.2 | COG Right Hand Grip |
| 20 | MRHG | 5 | 1 | 3 | 2 | 3.7 | 3.8 | 3.9 | 4.0 | MOT Right Hand Grip |
| 21 | TRHG | 5 | 1 | 3 | 2 | 3.7 | 3.8 | 3.9 | 4.3 | TOU Right Hand Grip |
| 22 | CRHH | 10 | 2 | 5 | 5 | 3.8 | 4.0 | - | - | COG Right Hand Hold |
| 23 | MRHH | 10 | 2 | 3 | 3 | 3.9 | 4.1 | - | - | MOT Right Hand Hold |
| 24 | CLK | 10 | 3 | 6 | 5 | 4.0 | 4.2 | 4.7 | 4.8 | COG Lift Kettle |
| 25 | MLK | 5 | 1 | 3 | 2 | 4.1 | 4.3 | 4.4 | 4.5 | MOT Lift Kettle |
| 26 | KKW | 5 | 1 | 3 | 3 | 4.2 | 4.4 | 4.5 | 4.6 | KIN Kettle Weight |
| 27 | VLK | 10 | 3 | 6 | 5 | 4.3 | 4.5 | 4.6 | 4.7 | VIS Lift Kettle |
| 28 | CD | 15 | 5 | 8 | 6 | 4.5 | 4.6 | 6.0 | 6.1 | COG Drainer |
| 29 | VD | 25 | 8 | 15 | 13 | 4.6 | 4.7 | 5.5 | 5.8 | VIS Drainer |
| 30 | CMKS | 25 | 5 | 15 | 12 | 4.7 | 4.8 | 6.6 | 6.7 | COG Move Kettle Sink |
| 31 | VMKS | 15 | 5 | 10 | 9 | 4.8 | 4.9 | 6.5 | 6.6 | VIS Move Kettle Sink |
| 32 | MMKS | 20 | 5 | 10 | 9 | 4.9 | 5.0 | 6.5 | 6.6 | MOT Move Kettle Sink |
| 33 | MLHTKL | 15 | 3 | 9 | 6 | 5.0 | 5.1 | 7.0 | 7.0 | MOT Left Hand Track Kettle Lid |
| 34 | KLHTKL | 10 | 2 | 6 | 5 | 5.1 | 5.2 | 7.8 | 7.8 | KIN Left Hand Track Kettle Lid |
| 35 | MSBS | 10 | 5 | 7 | 6 | 5.1 | 5.3 | 6.9 | 7.0 | MOT Shuffle Body Sink |
| 36 | CS | 5 | 2 | 4 | 3 | 6.5 | 6.7 | - | - | COG Sink |
| 37 | VS | 10 | 5 | 7 | 6 | 6.6 | 6.8 | - | - | VIS Sink |
| 38 | CLHRKL | 5 | 1 | 4 | 3 | 6.8 | 6.9 | 7.2 | 7.3 | COG Left Hand Remove Kettle Lid |

| | | | | | | | | | |
|---|---|---|---|---|---|---|---|---|---|
| 39 | VKL | 10 | 5 | 7 | 6 | 6.9 | 7.0 | 7.1 | 7.2 | VIS Kettle Lid |
| 40 | VLH | 10 | 5 | 7 | 6 | 6.9 | 7.0 | 7.1 | 7.2 | VIS Left Hand |
| 41 | MLHRKL | 7 | 2 | 6 | 5 | 7.0 | 7.1 | 7.7 | 7.7 | MOT Left Hand Remove Kettle Lid |
| 42 | VKWL | 10 | 5 | 7 | 6 | 7.1 | 7.2 | 7.3 | 7.4 | VIS Kettle Without Lid |
| 43 | CEK | 5 | 1 | 4 | 3 | 7.1 | 7.2 | 7.4 | 7.5 | COG Empty Kettle |
| 44 | MRHIK | 3 | 1 | 2 | 2 | 7.2 | 7.3 | 7.4 | 7.4 | MOT Right Hand Invert Kettle |
| 45 | VKE | 10 | 3 | 5 | 5 | 7.3 | 7.4 | 7.5 | 7.6 | VIS Kettle Empty |
| 46 | CKE | 3 | 1 | 2 | 2 | 7.4 | 7.5 | 7.6 | 7.6 | COG Kettle Empty |
| 47 | CRHOK | 5 | 1 | 4 | 3 | 7.5 | 7.6 | 7.8 | 7.9 | COG Right Hand Orientate Kettle |
| 48 | VRHOK | 10 | 5 | 7 | 6 | 7.5 | 7.6 | 7.9 | 8.0 | VIS Right Hand Orientate Kettle |
| 49 | MRHOK | 3 | 1 | 2 | 2 | 7.6 | 7.7 | 7.8 | 7.9 | MOT Right Hand Orientate Kettle |
| 50 | CRKLLH | 8 | 3 | 6 | 5 | 7.8 | 7.9 | 8.2 | 8.3 | COG Replace Kettle Lid Left Hand |
| 51 | VRKLLH | 10 | 5 | 7 | 6 | 7.8 | 7.9 | 8.2 | 8.3 | VIS Replace Kettle Lid Left Hand |
| 52 | MRKLLH | 10 | 3 | 7 | 6 | 7.9 | 8.0 | 8.1 | 8.2 | MOT Remove Kettle Lid Left Hand |
| 53 | CMKT | 15 | 5 | 10 | 9 | 8.1 | 8.2 | - | - | COG Move Kettle Tap |
| 54 | VT | 10 | 3 | 5 | 5 | 8.2 | 8.3 | 8.6 | 8.7 | VIS Tap |
| 55 | VK | 15 | 5 | 8 | 7 | 8.2 | 8.3 | - | - | VIS Kettle |
| 56 | MMKT | 15 | 5 | 10 | 8 | 8.3 | 8.4 | 8.6 | 8.6 | MOT Move Kettle Tap |
| 57 | MHKT | 6 | 1 | 3 | 3 | 8.4 | 8.5 | - | - | MOT Hold Kettle Tap |
| 58 | CMLHTS | 15 | 7 | 10 | 8 | 8.3 | 8.5 | 8.9 | 9.0 | COG Move Left Hand Tap Switch |
| 59 | VLHTS | 20 | 5 | 10 | 7 | 8.5 | 8.6 | - | - | VIS Left Hand to Tap Switch |
| 60 | VTS | 10 | 5 | 7 | 6 | 8.6 | 8.7 | - | - | VIS Tap Switch |
| 61 | MMLHTS | 15 | 5 | 8 | 7 | 8.7 | 8.7 | 8.9 | 9.0 | MOT Move Left Hand Tap Switch |
| 62 | TLHTS | 8 | 2 | 6 | 5 | 8.7 | 8.8 | - | - | TOU Left Hand Tap Switch |
| 63 | CFK | 10 | 3 | 7 | 6 | 8.8 | 8.9 | - | - | COG Fill Kettle |
| 64 | MPTSU | 5 | 1 | 3 | 3 | 8.9 | 9.0 | - | - | MOT Pull Tap Switch Up |
| 65 | CMC … | | | | | | | | | COG Make Coffee |

*Table 2 – The SCAM table.*

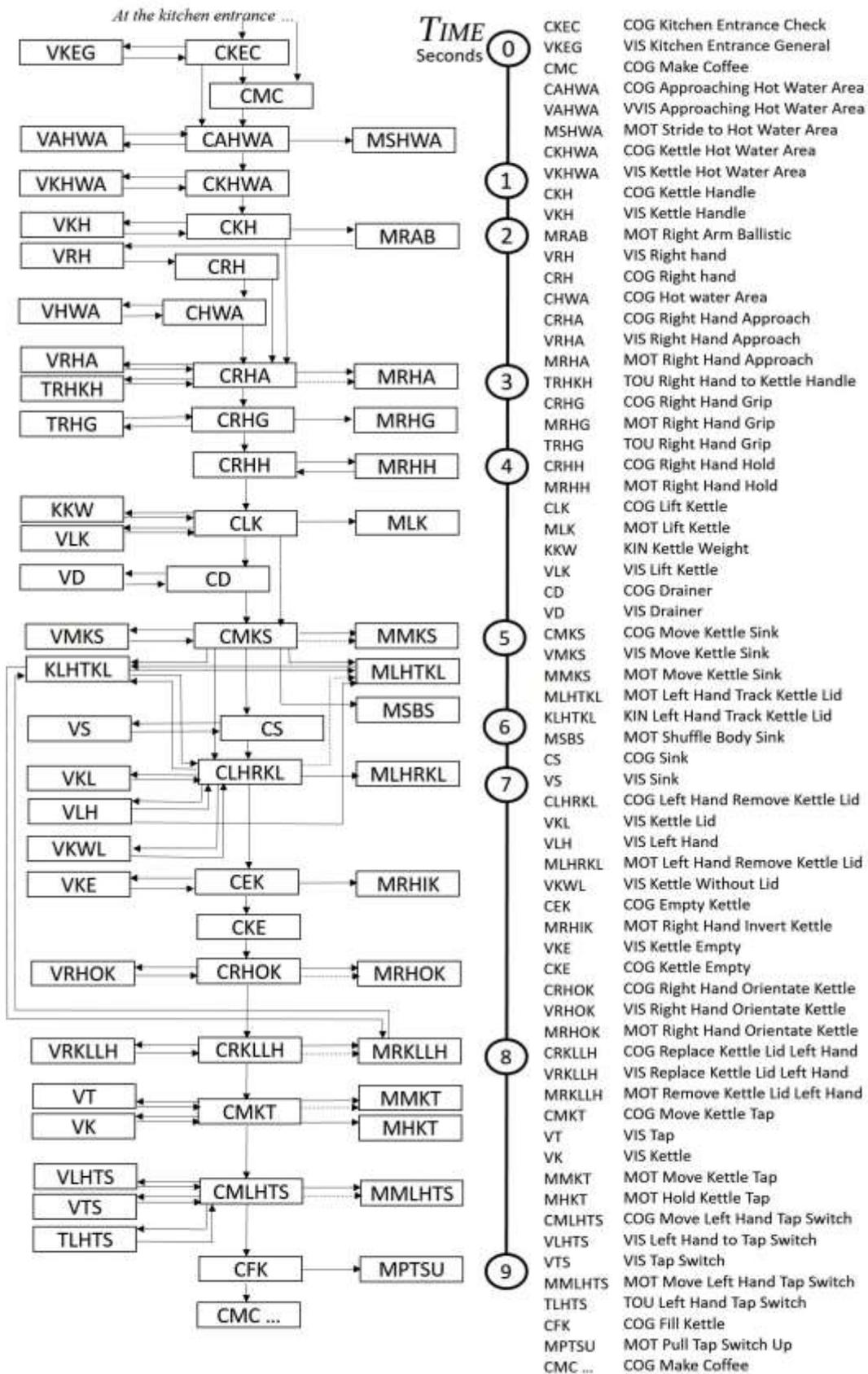

*Figure 9– The CAAR Diagram.*

## .4 Results

Everything in this results section is potentially nothing more than analyst artefacts. What these results present are the consequences of the decisions made by the analyst at a lower level of analysis, i.e. these results are the collective description of applying the TACAP analysis technique. Furthermore, and particularly because the analysis was iterative and decision consistency was a primary concern, then patterns in the data presented here have sometimes been deliberately imposed during analysis. For example, thresholds will tend to be larger with the larger CAs (PotN) so any correlation between the two is deliberate and therefore rather uninteresting.

On the other hand, at the very least these results demonstrate that the analysis has been applied in a tidy and consistent manner. They also give an insight into the detail and complexity of analysing at the low levels chosen, and hint at what more complete and relevant task examples would require.

### .4.1 *Time Results*

Timing data to the nearest 0.1 seconds was collected over several days using the stopwatch function on a mobile 'phone. From the kitchen entrance, data was collected from two easily identified steps in the task: (i) when the kettle handle is gripped and ready for the kettle's lift from its base (CA 23 – MRHH); and (ii) at the end of the analysed task portion when the kettle starts to fill (CA 64 – MPTSU). According to the main analysis, these times were 4.1 seconds and 9 seconds, respectively.

Time data is nearly always a problem in TAs, as it was in this study. As illustration of TAs typical problems with time data, the first measure at MRHH had a recorded range of 3.3 – 4.2 seconds. The first problem is that a first opportunity sample would tend to be around 4 seconds but if repeated half a dozen times then the times would decrease to around the 3.5 second mark, i.e. even highly practiced performance improves with several goes at the same task. Secondly, if only first times are considered then there is still half a second of variability, much of which depends on the state of the drainer and the concomitant complexity of the right hand's flight path to the kettle handle (section 3.3.3).

Generally, time data is far less important than sequence data in most TAs and it is one more craft skill of analysts to give a single time estimate to each task step. The estimates in the main analysis are, in this tradition, mostly interpolated, approximately correct and on the higher side of the range of times recorded.

### .4.2 *SCAM Results*

There were 64 CAs identified in the main analysis: 34.4% (22/64) were cognitive; 39.1% (25/64) were perceptual; and 26.6% (17/64) motor. Of the perceptual CAs, 31.3% (20/64) were visual and there were 5 other perceptual CAs: 4.7% (3/64) touch and 3.1% (2/64) kinaesthetic.

The general, as an average (arithmetic mean), CA from the main analysis can be drawn, as can the SCAM models for the three main types of CA: cognitive, visual and motor. To do so, the five non-visual sensory CAs (3 x touch, 2 x kinaesthetic) and those CAs that are still ignited at the end of the analysis, were removed from the data, leaving 48 CAs on which the following analysis is based. Table 3 gives such average data.

| CA Type | PotN | Thresh | IgMax | IgFat | P50% | IgTIg | IgTEx | D50% |
|---|---|---|---|---|---|---|---|---|
| All | 11.1 | 3.3 | 6.8 | 5.7 | *4.5* | *4.7* | *5.4* | *5.5* |
| Cognitive | 10.4 | 2.8 | 6.6 | 5.2 | *3.8* | *4.1* | *5.0* | *5.2* |
| Visual | 14.4 | 5.0 | 8.6 | 7.5 | *4.5* | *4.7* | *5.3* | *5.4* |
| Motor | 9.5 | 2.6 | 5.9 | 4.9 | *5.3* | *5.4* | *6.1* | *6.2* |

*Table 3 Average data for the 8 SCAM parameters.*

The four time metrics (in italics in Table 3) require a little manipulation before they can be used to draw versions of the SCAM diagrams. The details of this are included below because they provide an example of suboptimal analysis technique design, which is addressed in the Discussion (section 5.3.1).

The time parameters (t0 – t3) in Table 4 are calculated to correspond to the start of a CA, i.e. t0 = 0.0 seconds, and the priming time to ignition (t1), the duration of the ignition until extinction (t2), and the decay to zero (t3).

Since P50% is the time at which there is 50% of the neurons firing to reach threshold, then, for graphical purposes, the simplified linear priming in the SCAM requires P50% to be doubled for the average time to ignition, after subtracting from the data's time of ignition (IgTIg), i.e.

$$t1 = (IgTIg - P50\%) \times 2$$

The time a CA is ignited (t2) is simply the difference between its extinction minus its ignition time, with the elapsed priming time added for graphical purposes, i.e.

$$t2 = (IgTEx - IgTIg) + t1$$

As with t1, the full elapsed decay time requires D50% to be doubled, after subtraction from the extinction time (IgTEx), and then the elapsed time to extinction (t2) needs adding, i.e.

$$t3 = ((D50\% - IgTEx) \times 2) + t2$$

Table 4 shows the data as used to represent the average SCAM diagrams.

| CA Type | PotN | Thresh | IgMax | IgFat | t0 | t1 | t2 | t3 |
|---|---|---|---|---|---|---|---|---|
| All | 11.1 | 3.3 | 6.8 | 5.7 | 0.0 | 0.4 | 1.1 | 1.3 |
| Cognitive | 10.4 | 2.8 | 6.6 | 5.2 | 0.0 | 0.6 | 1.5 | 1.9 |
| Visual | 14.4 | 5.0 | 8.6 | 7.5 | 0.0 | 0.4 | 1.0 | 1.2 |
| Motor | 9.5 | 2.6 | 5.9 | 4.9 | 0.0 | 0.2 | 0.9 | 1.1 |

*Table 4 Average data for the 8 SCAM parameters as used to draw the average SCAM diagrams in Figure 10.*

From Table 4 are derived the following four SCAM diagrams in Figure 10.

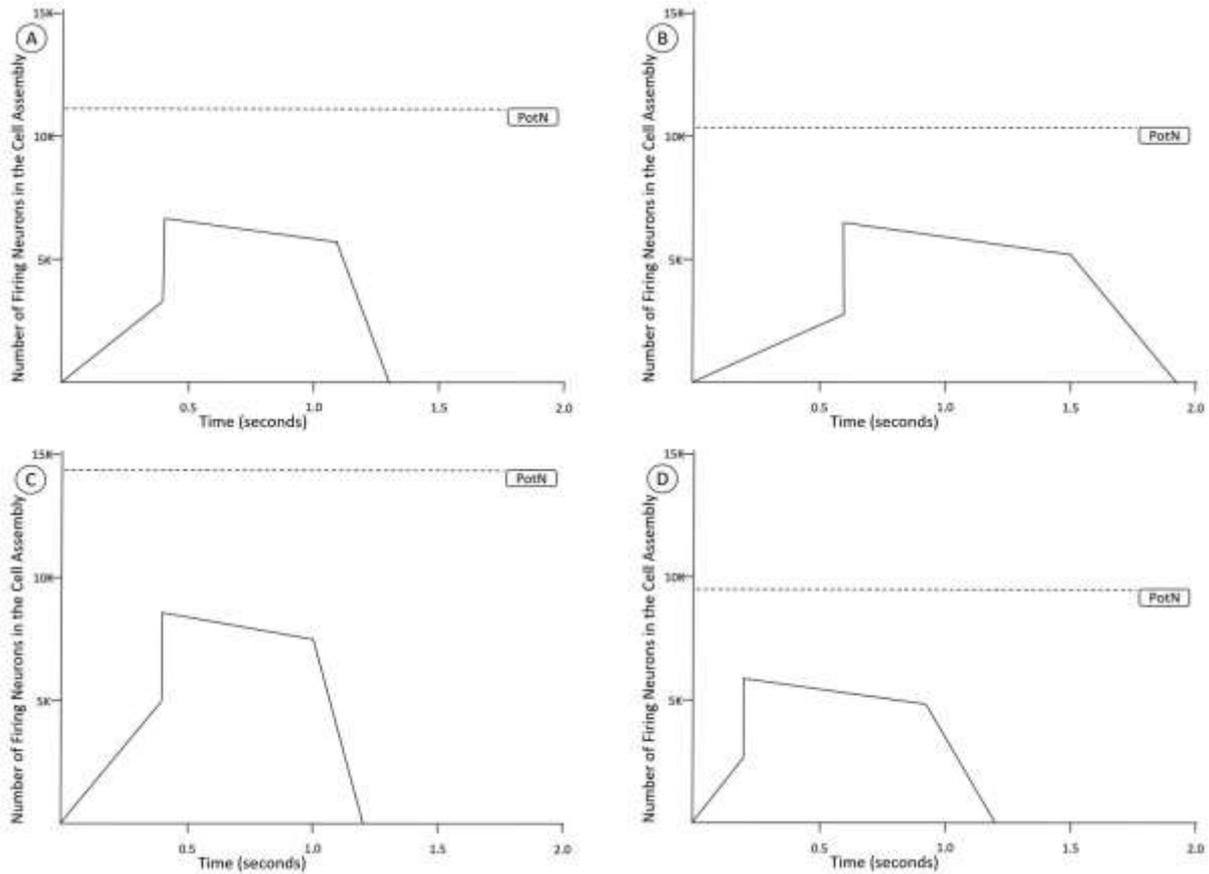

*Figure 10 Average SCAM diagrams: (A) all; (B) cognitive; (C) visual; and (D) motor.*

For the task analysed, Figure 10A shows the shape of the general CA, but this may involve inappropriate averaging whereas the differences between the three classes of CAs (B, C, and D) is of interest because, at the very least, the results show that the analyst's theoretical model has been successfully applied. This is a *post hoc* result in that it was possible that after the analysis the SCAM diagrams would not be as anticipated; the results, however, are as expected.

The number of neurons potentially in a CA (PotN) is highest for the visual CAs, and as can be seen in Table 5a, they are nearly 30% higher than the overall mean and nearly 40% higher than the cognitive CAs' mean and 50% higher than the motor CAs' mean.

|            | /All    | /Cognitive | /Motor |
|------------|---------|------------|--------|
| Visual/    | 29.7%   | 38.5%      | 51.6%  |
| Cognitive/ | -6.4%   | -          | 9.5%   |
| Motor/     | -14.4%  | -          | -      |

*Table 5a  Difference in means for PotN. The backslash represents how parameters are divided, i.e. vertical parameter divided by horizontal one.*

The results in Table 5a reflects the theoretical assumptions that the visual cortex is large and visual processes complicated, so visual CAs will be concomitantly large, particularly when compared to those of the motor cortex and, although a great deal of the human cortex appears unspecialised, it has a great deal to do at any moment, i.e. there will be many cognitive CAs ignited in parallel and not just those identified in a specific analysis.

The same pattern of results can be seen for differences in the means for both estimates of Threshold and IgMax, as can be seen in Tables 5b and 5c, respectively.

|            | /All   | /Cognitive | /Motor |
|------------|--------|------------|--------|
| Visual/    | 51.6%  | 78.6%      | 92.3%  |
| Cognitive/ | -15.1% | -          | 7.7%   |
| Motor/     | -21.2% | -          | -      |

*Table 5b Difference in means for Threshold. The backslash represents how parameters are divided, i.e. vertical parameter divided by horizontal one.*

|            | /All   | /Cognitive | /Motor |
|------------|--------|------------|--------|
| Visual/    | 26.5%  | 30.3%      | 45.8%  |
| Cognitive/ | -2.9%  | -          | 1.1%   |
| Motor/     | -13.2% | -          | -      |

*Table 5c Difference in means for IgMax. The backslash represents how parameters are divided, i.e. vertical parameter divided by horizontal one.*

Again, these results confirm that the theories have been successfully applied, in this case, that CAs, which may involve many neurons, i.e. a large PotN, will also tend to be large (IgMax) and with a relatively high Threshold to match.

The raw data summarised in Tables 5a-c could be subjected to statistical analysis, but it is not done so in this paper because: (a) most differences would not be significant, given the sample sizes and even using non-parametric tests; (b) such analyses would be *post hoc* and therefore statistically weak; and (c) we would be guilty of data hunting and significance chasing. On the other hand, clearly the potential is there for later, better planned research, to use decent analytical statistics.

.4.2.1  Fatigue Results

CAs are not simple negative feedback circuits in that the model of brain CA ignition is that they will fatigue, even with recruiting additional neurons from their potential pool (PotN), unless post ignition activity from other CAs adds to a CA's activity. N.B. the possibilities are for: (a) functionally just replacement neurons to maintain the current CA; or (b) similar, functionally related neurons, which might, for example, be involved in learning, even just up-dating one of one's Grandmother CAs when one visits her (see Introduction). Otherwise, a CA will fatigue and extinguish, "naturally", i.e. they have a "life-expectancy", without CA external neural support.

Fatigue, in terms of the number of K neurons, is simply: IgMax – IgFat. To compensate for different numbers of CAs in the three types analysed (N = All 48; Cognitive 18; Visual 16; Motor 14) Fatigue% is Fatigue divided by the size of the CA at ignition, i.e. ((IgMax – IgFat) / IgMax) x 100.

The fatigue data has been analysed in some detail. The overall view is given in Table 6.

|  | **Fatigue** | **IgMax** | **Fatigue%** |
|---|---|---|---|
| **All** | 1.1 | 6.8 | 16.2% |
| **Cognitive** | 1.4 | 6.6 | 21.2% |
| **Visual** | 1.1 | 8.6 | 12.8% |
| **Motor** | 1.0 | 5.9 | 17.0% |

*Table 6 Fatigue and percentage Fatigue, i.e. the latter corrected for differing numbers of CA types.*

Stressing that there can be no hope of statistically significant results, it was hypothesised that the 8.4% difference in Fatigue% between cognitive and visual CAs could be interpreted as: (a) a difference between types of CA; or (b) due to time, that cognitive CAs last longer (Figure 10). Data for the duration of ignition (IgTEx – IgTIg) and Fatigue (IgMax – IgFat) were examined in detail but all attempts at even the most speculative hypothesis testing was thwarted by Fatigue's range (0-3 K neurons for Cognitive and Motor CAs and 0-4 for Visual ones) and that the large majority of CAs had a Fatigue value of one.

### .4.2.2 Ignition Duration Results

Following the above, failed, analysis, the CA ignition duration data (IgTEx – IgTIg) was examined further  The investigation was driven by a desire to understand the distribution of data that underlies, and thus causes, the arithmetical average values used in the SCAM diagrams (Figure 10). The duration of a CA is one of its two primary features, and it can be argued its most important, not merely theoretically, but, critically, ignition duration is a measure in time (seconds), and time is linear. The estimates of the size of CAs may be wildly incorrect (Introduction), but whatever the caveats about timing tasks expressed in section 4.1, one can have more confidence about sequence; the time estimates in the analysis can only be in error by a couple of tenths of a second, because the times must fit the sequence. Furthermore, and consequentially, examination of the ignition duration data is less open to analyst bias, and thus of great potential value.

Using bins of half a second, Figure 11 summarises the ignition duration of the types of CA. This figure represents the same data in two ways, as a histogram and a line graph.

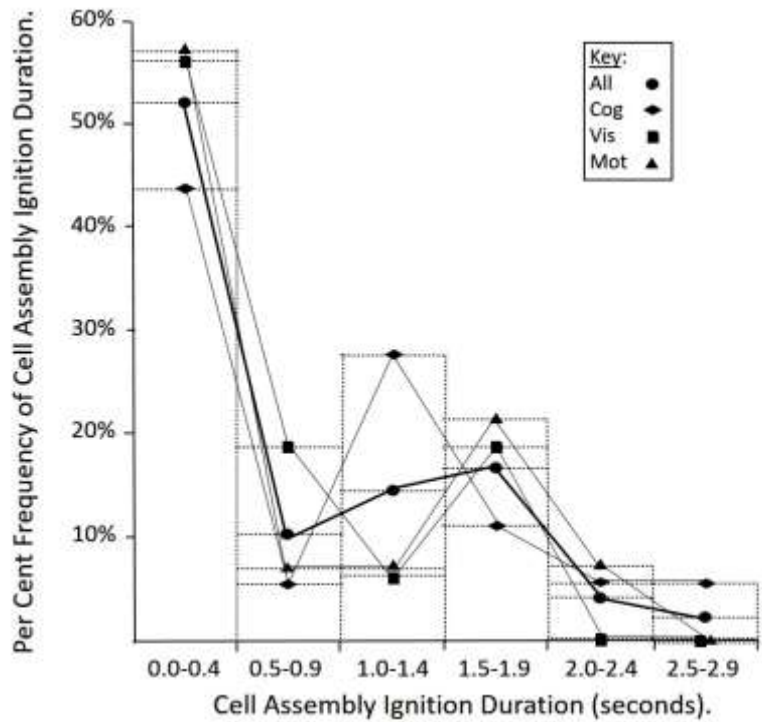

*Figure 11 Ignition durations of CA types presented as both line graphs and histograms.*

One would need a lot more data, but there is a hint that these CA ignition duration results are bi-modal, i.e. half the CAs last for less than half a second and most of the remainder last for over a second, with a few lasting over two seconds. It would not be implausible that, in the task, that there are two types of CA: short lasting ones and persisting ones.

.4.2.3 *Priming and Decay Results*

There is background activity in brains caused by neurons firing that appears random (section 2.1). The amount of such background activity may vary. For example, in the visual system there is, overall, more activity in the optic nerve in darkness than under normal viewing conditions, because retinal processes use lateral inhibition, but this background lacks the highly organised transmission of spike trains down the optic nerve bundles that signal retinal receptive field stimulation of varying spatial frequencies, and their location. What happens when disorganised activity reaches the visual cortex? The various forms of pattern recognition CAs are not ignited, although people do report fleeting and vague visual experiences in darkness (phosphines). We hypothesise that in such circumstances the overall background activity in the visual cortex may be quite high, but insufficient to ignite any of the vast number of potential visual CAs.

Little is really known about the relationship between background neural activity and potential CA ignition and the same is so for both priming and decay: see the QPID model (Introduction). For example, with higher levels of background activity, would a CA need more, the same, or less priming to reach ignition? Theoretically all are possible. Similarly, are CA thresholds changed by an elevated background?

When a CA extinguishes, the evidence is that there is an initial rapid decay of member neurons, but what is less clear is whether the later stages of decay return to whatever is the background level, or remain above this level for an appreciable time, or suffer a refractory period where the CA is harder to re-ignite. Furthermore, different CAs, and in different circumstances, may behave differently.

Perhaps the most unsatisfactory aspect of the SCAM used concerns priming and decay. Just looking at the SCAM diagrams (Figure 10), the priming and decay functions look exaggerated. This is undoubtedly caused by the single linear parameter used for each (P50% and D50%). Figure 12 shows a redrawn general SCAM diagram with more plausible priming and decay functions. These issues are returned to in the Discussion (section 5.3.1).

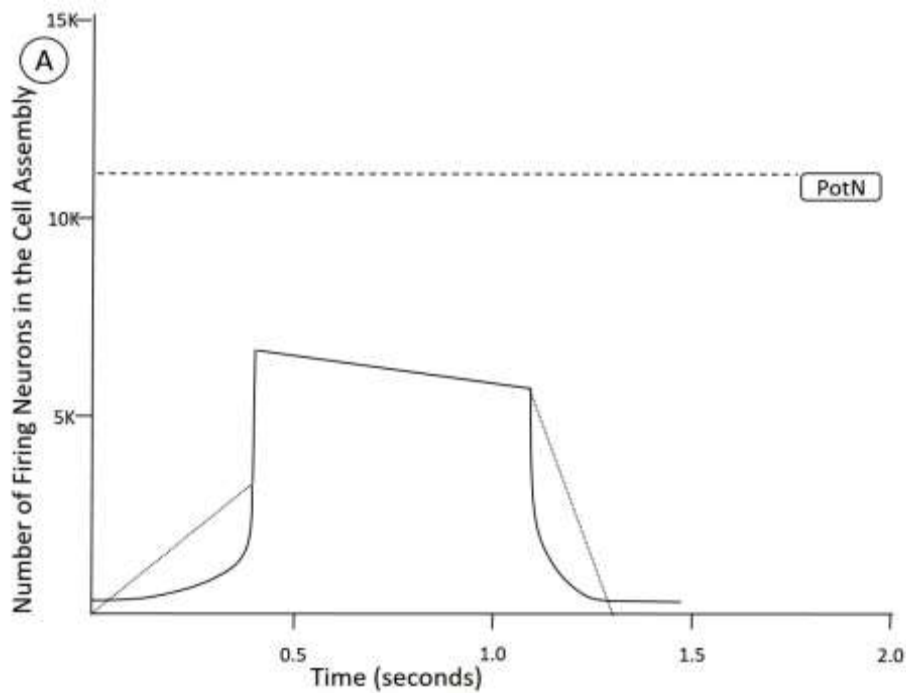

*Figure12  Redrawn general SCAM diagram with original Figure 1 shown with dotted lines where these two figures differ.*

### .4.3    CAAR Results

In the SCAM, which does not model internal CA processes, for every output from a CA there is its equivalent input to another CA or to a motor output that goes outside the system studied. Therefore, one can either model CAAR inputs or outputs as the results of one simply mirroring the other. The following analysis models outputs from CAs. Due to lack of data, the following results are ignored: (a) the five non-visual perceptual CAs (touch and kinaesthetic); (b) the seven inhibitory relationships (all between cognitive and motor CAs); and (c) system external motor outputs.

There were 89 relationships identified from the main analysis' CAAR diagram (Figure 9) and their outputs, and to where these outputs go, is summarised in Table 7.

|              | Visual → | Cognitive → | Motor → |
|--------------|----------|-------------|---------|
| → Visual     | 0        | 20          | 3       |
| → Cognitive  | 21       | 26          | 1       |
| → Motor      | 1        | 17          | 0       |

*Table 7 Input-Output numbers between CAs of different types from the CAAR Diagram (Figure 9); horizontal output to vertical input.*

The same results can be represented graphically (Figure 13), where it is easier to see the cognitive architecture that was used during the main analysis (Appendix I).

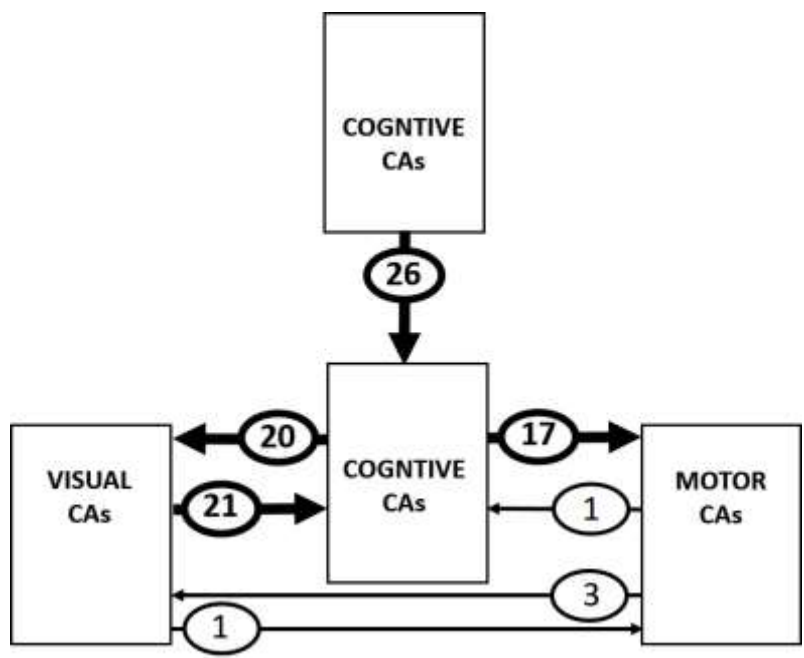

*Figure 13 Graphical representation of Table 7's Input-Output numbers between CAs of different types from the CAAR Diagram (Figure 9); main relationships in bold.*

Figure 13 confirms that in the vast majority of cases the Generic CAAR model (see section 2.3 and its Figure 2) was adhered to successfully during analysis. The centre portion of Figure 13 represents the basic chain of cognitive CAs, although noting that while there were 18 cognitive CAs, there were 26 outputs from one cognitive CA to another because some cognitive CAs may output to more than one CA of this type, i.e. the "basic chain" does have some branches or overtakes.

The intended, tight binding between cognitive CAs and visual ones (N=16) is well illustrated in Figure 13. That the outputs between cognitive and visual CAs is not equal (20 versus 21 relationships) is caused mostly by the occasional tight binding of motor and visual CAs. For example, the ballistic movement of the right arm (CA 11 MRAB) directly primes the visual system to expect the appearance of the right hand (CA 12 VRH) without going through an intermediate cognitive CA. Less than 5% (4/89) of the relationships analysed show such direct binding of motor and visual processes.

With one exception, inputs to the 14 motor CAs are from cognitive ones (N=17). There are only four outputs from the motor CAs as most of their outputs will be to the mid or hind brain and body movement systems. Many of these system external motor outputs will have inputs back into the system via sensory inputs. For example, when a hand is under negative feedback control then there is a cycle of: motor CA output → motor behaviour → optical input → visual CAs → cognitive CAs → motor CAs → motor CA output … .

## .5 Discussion

The authors consider the research reported to be fantastically successful, *for a first demonstration!* This section therefore starts with the positives, first at the level of TA (5.1), and then at a more rarefied, philosophical level concerning the integration in a single model of both brain and mental function (5.2). The final sub-section (5.3) suggests possible future developments of the work, including: development of a CA based TA technique; AI implementation of CAs; more general theoretical considerations; and some practical near term potential developments by the authors, and, they hope, others.

### .5.1 *Task Analysis with a Cell Assembly Perspective*

That it is possible to carry out a TA using a CA perspective is itself a success. The authors have worked for some years, together and independently, developing CA-based models and by exploiting TA's applied psychological approach, it is perhaps not surprising that they could identify putative CAs to associate with the task analysed. In terms of difficulty this is perhaps akin to attempting a *tabula rasa* GOMS analysis where every module decomposed must be invented from scratch, i.e. without reference to any previous GOMS analyses.

A more impressive success is the development of the first TACAP technique. The authors claim that their main analysis in Appendix I is their main result and the technique's success can be judged by the difference between the first third of the analysis, when they were in an iterative development mode, and the latter two thirds, which went quite smoothly and, relative to other TA approaches, quite quickly. While they are very cautious with the results (section 4), these generally indicate that they applied the various theories about mind, brain and CAs in a consistent manner.

In the end, the three representations developed, the SCAM diagrams, table and the CAAR diagram, were not only effective alone but were well integrated in that changes to one were usually relatively easy to propagate to the others, even though done manually (section 3.2). Naturally, we take Diaper's (2001) point that complex method development, and particularly method specification, must be done with analysts' software tool support. This topic is continued in section 5.3.1.

Acknowledging that the initial, main analysis covered but 9 seconds of elapsed task time, it is possible that any CA-based TA will always be at a low level of analysis and would therefore be unsuitable for analysing task of more than a few minutes. On the other hand, even if this were so, there are many tasks or subtasks which are super safety critical, and therefore worthy of detailed, if expensive, analysis, e.g. the time between V0, when an aircraft is committed to take-off, and rotation, when the aircraft has sufficient airspeed and height above ground that it can safely start to climb; or during an aircraft's handover from one sector to another by air traffic controllers; and numerous similar situations. Furthermore, first a CA-based technique might be used only on especially important subtasks and other TA methods used for the bigger task and, second, a library of CAs might allow overall task description at some meta-level that would then require only occasional descent to more detailed levels when appropriate.

Beyond the scope of this paper, a meta-cognitive architecture at the CA level needs developing and specifying. While such an architecture might include relatively distant brain areas, the expected focus would be within a localised brain area where, for example, two spatially adjacent, ignited CAs might already, or start, to share neurons and such sharing increases so as to create a super-CA; on subsequent ignitions, ignition of either will ignite the other. Such a model hypothesises a tighter binding between CAs than that of two interacting with each other, but which don't share any, or not very many, neurons. It might be possible to distinguish super-CAs from separate, interacting ones behaviourally in that ignition of one component CA (nearly) always causes ignition of the other(s) in the super-CA, whereas with separate CAs, then in some circumstances one CA igniting does not cause its sometimes related one(s) to ignite. There aren't great problems on the TA side about this since the levels concept is ubiquitous in TA, but a great deal remains to be done on CA meta-architectures, in the brain and in CA-based AIs. Much of the cognitive psychology literature, e.g. on selected and divided attention, may also need some redrafting to fit better at a CA level of analysis.

### .5.2   *Psychology, Neuroscience and Artificial Intelligence*

The relationship between brain and mind remains one of the great scientific puzzles. Neuroscience involves describing the physiology and biochemistry of the brain whereas scientific cognitive psychology describes the mind as an information processing device (see Introduction). At best for such models of brain and mind, they represent two different descriptions of the same thing, a physical one and a functional one, respectively. Such different descriptions of a thing are often conflated, for example, describing the heart as a "muscular fluid pump" combines its physical physiology with its function as a pump; for further discussion see Scott- Phillips *et al*. (2011) in the context of their distinction between proximate and ultimate explanations: the former correspond to physical, brain, descriptions and the later to mental, functional ones.

There are a number of problems with careless conflation of different descriptions. An obvious one concerns establishing functionality. For example, one might describe an electric hand drill as a device for making holes, but if it is considered as a spike rotator, then its functionality can be extended to sanding and polishing and, using a crank, such a drill can perform tasks involving linear reciprocating motion, e.g. sawing. Furthermore, multiple functionality is common in biology, e.g. that bones provide structural support and the production of red blood cells. The brain is particularly complicated because a great deal of the cortex is unspecialised, as far as currently known, and can be involved in many and apparently very different tasks. Such a property is central to the SCAM and its PotN conception.

There are areas of the cortex that do have a specialised functionality, but just what this might be is difficult to establish with complete certainty. Whatever physiological methods are used, the basic problem is the range of tasks tested. As a hypothetical illustration, one might find a brain area that is always active during language tasks, and careful experimentation might show this area is only active during parsing, but whether it is a specialised language parser, or part of one, would remain moot. Apart from the problems of specifying functionality, it is always possible that the same area may be active in tasks that are untested, say when riding a bicycle or listening to music, and the range of untested tasks is effectively infinite.

The logical problems remain at whatever physiological level of detailed studied, from single cell recording to what are quite large brain areas, i.e. relative to the size and number of neurons involved, and this is also the case with CA-based models. Indeed, it might seem that the problems are hardest at the CA level, but they do have a subtle advantage in that ignited CAs exist only temporarily and so searching for fixed brain neuron or area functionality will often be bootless. A further, more important advantage to using CAs to model both brain and mind is that there is a tight binding between the two such that the physical properties of a CA closely match their functional, information processing ones. No such tight binding exists for the physiology of larger brain units and traditional cognitive psychology, and while there is a similar tight binding at the level of single cells, we hypothesised in the Introduction that a Grandmother neuron might be better understood as being a frequent member of a Grandmother CA, which also solves the problem of what happens if such a cell dies.

CA-based ANNs also suffer the same logical problems in that once they have been running, and learning, for some time, then the function of a particular CA is difficult to infer, even though the state of the whole system is open to inspection. In contrast, with symbolic AIs such as ACT-R, the function of each of its software modules is well understood as these are programmed using traditional software methods, i.e. the functionality is as well understood as for that of any piece of correctly running software code. Although the authors are confident they could do so, with sufficient resources, they have not attempted to implement anything from their first TACAP analysis as a CA-based ANN. Their plans on this are discussed further in section 5.3.4.

The authors' view is that a major benefit of this first TACAP analysis is that of a precursor to a General Theory of both brain and mind. This is discussed further in section 5.3.3. TACAP is intended to encourage cognitive scientists of all sorts to consider both the neural and cognitive at the CA level and, by exploiting the applied cognitive approach of TA, facilitate creative, sensible proposals about CAs and their architecture. When TA is done well, then it places quite severe constraints on what is "sensible" and, as illustrated throughout Appendix I, a considerable amount of psychology is involved; and with TACAP, some neuroscience as well.

### .5.3  *Future Developments*

This TACAP paper is the start of a story. While research on both TA and CAs has been going on for decades, it is their combination that makes TACAP unique. The following subsections outline work that needs doing to further develop TACAP (sections 5.3.1 and 5.3.2), how it might have substantial theoretical consequences (5.3.3), and the authors' near term plans for TACAP development (5.3.4).

*.5.3.1 Method and Software*

Continuing from section 5.1 and the essential requirement to develop analyst support tools, Figure 14 shows one high level, user perspective of the suite of tools that need developing for this paper's TACAP technique.

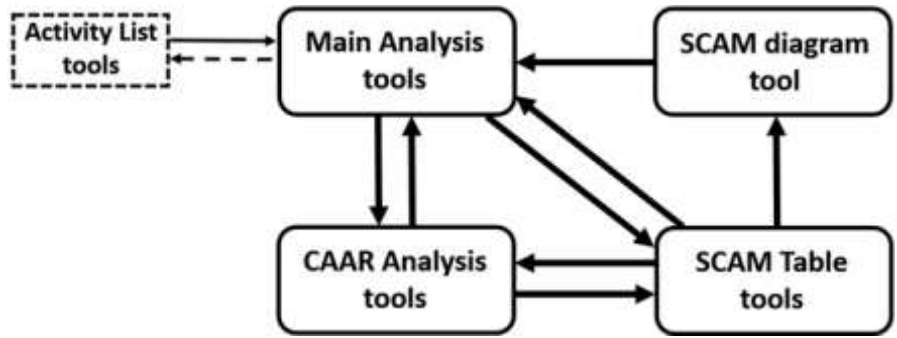

*Figure 14 Software tools suite required to automate the TACAP technique.*

It is assumed that existing or new tools would support analysts working with various types of task performance data and that AL lines would be imported into the Main Analysis tools.  It is envisaged that the latter is the analyst user's main interface that, apart from free text entries, would automate the decisions made and, of course, test and flag inconsistencies, a.k.a. current errors, in an ongoing analysis.  From the early stages of a TACAP analysis, as CAs are identified they would create SCAM Table entries and, as the parameters are filled in, then there may be feedback to the Main Analysis tools.  Once each CA's SCAM table's row of data are all filled in, then a SCAM diagram is created for that CA and made available in the Main Analysis tools.  The SCAM Table tools also seed the CAAR Analysis tools with both identified CAs and their location on the task timeline.  The analyst user must still specify relationships between CAs, but producing the CAAR diagram should be at least semi-automated. Furthermore, much more sophisticated relationships between CAs could be relatively easy to specify than was realistic with the first, manual analysis, e.g. cycles of feedback between CAs could be indicated, say by multiple arrow heads, and types of input/output could also be coded beyond the simple excitatory or inhibitory relationships used in this first analysis.

Noting the priming and decay parts of the SCAM, P50% and D50%, were clumsy for producing the SCAM diagrams manually (section 4.2.3 and Figure 12), a simple power function would could easily be applied in a SCAM diagram production tool.

Similarly, the sub-optimal entries to the SCAM table with respect to generating SCAM diagrams in a manual analysis (section 4.2, Tables 3 and 4 and Figure 10) involve trivial software calculation, allowing future tools to optimise the user analyst's ease of input as the simple backend software would take care of the rest.  These are examples that emphasise the importance of software tools to support the development and specification of complex methods.

While a design feature of the first TACAP analysis was to include various capabilities to cross-check within and across the main representations, the suspicion is that the analysis is not entirely error free, notwithstanding many hours of testing.  As an example, only after the first draft of this paper was completed was it discovered that CA VHWA (Visual Hot water Area) was correctly present in the CAAR diagram but entirely absent from the SCAM table and Appendix I; most of the testing had been done between the latter two. The belief is that a reasonable analysts software suite would not only make analyses better, and nigh error free, but would reduce analysis time to a third or a quarter of what it might take to do manually.

Experience with developing such tools, e.g. Diaper's (e.g. 2001) LUTAKD toolkit, suggests that in addition to being essential for method specification, such tools are also likely to change the method itself, not least because what was implausible effort in a manual analysis becomes easy with appropriate software. Nigh impossible to predict in advance, as an example, one candidate would be the animation of the CAAR diagram. For the initial TACAP analysis, the CAAR diagram was done in PowerPoint (section 3.2) and for the expert user it is relatively easy to animate the timeline and the CA boxes and the arrows. Like envisaging the SCAM diagrams without drawing most of them (section 3.2), the CAAR diagram was only animated in the analyst's mind during analysis. The animation (Appendix III) was only done after the main analyses were completed. On the other hand, for less visually adept analysts, they might well find an automatically animated CAAR diagram of considerable help. It should certainly help when presenting such work to conference or seminar audiences.

*.5.3.2 Artificial Intelligence*

The evidence is that CAs do exist in the brain (Harris, 2005; Huyck and Passmore, 2013; and Introduction), although a great deal of our understanding of CAs has arisen from AI work with ANNs. No doubt there are interesting scientific research opportunities involving the mimicry of brains and minds (section 5.3.3), but future, practical applications of CA-based AIs depends on identifying roles and functions. One CABot, for example (Huyck *et al*., 2011*)*, was implemented as a robot in a virtual, simple games-like environment with a general role of operating as a user's assistant. TA is rarely done frivolously because it is expensive in time, money, and human resources, of expert analysts and task performers. Monitoring and assisting users in complex, safety critical tasks, particularly when tasks and their environments are variable and require rapid decision making, for example in aviation as mentioned in section 5.1, would seem to provide appropriate and useful application areas for development of versions of the TACAP approach and their useful implementation as CA-based AI assistants.

Building the CABot systems provides confidence that such CA-based AIs, with only minimal initial programming, are able to learn to carry out tasks. They will develop their own CAs by unsupervised learning, by trial and error. As discussed in section 5.2, it is difficult to infer such CAs' functionality even though there is the potential to inspect every state in every program cycle. Unless particular CAs are forced on a system, then it is unlikely that AI CAs will coincide with brain and mind CAs, i.e. both AIs and people can learn to perform the notional "same" task but the fine details at the CA-level will differ. The same is true between any two people and, anyway, even frequently repeated tasks by the same person will not use quite the same CAs each time. We cope with these within and between differences in people and it will be necessary to extend the same coping strategies to genuinely intelligent, flexible, self-learning AIs.

We believe that CA based AIs will become increasingly popular. They are capable of learning new domains and while all AI systems are currently domain specific, CA-based systems will be more flexible than Expert/Knowledge Based Systems or symbolic ones. A virtual agent with a simulated neural brain will function in an environment, and learn significant aspects of that environment. Upfront programming effort required in symbolic AI development, and maintenance, will be replaced by the self-programming capabilities of CA-based systems, although there may be a cost if it is necessary to provide learning nurseries for new CA-based AIs. Perhaps within only a few decades, but after the emancipation of the early CA-based AIs, people will have another highly intelligent species with which to share planet Earth; and one that can talk to us in our own languages.

*.5.3.3 General Theories of Psychology and Neuroscience*

A General Theory is, within its scope, a theory of everything. General Theories are quite common in psychology, even if below cognitive psychology's axiom concerning the mind as an information device, and they are often quite simple. What makes a CA-based General Theory attractive is the "tight binding" (section 5.2) between psychology and physiology. A possible future development might be the deliberate conflation of description of brain and mind, producing descriptions where a CA has physical properties, presumably an improvement of the SCAM table, and functional ones, what the CA does and its relationships to other CAs.

Traditionally, psychology has borrowed from other technologies, from Victorian hydraulics, e.g. people feel pressure, to computing, and even changing psychological models as technologies improve, e.g. Diaper's (1989b) PDP8 versus PDP11 models of cognition (the PDP8 models do operations in registers whereas the PDP11 ones dispense with registers altogether). With CAs, for once the direction might be opposite, in that there is a chance for such a psychology to focus physiological studies, i.e. having posited the existence of one or more CAs, then the physiologists might try and find them.

Such possibilities may be some considerable time away as at the moment too little is known about CAs, in brains, minds and in AIs. Indeed, the TACAP development was explicitly intended to encourage cognitive scientists to think and work at the CA-level and, over time, thus might an international community become established.

*.5.3.4 Practical Near Term Developments*

While the authors wish to enthuse others with a practical approach to CA-orientated thinking, they have some near term plans following this paper's publication. They will offer seminars and conference presentations focusing on special aspects of the TACAP research suitable for different audiences. The full animation of the CAAR diagram (Appendix III) might be particularly useful for these (section 5.2). At least one on-line presentation will also be developed.

We are also in the process of developing a proto-neural cognitive architecture. We can currently implement simple associative memories, and generic rule based systems in simulated spiking neurons. Combining these will make a proto-neural cognitive architecture, which could be used for executing tasks to simulate, at a neural level, task execution. An obvious extension would be to extend our existing binary CAs to more complex ones that behaved as those described in the analysis (Appendix I). This would enable us to develop the TA mechanism in step with a neural cognitive architecture.

## .6 Conclusions

The authors believe that this is a 'John the Baptist' paper that starts a new chapter in the combination of psychology, neuroscience and AI. In the end, it is probably not what they have done that is important, but how they did it. The TACAP provides an easy entrance for others to learn to think at the CA level. Appendix I, the main analysis, is crucial for such a purpose as it contains 65 examples of CAs which others can study and use as a basis for identifying CAs in more appropriate tasks.

Although the trend in ergonomics is to study general systems above the level of tasks, as Sociotechnical Systems (e.g. Stanton and Harvey, 2017), and sometimes called Systems-of-Systems (Harvey and Stanton, 2014), the essential need for the detailed study of some tasks will remain. Recently the terms "Artificial Intelligence" and "AI" have entered popular awareness, although, like "psychology" for much longer, the general public may know little beyond the terms themselves. Just how intelligent, if at all, some of the systems that these days claim to be AI is open to question, but the AI cat is now out of the bag and genuinely intelligent systems may result from AI's commercialisation.

TACAP is at least paddling hard to catch the crest of the coming AI wave. As a new approach it lacks much of the baggage of older TA approaches, which might further commend it for development.

**Acknowledgement**

The authors thank Professor Tom Dickens of Middlesex University for his comments on a draft of this paper.

**APPENDIX**

I           Cell Assembly Descriptions – The First Steps to Making Coffee

II          Acronym Glossary

III         Animated CAAR Diagram: http://www.cwa.mdx.ac.uk/chris/hebb/TACAP.html

# APPENDIX I

## The 'First Steps to Making Coffee' TACAP Main Analysis Descriptions.

At the kitchen entrance …

**01    CA: COGNITIVE – Kitchen Entrance Check (CKEC)**

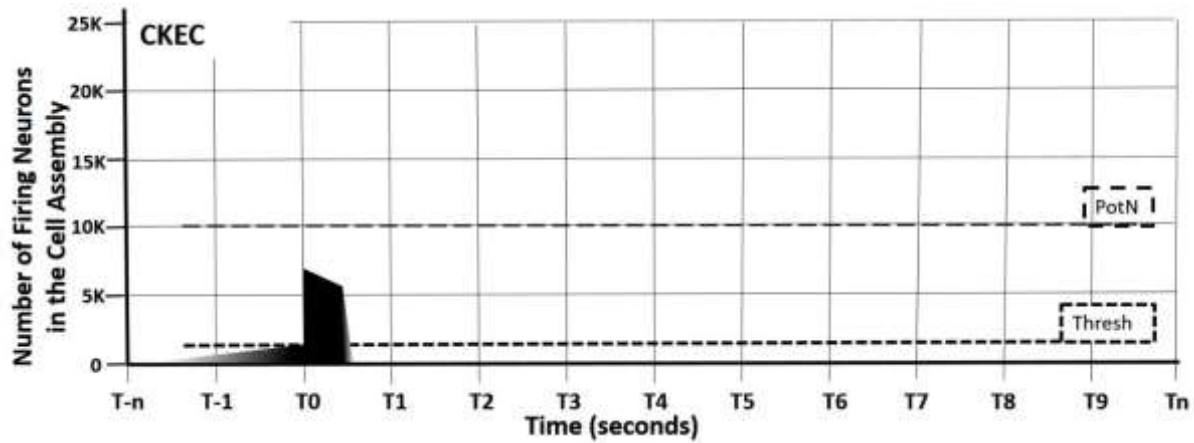

| ID   | PotN | Thresh | IgMax | IgFat | P50% | IgTIg | IgTEx | D50% |
|------|------|--------|-------|-------|------|-------|-------|------|
| CKEC | 10   | 2      | 7     | 6     | -1.0 | 0.0   | 0.4   | 0.5  |

|  |  |
|--|--|
| INPUTS: | "… at kitchen entrance". |
|  | CA: VISUAL – Kitchen Entrance General (VKEG). |
| OUTPUTS: | CA: VISUAL – Kitchen Entrance General (VKEG), |
|  | CA: COGNITIVE – Make Coffee (CMC), |
|  | CA: COGNITIVE – Approach Hot Water Area (CAHWA). |

Primed by the various CAs that have bought the subject to the kitchen entrance, the CA is ignited at T0, or just before, and represents the expectation of what, in general, the kitchen should look like. It checks for major disasters: fire, smoke, steam, flooding, major damage to cabinets and window, but not details such as whether the cooker is on. It also checks that there is no one else in the kitchen and that the floor is clear of obstructions, e.g. shopping not yet unpacked.

This CA, or something similar, must rationally exist because if there is a major problem with the kitchen then it will be immediately detected at the entrance. For a cognitive CA this one is modelled as being fairly large (PotN 10K) because a general view of the kitchen is a complicated one, so its expectation CA must also be fairly large. Its threshold (2K) is fairly low and most (IgMax 7K) of its potential neuron membership are modelled as firing after ignition as the CA will nearly always last only very briefly, whether the kitchen is judged satisfactory or not.

Post ignition it then takes input from 'CA: VISUAL – Kitchen Entrance General' (VKEG) and makes a match comparison of expectation to visual input. Note, the comparison process is here modelled as part of CKEC but an alternative would be to have a CA that took inputs from both cognitive and visual CAs and it then makes the comparison. This sort of general comparison of expectations to visual input must be a fairly common type of operation. Whatever CA Architecture (CAA) chosen, however, the effect of the visual input is basically inhibitory, the cognitive CA is turned off either because the kitchen is judged as satisfactory or other emergency dealing CAs are ignited. If satisfactory, the cognitive CA to Make Coffee (CMC) is reignited. This must precede the striding into the kitchen as alternatives at this point involve going to other kitchen locations, and such movements are all highly practiced and would have similar CAs to the making coffee one.

There is a CAA issue concerning how tasks might share common CAs, for example, the early stages of making either coffee or tea are behaviourally identical, but still might use different CAs, or, perhaps more likely, neuron membership may overlap between coffee and tea making CAs, if exactly the same CAs are not used, which may be simplest option for analysis purposes.

## 02       CA: VISUAL – Kitchen Entrance General (VKEG)

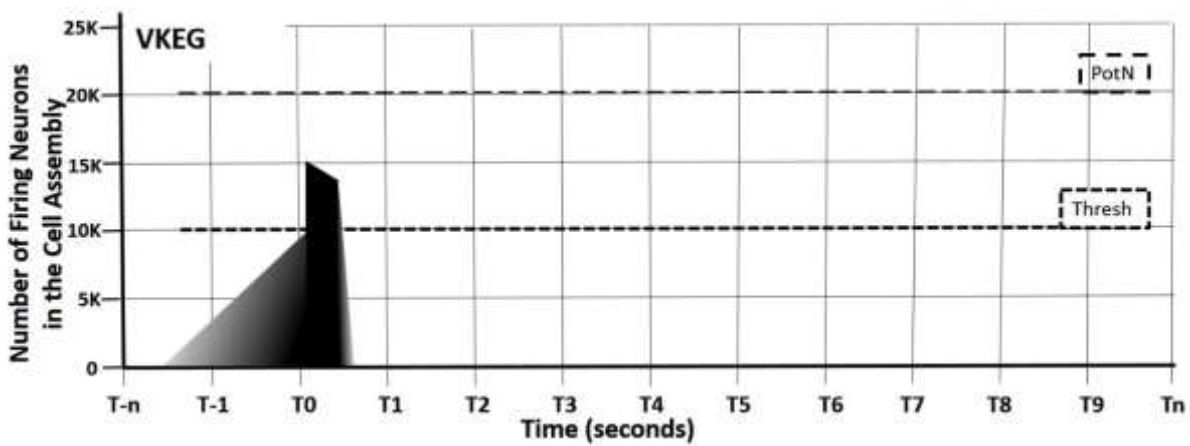

| ID | PotN | Thresh | IgMax | IgFat | P50% | IgTIg | IgTEx | D50% |
|---|---|---|---|---|---|---|---|---|
| VKEG | 20 | 10 | 15 | 14 | -0.8 | 0.1 | 0.3 | 0.4 |

   INPUTS:     CA: COGNTIVE – Kitchen Entrance Check (CKEC).

   OUTPUTS:    CA: COGNTIVE – Kitchen Entrance Check (CKEC).

Typically visual CAs are large (PotN 20K for VKEG) because the visual cortex is large and with complex scenes then thresholds need to be relatively high, but if CAs are to persist then there must also be a sufficiency of neurons that can fire as some fatigue and so CA ignition can be maintained.

A saccade takes about a quarter of a second and during such eye movements retinal output to the optic nerve is suppressed. Thus this CA cannot ignite until after the kitchen entrance is reached (T0), and the prior visual CAs are suppressed. Its function is primarily as the data provider for CKEC.

The CA will be suppressed (overwritten) by following visual input, although if the cognitive check fails then it may persists for several saccades as the problem is generally inspected.

### 03      CA: COGNITIVE – Make Coffee (CMC)

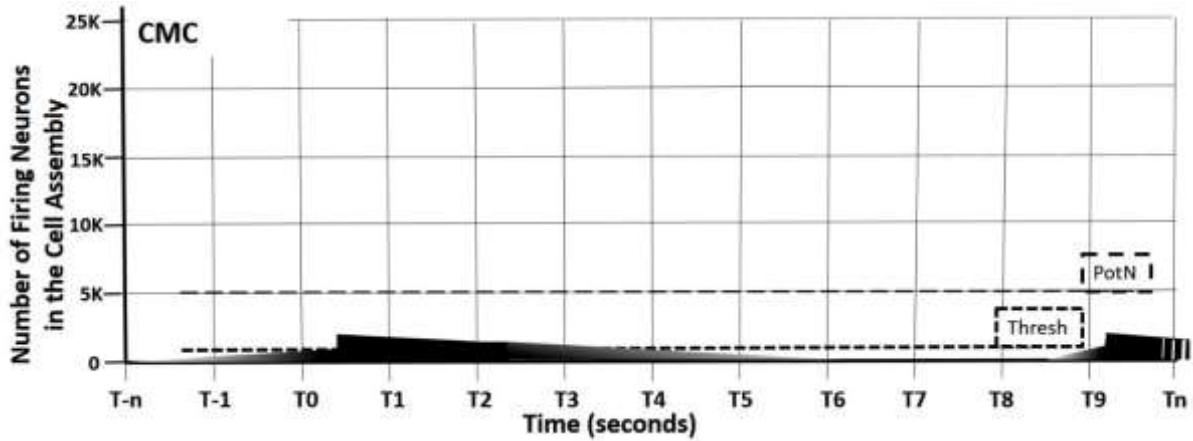

| ID | PotN | Thresh | IgMax | IgFat | P50% | IgTIg | IgTEx | D50% |
|---|---|---|---|---|---|---|---|---|
| CMC | 5 | 1 | 2 | 1.5 | -1.0 | 0.4 | 2.5 | 4.0 |

INPUTS:     "… at kitchen entrance".

              CA: COGNITIVE – Kitchen Entrance Check (CKEC).

OUTPUTS:    CA: COGNITIVE – Approach Hot Water Area (CAHWA).

Discussed in general (Sections 3.3.1 and 3.3.2), the Make Coffee CA is already primed, and probably more so at the kitchen entrance, and must ignite when the general kitchen checking CA (CKEC) extinguishes as there are several possible destinations within the kitchen, including, for example, curvetting through 130 degrees to go to the fridge (section 3.3.2).

In its minimal decision making form where the CA does not contain a plan for making coffee, the CA is quite small (PotN 5K) and post-ignition, after directing the subject to the hot water making area it decays until it is below threshold. It remains primed, however, as it needs to be re-ignited when water is added to the empty kettle as the amount added depends on what hot beverage, in what sized mug or cup, is being prepared, e.g. a count of 15 (seconds) for a small mug of coffee versus 20 for a large mug (N.B. The kettle has no external indicator of how much water is in it).

## 04    CA: COGNITIVE – Approach Hot Water Area (CAHWA)

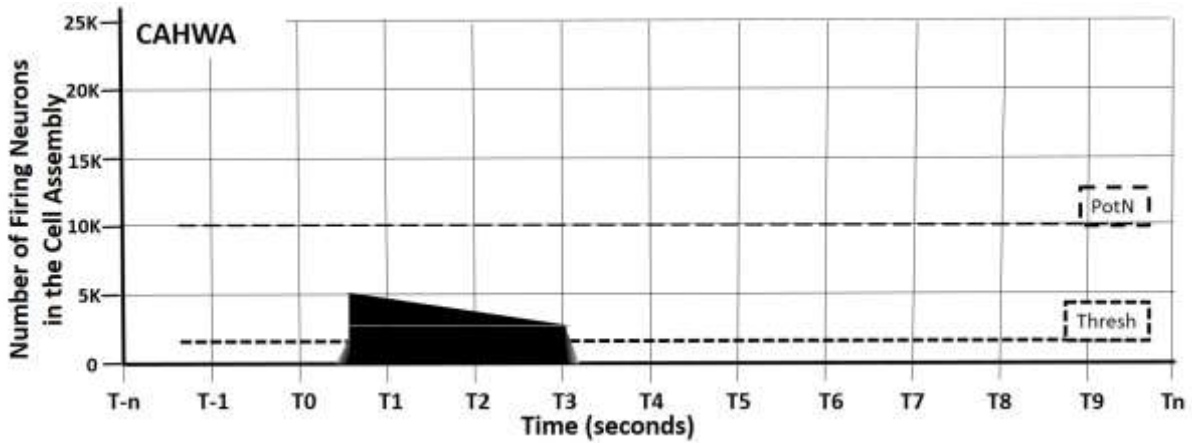

| ID | PotN | Thresh | IgMax | IgFat | P50% | IgTIg | IgTEx | D50% |
|---|---|---|---|---|---|---|---|---|
| CAHWA | 10 | 2 | 5 | 3 | 0.5 | 0.6 | 3.1 | 3.2 |

INPUTS:    CA: COGNITIVE – Kitchen Entrance Check (CKEC),

                CA: COGNITIVE – Make Coffee (CMC).

                CA: VISUAL – Approach Hot Water Area (VAHWA).

OUTPUTS:   CA: VISUAL – Approach Hot Water Area (VAHWA).

                CA: COGNITIVE – Kettle in Hot Water Area (CKHWA);

                CA: MOTOR – Stride to Hot Water Area (MSHWA).

Apart from flow-field related visual inputs, the CA operates, like the kitchen entrance check (CKEC), as an expectation, checking the foveal input against what should be in the hot water area, how it is organised (the strong expectation is "neatly"); if the kettle were missing then this would certainly cause a "pause & consider" CA; output from the CA causes ignition of the kettle search and identify CA (CKHWA)

This CAHWA CA will persist the longest of the three related approach CAs (motor, visual and cognitive), i.e. until after movement to the hot water area has stopped (MSHWA); the visual (VAHWA) CA extinguishes even earlier as the flow fields become increasingly peripheral close to the hot water area.

**05  CA: VISUAL – Approach Hot Water Area (VAHWA)**

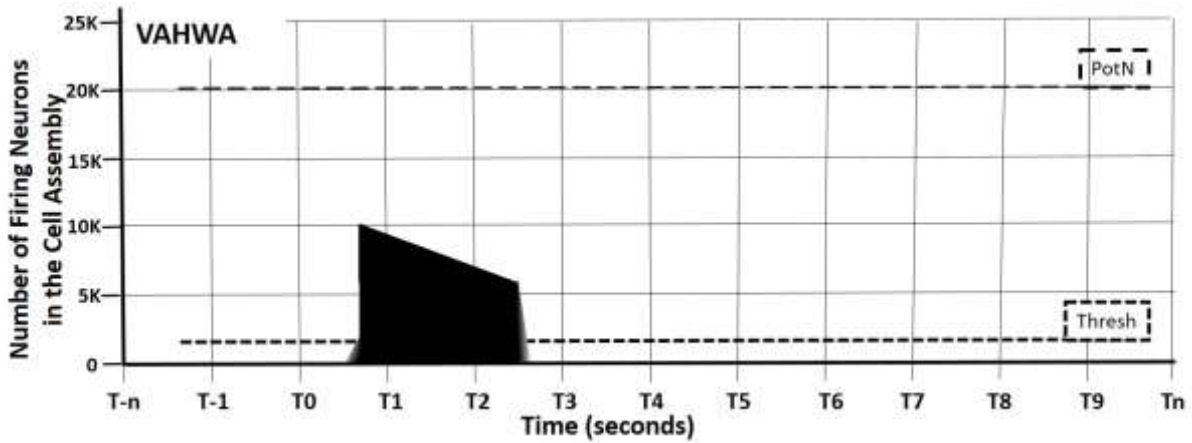

| ID | PotN | Thresh | IgMax | IgFat | P50% | IgTIg | IgTEx | D50% |
|---|---|---|---|---|---|---|---|---|
| VAHWA | 20 | 2 | 10 | 6 | 0.6 | 0.7 | 2.5 | 2.6 |

INPUTS:     CA: COGNITIVE – Approach Hot water Area (CAHWA).

OUTPUTS:   CA: COGNITIVE – Approach Hot Water Area (CAHWA).

The CA is part of the specialised visual processing involved with moving through an environment. Interest in visual flow fields (Gibson, 1950) was rekindled with Marr's (1982) computational approach to vision; the theory remains that flow fields are handled separately from other, more integrated, visual processes.

A lot of neurons (PotN 20K) are potentially involved and a low threshold of 2K is set since this sort of processing is used constantly and can be for many hours (e.g. car driving. N.B. different CAs are ignited as visually the road ahead (and behind one hopes for safety reasons) changes). On the other hand, in this highly practiced task of about 3 seconds the proposed CAA is that VAHWA is a self-terminating CA and that the neurons at ignition are not much replaced, hence fatigue is relatively high (IgMax – IgFat = 10K – 6K = 4K), i.e. 40% of the neurons have fatigued but sufficient survive to maintain ignition above threshold (2K). P50% & D50% are very fast as part of this type of visual processing: an on-demand, switch on-off facility.

## 06    CA: MOTOR – Stride To Hot Water Area (MSHWA)

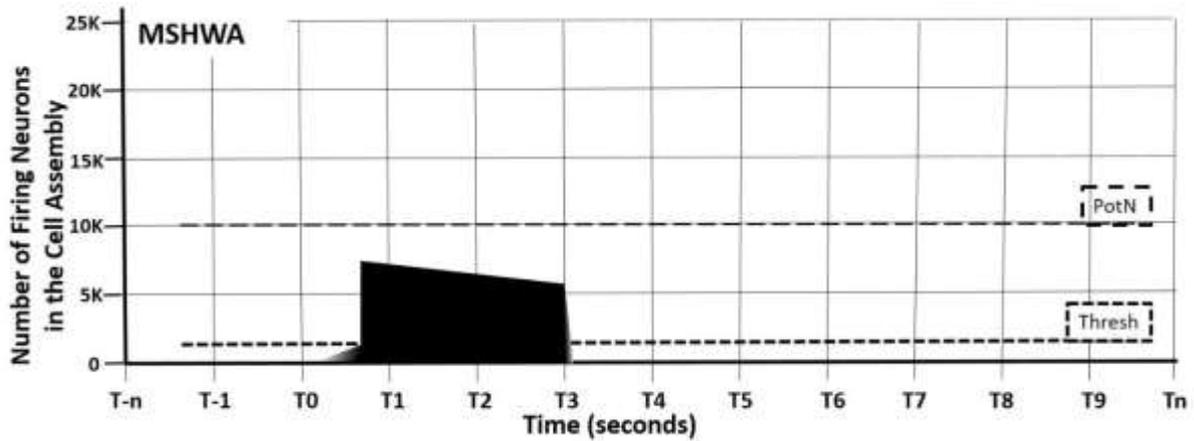

| ID | PotN | Thresh | IgMax | IgFat | P50% | IgTIg | IgTEx | D50% |
|---|---|---|---|---|---|---|---|---|
| MSHWA | 10 | 2 | 7 | 6 | 0.6 | 0.7 | 3.0 | 3.1 |

INPUTS:    CA: COGNITIVE – Approaching Hot Water Area (CAHWA).

OUTPUTS:    *motor behaviours …*

From the shuffle zone outside the kitchen entrance, the right foot is planted in the centre of the entrance as described above (section 3.3.2). The CA is ignited by CAHWA once the general kitchen check CA confirms the kitchen is in a suitable state. There is no observable behavioural pause at the entrance and from detailed analysis the main subject always approaches the hot water preparation area with three strides (left, right, left) and then a right footed half stride that curves the right foot so it ends up next to the left (Figure 3). The strides are longer than a usual walking step around the house and the whole behaviour is very precise in that it ends with the body close, but not touching, the hot water preparation area; toes are never stubbed or the knees hit the cabinet beneath the work surface, although the knees come within a few centimetres of this vertical surface.

The other three resident adults have also been observed approaching the hot water area. The subject's daughter, in her early 30s and nearly as tall as her father, takes the same three strides and the final right foot movement in a manner indistinguishable from those described above. In contrast, the wife, in her early 70s, takes five steps, not strides, as she is considerably shorter, but repeated observation suggests that a similar behavioural invariance is present. The fourth resident, in his early 30s, had only lived in the house for about 6 months and doesn't use the kitchen that much. Observed from his approach to the kitchen down the corridor, his behaviour was inconsistent, e.g. either foot could be the launch one, and, indeed, he was much less accurate at reaching the hot water area, a final shuffle being required. The obvious conclusion is that the family who have all lived in the house for over twenty years have a CA for approach that the new lodger does not.

As a learned and highly practiced behaviour, the MSHWA one need be of only modest size (PotN 10K), with a low threshold (2K) and most of its neurons firing on ignition since it cannot persist meaningfully beyond the completion of the behaviour. The CA does, however, have to be of sufficient size to take cognitive approach inputs from CAHWA based on that CA's visual inputs from VAHWA as the three strides need to compensate for the location of the launching right foot, which may vary up to 30cm in front of or behind the bar on the floor of the kitchen entrance.

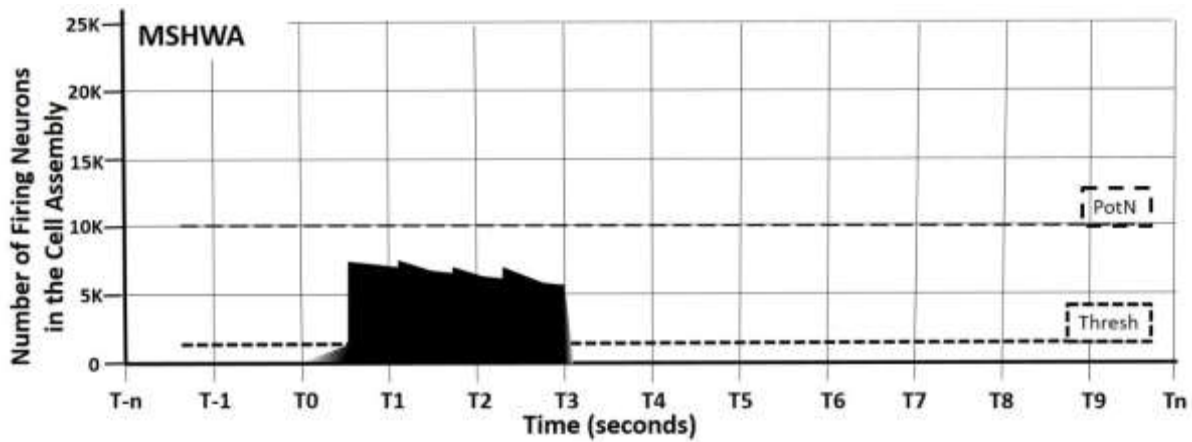

It is likely that the basic SCAM diagram is not an adequate representation of this CA, which, for example, might have internal processes representing the strides and terminal shuffle as shown above.

A number of alternative CA Architectures (CAAs) were considered for MSHWA, notably a CAA where this motor CA might ignite its associated visual CA (VAHWA) and receive feedback from this, rather than being mediated by the cognitive CA (CAHWA).

As a codicil to the above concerning the invariant striding behaviour, this occurs when the subject is not carrying something into the kitchen, most probably an empty coffee mug. In a more complete analysis an alternative CA involving striding to the sink to deposit an empty mug to the right of the sink in preparation for washing needs specifying, although the CAs involved are similar to the ones described above; there is a sidestep from sink to hot water area after mug deposition.

A further CAA issue concerns the extent that CAs are common in different tasks. Behaviourally there is no difference between making tea rather than coffee when going to the hot water area and filling the kettle. There is a difference as to how much water is put in the kettle (25% less for a small mug).

## 07  CA: COGNITIVE – Kettle in Hot Water Area (CKHWA)

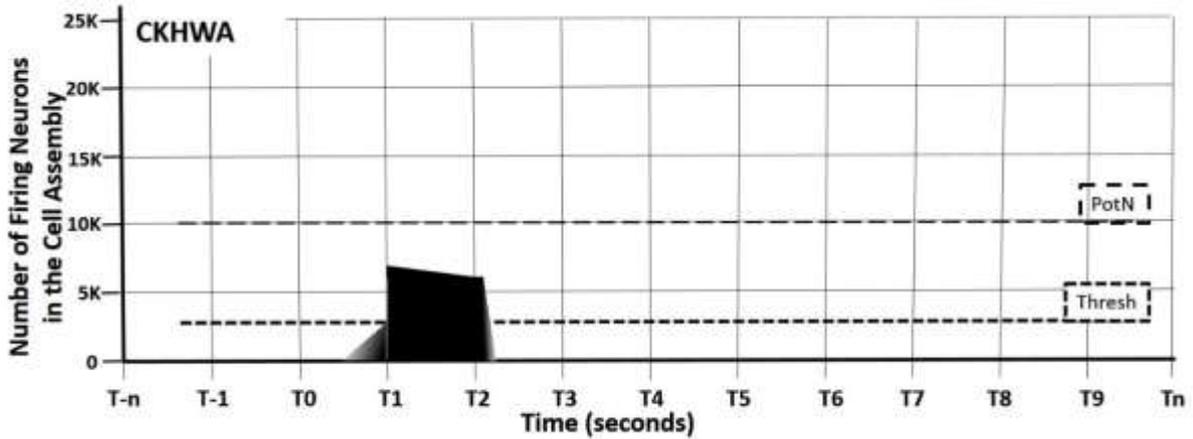

| ID | PotN | Thresh | IgMax | IgFat | P50% | IgTIg | IgTEx | D50% |
|---|---|---|---|---|---|---|---|---|
| CKHWA | 10 | 3 | 7 | 6 | 0.8 | 1.0 | 2.1 | 2.2 |

INPUTS:   CA: COGNITIVE – Approach Hot Water Area (CAHWA),

CA: VISUAL – Kettle in Hot Water Area (VKHWA).

OUTPUTS:  CA: VISUAL – Kettle in Hot Water Area (VKHWA).

CA: COGNITIVE – Kettle Handle (CKH).

The kitchen's hot water area is a complex of small and medium sized objects which are nearly all in standard locations, although the kettle and circular tray may lay within an area of about 5cm radius beyond their footprints. The CA therefore needs to be reasonably sized (PotN 10K), although the threshold is low (3K). Ignition lasts about a second before being replaced by the more detailed target, the kettle handle (CKH).

## 08  CA: VISUAL – Kettle In Hot Water Area (VKHWA)

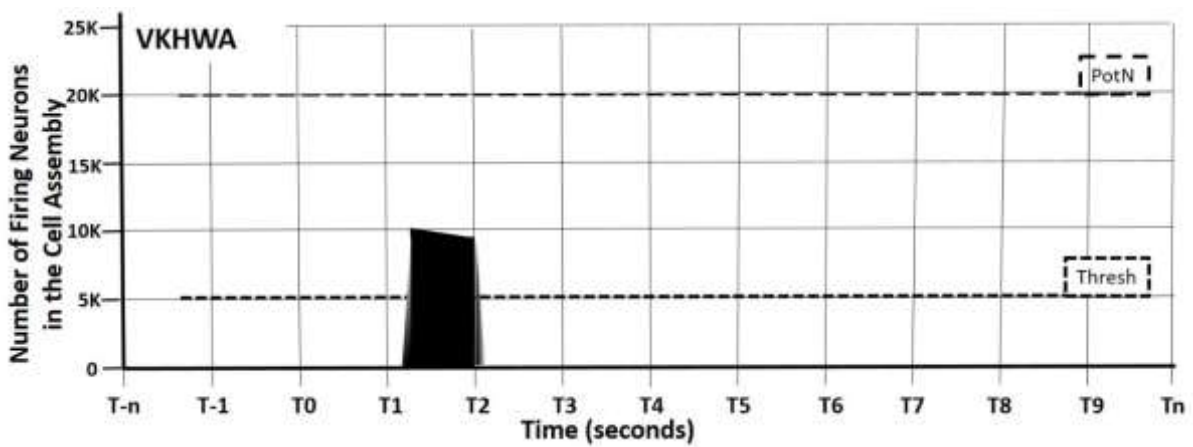

| ID    | PotN | Thresh | IgMax | IgFat | P50% | IgTIg | IgTEx | D50% |
|-------|------|--------|-------|-------|------|-------|-------|------|
| VKHWA | 20   | 5      | 10    | 9     | 1.2  | 1.3   | 2.0   | 2.1  |

   INPUTS:    CA: COGNITIVE – Kettle In Hot Water Area (CKHWA),

   OUTPUTS:   CA: COGNITIVE – Kettle In Hot Water Area (CKHWA.

Primed and ignited from inputs from CKHWA, feedback between the two CAs directs and identifies the kettle's location within the cluttered hot water area. The CA gradually decays post-ignition as the more specific kettle handle target is acquired in the next two CAs (CKH and VKH).

**09    CA: COGNITIVE – Kettle Handle (CKH)**

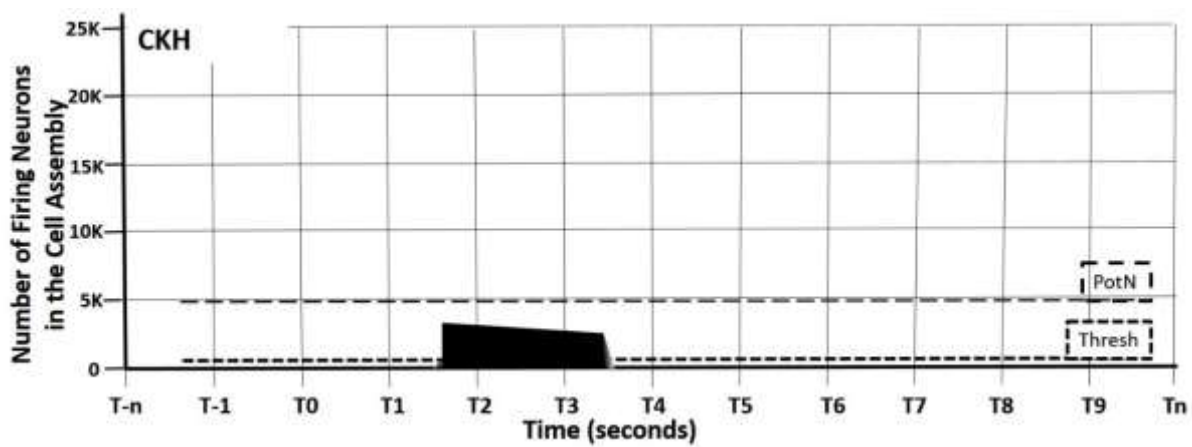

| ID  | PotN | Thresh | IgMax | IgFat | P50% | IgTIg | IgTEx | D50% |
|-----|------|--------|-------|-------|------|-------|-------|------|
| CKH | 5    | 1      | 3     | 2     | 1.5  | 1.6   | 3.5   | 3.6  |

   INPUTS:    CA: COGNITIVE – Kettle In Hot Water Area (CKHWA).

              CA: VISUAL – Kettle Handle (VKH).

   OUTPUTS:   CA: VISUAL – Kettle Handle (VKH).

              CA: MOTOR – Right Arm Ballistic (MRAB).

              CA: COGNITIVE – Right Hand Approach (CRHA).

As an object the kettle's handle is very simple, being a uniform, matt dark grey/black and smoothly shaped. Thus it does not need a large CA (PotN 5K) to be identified as the critical task target for control of the right hand approach to the handle. The CA does have to represent the current orientation of the handle, but the corner location of the hot water area means that the handle will virtually always be to the right within an arc of less than 90 degrees.

The CA persists for about two seconds and then decays quickly and before the right hand actually grips the handle because the hand obscures its target in the final approach stage. N.B. general introspective experience suggests that once part of an object is gripped so as to transport the object, the gripped part of the object itself is ignored, whether it be a kettle handle, a book, a bag or whatever; a CA for the object itself must still be ignited as different objects are treated differently while being transported, e.g. I wouldn't try and empty a book over the kitchen sink (below this is the CA 'Lift Kettle' (CLK) to indicate its difference from when the kettle is, for example, located on its base unit).

## 10    CA: VISUAL – Kettle Handle (VKH)

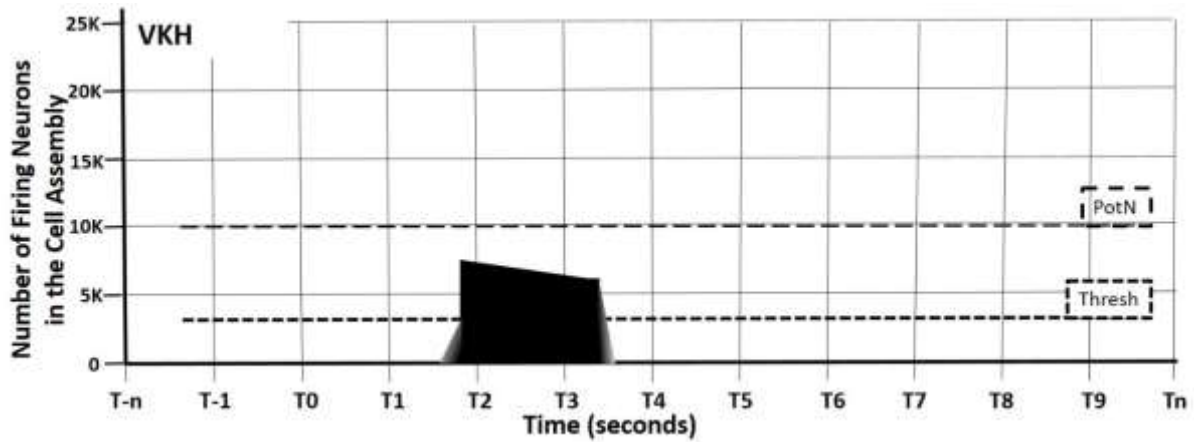

| ID | PotN | Thresh | IgMax | IgFat | P50% | IgTIg | IgTEx | D50% |
|---|---|---|---|---|---|---|---|---|
| VKH | 10 | 3 | 7 | 6 | 1.6 | 1.8 | 3.3 | 3.4 |

    INPUTS:    CA: COGNITIVE – Kettle Handle (CKH).

    OUTPUTS:    CA: COGNITIVE – Kettle Handle (CKH).

Like CKH, which primes and ignites this CA (Threshold 3K), VKH is smaller than many other visual CAs (PotN 10K). It provides feedback to CKH which it pre-extinguishes.

## 11    CA: MOTOR – Right Arm Ballistic (MRAB)

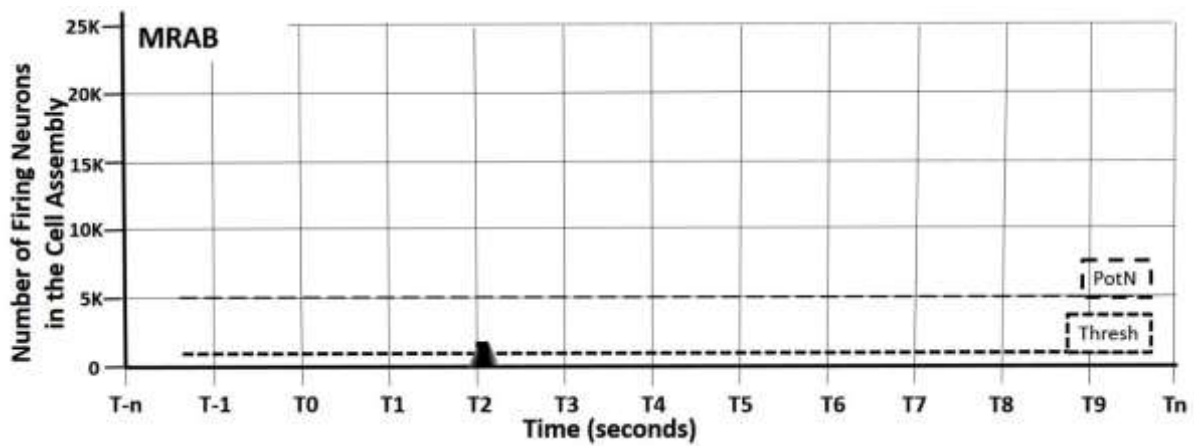

| ID | PotN | Thresh | IgMax | IgFat | P50% | IgTIg | IgTEx | D50% |
|---|---|---|---|---|---|---|---|---|
| MRAB | 5 | 1 | 2 | 2 | 1.9 | 2.0 | 2.1 | 2.2 |

INPUTS:     CA: COGNITIVE – Kettle Handle (CKH)

OUTPUTS:    CA: VISUAL – Right Hand (VRH)

This is the first of the two parts of normal human reaching behaviour. Visually it is open-loop control, i.e. without feedback, although there must be some kinaesthetic feedback, not least the position of the arm when the hand is launched towards its target. It is ignited by CKH when feedback from VKH to CKH establishes that the target kettle handle has entered reach.

It's assumed in the model to be a small CA (PotN 5K) that exists for between, say, 50 and 150ms.

In the CAA described here it is assumed that this CA primes and ignites a visual CA (VRH), rather than a cognitive one, as the right hand, as expected, enters view.

## 12    CA: VISUAL – Right hand (VRH)

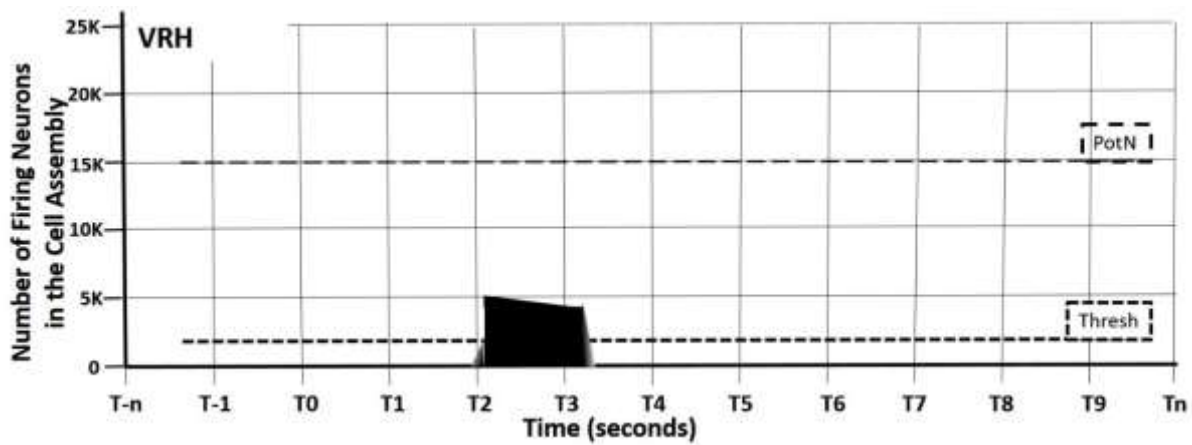

| ID | PotN | Thresh | IgMax | IgFat | P50% | IgTIg | IgTEx | D50% |
|---|---|---|---|---|---|---|---|---|
| VRH | 15 | 2 | 5 | 4 | 2.0 | 2.1 | 3.2 | 3.3 |

INPUTS:     CA: MOTOR – Right Arm Ballistic (MRAB).

OUTPUTS:    CA: COGNITIVE – Right Hand (CRH).

Ignited by MRAB, the CA predicts where the right hand will appear and then identifies its position and general configuration.

Note, human babies acquire visual tracking & the concept of object permanence fairly early in development. Also, we do often look at our hands, probably because kinaesthetic feedback is less precise than vision, and touch.

It's relatively small for a visual CA (PotN 15K) and has a low threshold (2K), strong ignition (IgMax 5K) and relatively little fatigue (IgFat 4K) because although ignition is only about a second here, it may have to persist for much long periods of time so must have a structure that facilitates neuron rotation to counter fatigue.

The CA is different from those used in manipulative tasks, but often precedes and initiates such tasks and subtasks.

## 13 CA: COGNITIVE – Right hand (CRH)

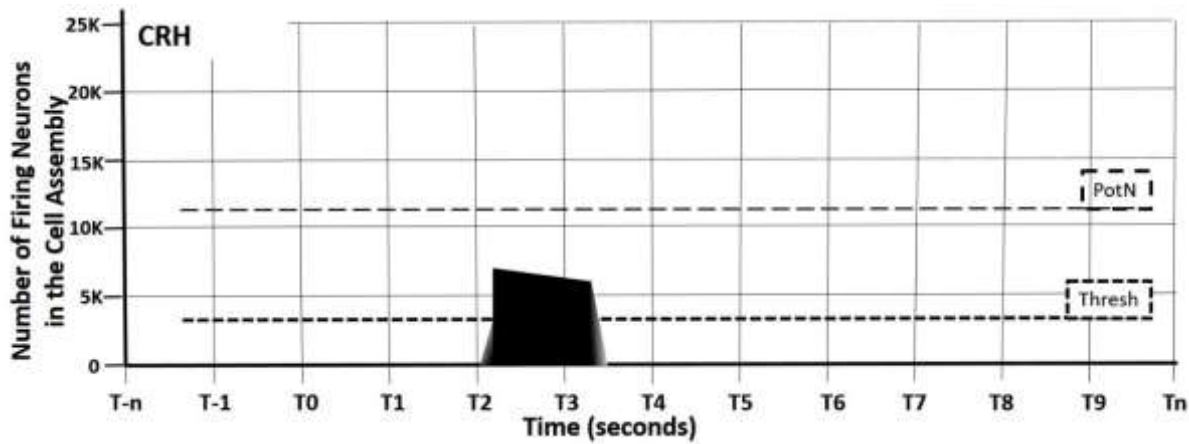

| ID  | PotN | Thresh | IgMax | IgFat | P50% | IgTIg | IgTEx | D50% |
|-----|------|--------|-------|-------|------|-------|-------|------|
| CRH | 12   | 3      | 7     | 6     | 2.1  | 2.2   | 3.4   | 3.5  |

INPUTS:    CA: VISUAL – Right Hand (VRH)

OUTPUTS:  CA: COGNITIVE – Hot water Area (CHWA)

                   CA: COGNITIVE – Right Hand Approach (CRHA)

Representing a general model of the hand, the CA, like VRH, needs to be of sufficient size (PotN 12K) to counter fatigue (IgFat 6K), and ditto w.r.t to threshold (3K) and IgMax (7K).

It causes CHWA to ignite so that the hand can be placed in its context relative to itself and its target, the kettle handle (CKH); these three CAs will be used as inputs by CRHA to control the right hand's final approach to the kettle handle.

## 14 CA: COGNITIVE – Hot water Area (CHWA)

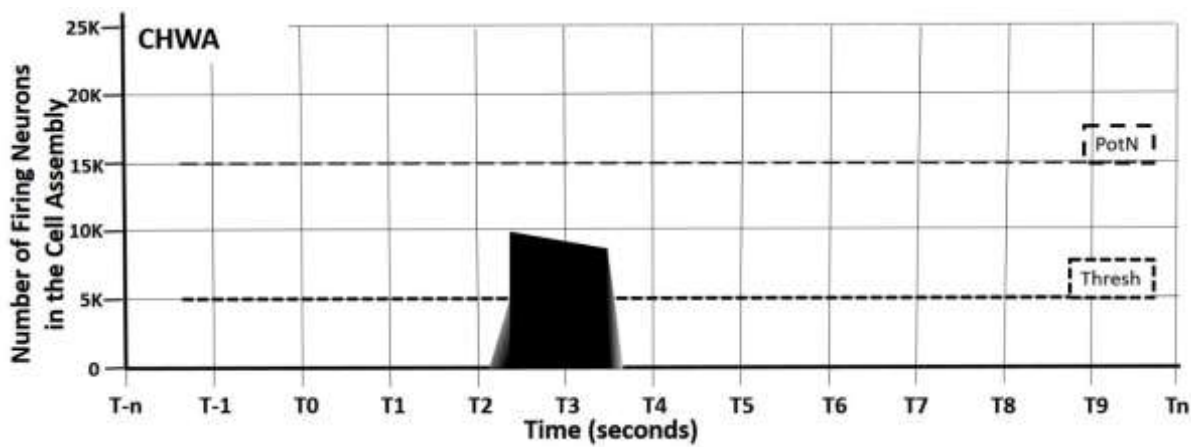

| ID   | PotN | Thresh | IgMax | IgFat | P50% | IgTIg | IgTEx | D50% |
|------|------|--------|-------|-------|------|-------|-------|------|
| CHWA | 15   | 5      | 10    | 8     | 2.2  | 2.4   | 3.5   | 3.7  |

INPUTS:     CA: COGNITIVE – Right Hand (CRH),

CA: VISUAL – Hot Water Area (VHWA).

OUTPUTS:   CA: VISUAL – Hot Water Area (VHWA),

CA: COGNITIVE – Right Hand Approach (CRHA).

This CA supplies a specialised representation of the hot water area, basically ignoring expected, static objects except for those that might interfere with the right hand's approach to the kettle handle. It provides the context for the hand's "flight path", in effect the tunnel of clear, relevant space between the tray holding the coffee cone (which is a potential flight hazard on the left) and the left side of the drainer, which could mean a wall on the right of over 20cm if large pots and their lids are draining, and which considerably narrows the hand's possible path to the kettle handle.

It is big for a cognitive CA (PotN 15K) because it not only deals with a complex visual input, but a specialised one that provides the critical input for CRHA to plan the hand-to-kettle flight path which CRHA then controls. In CAA terms it is here modelled as one of number of hot water area CAs. An alternative CAA would be to have a sufficiently general hot water area visual CA that it could be directed to different aspects of its input (visual attention). The preference here is due to it being a highly practice task so CAs will be relatively specialised, which is not to say that neurons in the CHWA at one time could not be part of other hot water related CAs at other times.

## 15    CA: COGNITIVE – Right Hand Approach (CRHA)

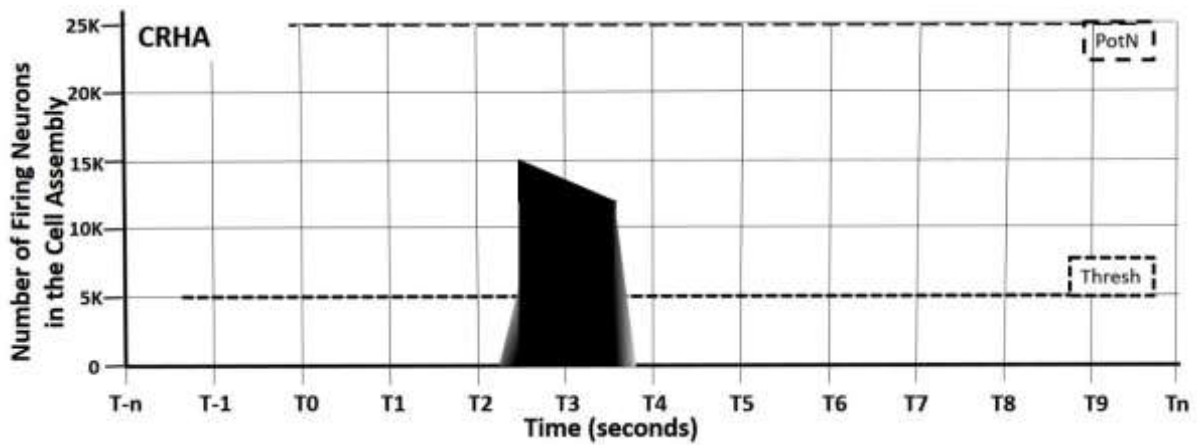

| ID | PotN | Thresh | IgMax | IgFat | P50% | IgTIg | IgTEx | D50% |
|---|---|---|---|---|---|---|---|---|
| CRHA | 25 | 5 | 15 | 12 | 2.3 | 2.5 | 3.6 | 3.7 |

INPUTS:     CA: COGNITIVE – Kettle Handle (CKH)

CA: COGNITIVE – Right Hand (CRH)

CA: COGNITIVE – Hot water Area (CHWA),

                CA: VISUAL – Right Hand Approach (VRHA),

                CA: TOUCH – Right Hand on Kettle Handle (TRHKH).

OUTPUTS:    CA: VISUAL – Right Hand Approach (VRHA)

                CA MOTOR – Right Hand Approach (MRHA),

                CA: TOUCH – Right Hand on Kettle Handle (TRHKH),

                CA: COGNITIVE – Right Hand Grip (CRHG)

This is big for a, still task specialised, cognitive CA (PotN 25K) and it undoubtedly is composed of a number of CAs below the level of this analysis. It's main functions are to: (i) integrate inputs from cognitive CAs concerning the kettle handle, right hand and the hot water area; (ii) compute the right hand's path to the kettle handle, avoiding obstructions, and (iii) control that path under visual negative feedback control, including (iv) adjustments to the hand and wrist in preparation to gripping the kettle handle at the trajectory's termination; and (v) it's final function before self-extinction is to supress MRHA and so halt the reaching behaviour once the handle is touched (TRHKH) and ignite the cognitive CA for gripping the kettle handle (CRHG).

It is well primed by its cognitive inputs and has a low threshold (5K) and a high IgMax (15K) while still having sufficient potential neurons to cope with both fatigue and the internal inhibition of some of its own neurons during processing (IgFat 12K). It may only last a second or so, but it is a cognitively complex, active second.

## 16    CA: VISUAL – Right Hand Approach (VRHA)

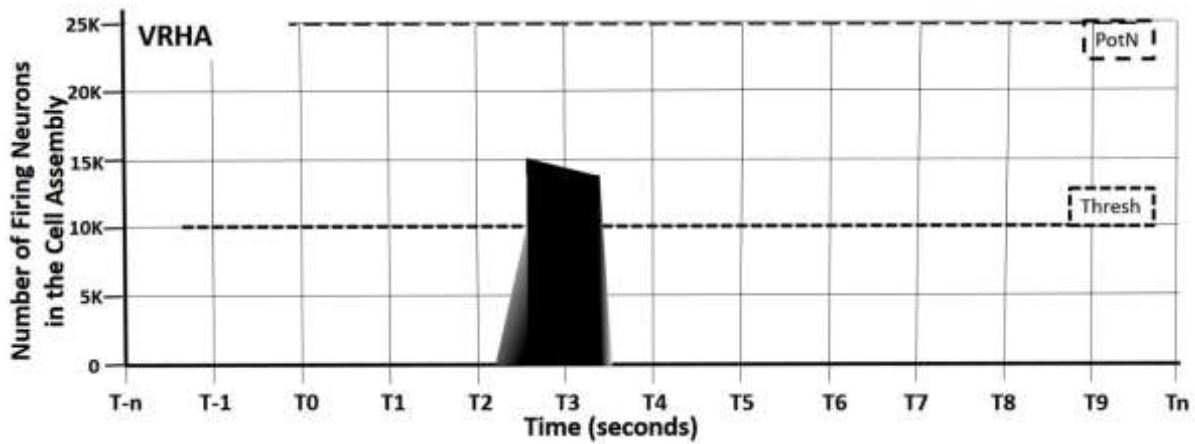

| ID | PotN | Thresh | IgMax | IgFat | P50% | IgTIg | IgTEx | D50% |
|---|---|---|---|---|---|---|---|---|
| VRHA | 25 | 10 | 15 | 14 | 2.3 | 2.6 | 3.3 | 3.4 |

        INPUTS:    CA: COGNITIVE – Right Hand Approach (CRHA).

        OUTPUTS:    CA: COGNITIVE – Right Hand Approach (CRHA).

This CA provides the visual input to CRHA that allows visual negative feedback control of the right hand approaching the kettle handle. It is fairly large, even for a visual CA (PotN 25K) and is well primed and finally ignited by CRHA. Although here lasting less than a second, it must have fatigue resisting capabilities by neuron rotation as in other tasks it may have to remain ignited for much longer. It extinguishes before CRHA when the hand obscures the target kettle handle in the final approach stage.

## 17     CA: MOTOR – Right Hand Approach (MRHA)

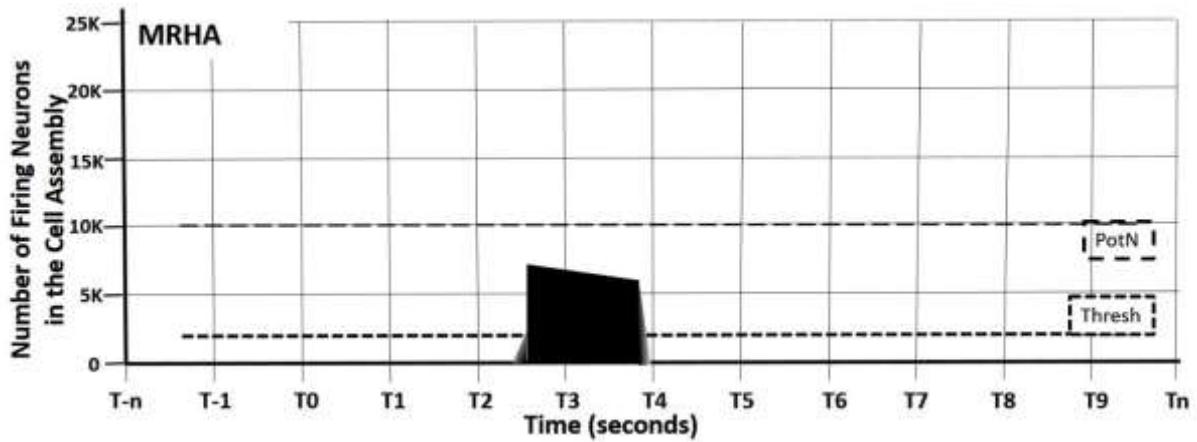

| ID | PotN | Thresh | IgMax | IgFat | P50% | IgTIg | IgTEx | D50% |
|---|---|---|---|---|---|---|---|---|
| MRHA | 10 | 2 | 7 | 6 | 2.4 | 2.7 | 3.7 | 3.8 |

INPUTS:      CA: COGNITIVE – Right Hand Approach (CRHA).

OUTPUTS:   *motor behaviour …*

The CA provides the motor component to CRHA's control of the hand approaching the kettle handle and also for configuring the hand so as to be ready to grasp the kettle handle. N.B. In the CAA used in the analysis, here there is no direct I/O between the visual and motor systems except via the cognitive one (CRHA); an alternative would be I/O between VRHA and MRHA, which may be plausible for fine control; similarly when the kettle handle is touched and TRHKH is ignited, it could be used to extinguish MRHA rather than, as modelled, extinction is via CHRA suppressing it.

## 18  CA: TOUCH – Right Hand to Kettle Handle (TRHKH)

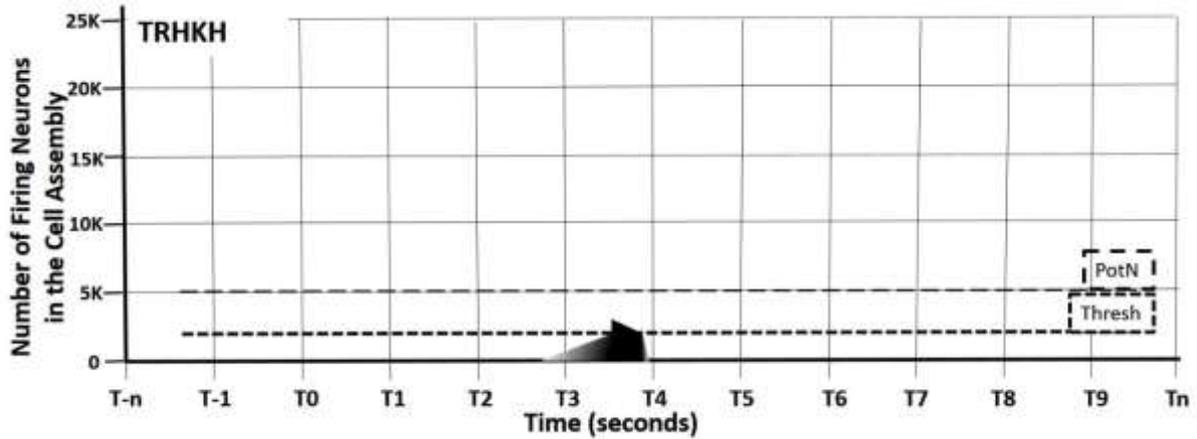

| ID | PotN | Thresh | IgMax | IgFat | P50% | IgTIg | IgTEx | D50% |
|---|---|---|---|---|---|---|---|---|
| TRHKH | 5 | 2 | 3 | 2 | 3.0 | 3.5 | 3.8 | 3.9 |

INPUTS:   CA: COGNITIVE – Right Hand Approach (CRHA)

OUTPUTS:   CA: COGNITIVE – Right Hand Approach (CRHA)

This CA signals the end of the right hand's kettle approaching behaviour. Although a small CA (PotN 5K), it has a low threshold (2K) and will have been extensively primed by CRHA (IgTIg – P50% = 0.5 seconds) because it is so critical that the reaching behaviour is neatly halted, even if CRHA slows the approach in the final fractions of a second.

## 19  CA: COGNITIVE – Right Hand Grip (CRHG)

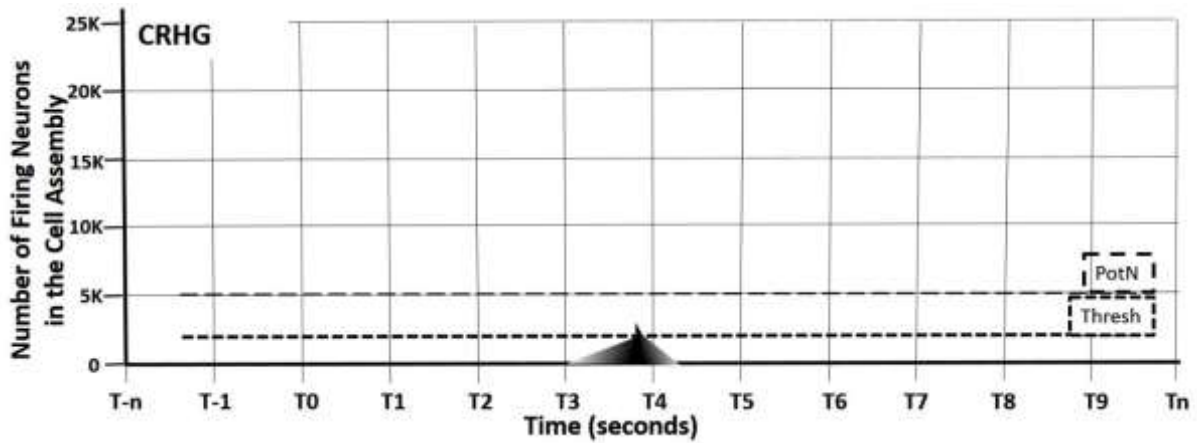

| ID | PotN | Thresh | IgMax | IgFat | P50% | IgTIg | IgTEx | D50% |
|---|---|---|---|---|---|---|---|---|
| CRHG | 5 | 2 | 3 | 2 | 3.2 | 3.7 | 3.8 | 4.2 |

INPUTS:   CA: COGNITIVE – Right Hand Approach (CRHA).

|  | CA: TOUCH – Right Hand Grip (TRHG). |
|---|---|
| OUTPUTS: | CA: TOUCH – Right Hand Grip (TRHG). |
|  | CA: MOTOR – Right Hand Grip (MRHG). |
|  | CA: COGNITIVE – Right Hand Hold (CRHH). |

Like TRHKH this CA is well primed (IgTIg – P50% = 0.5 seconds) and then ignited as CRHA's final function. It needs only to be a small CA (PotN 5K) since its only concern is the actual closing of the right hand on the kettle handle. It doesn't last long, just sufficient to ignite its motor CA (MRHG). There is also negative feedback from TRHG relating to the force of the gripping behaviour.

Before extinction CRHG ignites the right hand holding of the kettle (CRHH). It is modelled as decaying quite slowly (IgTEx – D50% = 0.4 seconds) so as to allow re-ignition if there is a problem with holding the kettle, howsoever rare.

In a very early SCAM analysis the difference between gripping the kettle handle and then holding it were not differentiated. Subsequently it became clear that this resulted in a SCAM diagram that could not be described using the SCAM parameters because what was needed was an initial ignition to represent the grasp and then a steady holding-the-kettle-handle state. Just as there are two answers to Popper's Black Swan problem (either the theory's wrong or, by definition, it is not a swan), so we have preferred the latter option, i.e. to separate the initial grip from the subsequent, long term holding of the kettle handle. Some psychological justification for this is offered below concerning CRHH.

## 20    CA: MOTOR – Right Hand Grip (MRHG)

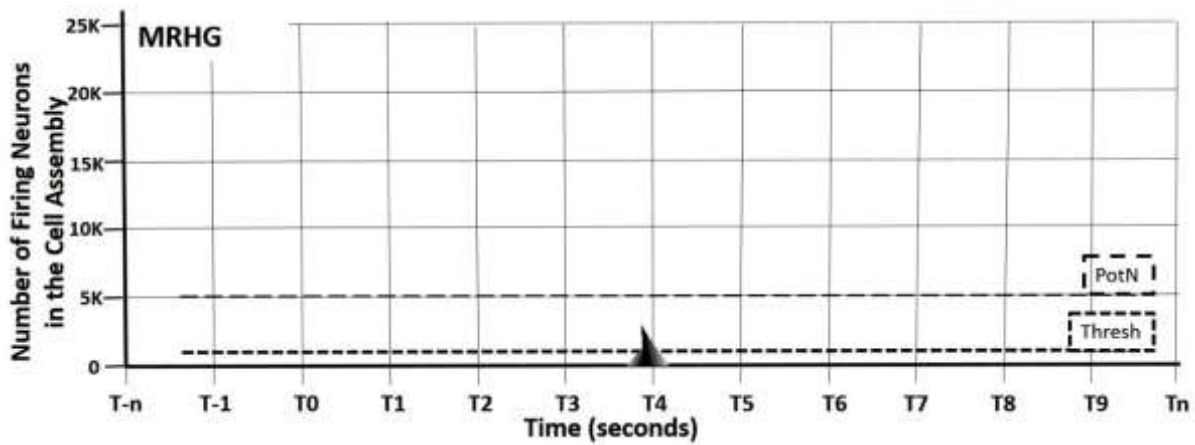

| ID | PotN | Thresh | IgMax | IgFat | P50% | IgTIg | IgTEx | D50% |
|---|---|---|---|---|---|---|---|---|
| MRHG | 5 | 1 | 3 | 2 | 3.7 | 3.8 | 3.9 | 4.0 |

INPUTS:      CA: COGNITIVE – Right Hand Grip (CRHG).

OUTPUTS:   *motor behaviour …*

This is a fairly standard small motor CA (PotN 5K), specialised for the task but one of (tens of?) thousands of other similarly specialised ones (e.g. gripping one's coffee cup before drinking from it and, indeed, picking up any "well known" object). It is ignited by CRHG and extinguishes itself as the grip is transformed into the stable holding behaviour of MRHH.

## 21    CA: TOUCH – Right Hand Grip (TRHG)

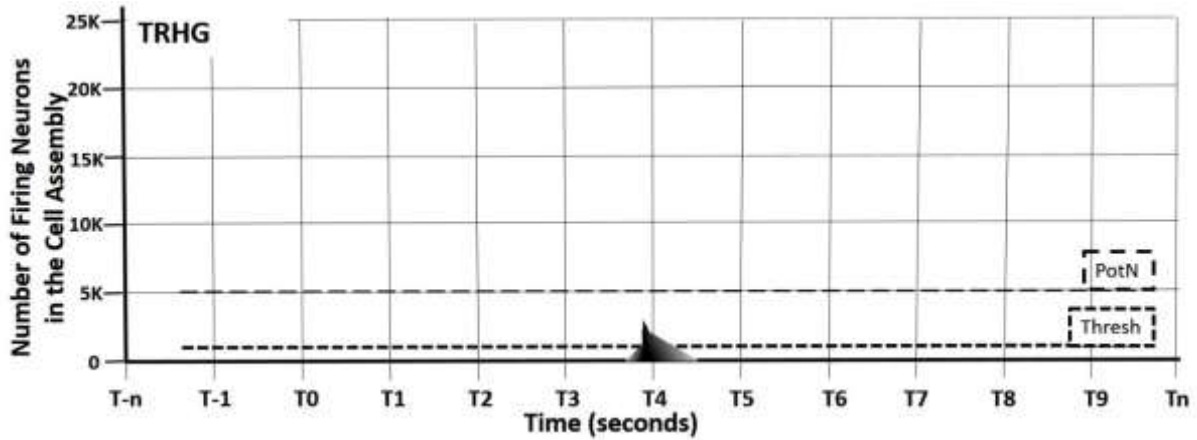

| ID   | PotN | Thresh | IgMax | IgFat | P50% | IgTIg | IgTEx | D50% |
|------|------|--------|-------|-------|------|-------|-------|------|
| TRHG | 5    | 1      | 3     | 2     | 3.7  | 3.8   | 3.9   | 4.3  |

INPUTS:     CA: COGNITIVE – Right Hand Grip (CRHG)

OUTPUTS:   CA: COGNITIVE – Right Hand Grip (CRHG)

Negative feedback control here is crude in that as soon as this CA is ignited, along with its motor complement, it simply confirms that there is adequate, expected grip (e.g. the kettle handle is not damp and friction poor) and sends output to CRHG. It is modelled as decaying slowly (IgTEx – D50% = 0.4 seconds) in case of early "gripping errors".

## 22  CA: COGNITIVE – Right Hand Hold (CRHH)

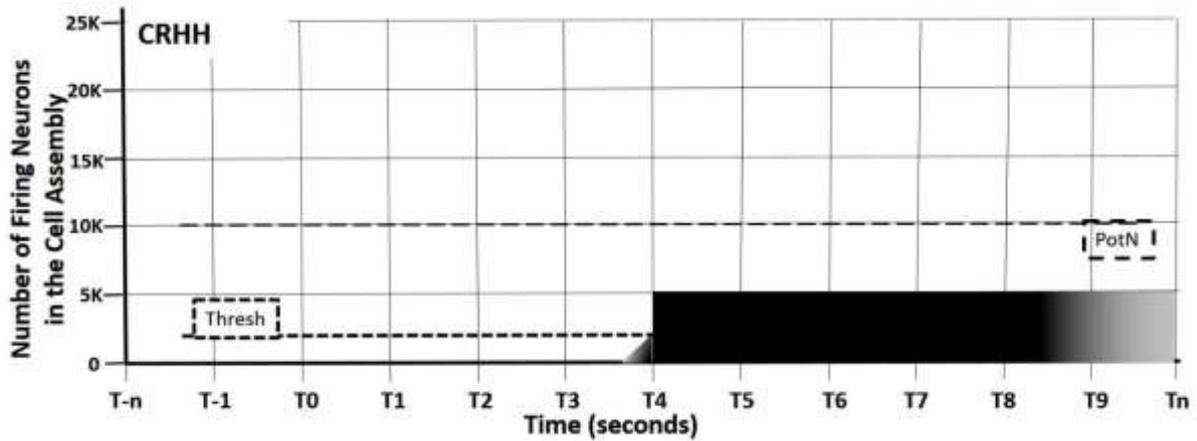

| ID | PotN | Thresh | IgMax | IgFat | P50% | IgTIg | IgTEx | D50% |
|---|---|---|---|---|---|---|---|---|
| CRHH | 10 | 2 | 5 | 5 | 3.8 | 4.0 | - | - |

INPUTS:   CA: COGNITIVE – Right Hand Grip (CRHG).

CA: MOTOR – Right Hand Hold (MRHH).

OUTPUTS:   CA: MOTOR – Right Hand Hold (CRHH),

CA: COGNITIVE – Lift Kettle (CLK).

Unusually in this highly practiced task, this CA is a fairly general one, hence its size (PotN 10K). It has a low threshold (2K), strong relative ignition (5K) and effectively no fatigue. The CA continues ignited beyond the duration of this analysis.

The experiential/introspective psychology, at least, is quite odd about holding objects as once they are held it seems we forget what we are holding. As evidence, often one looks at one's hand during a task to see just what is in it. Obviously different objects are treated differently, but it seems that once a hold is established, it is one or more CAs associated with the object, rather than the hold on it, which remain task relevant, i.e. ignited.

## 23 CA: MOTOR – Right Hand Hold (MRHH)

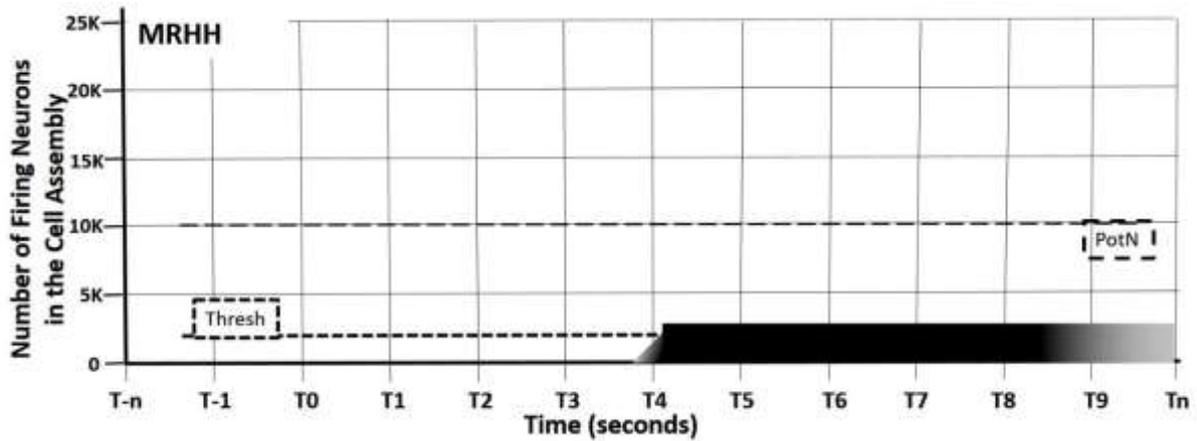

| ID | PotN | Thresh | IgMax | IgFat | P50% | IgTIg | IgTEx | D50% |
|---|---|---|---|---|---|---|---|---|
| MRHH | 10 | 2 | 3 | 3 | 3.9 | 4.1 | - | - |

INPUTS: CA: COGNITIVE – Right Hand Hold (CRHH).

OUTPUTS: CA: COGNITIVE – Right Hand Hold (CRHH)

Following CRHH, it just ignites, persists, and unless there is imperfect performance, e.g. the kettle over the drainer "in flight" hits an obstruction, as a motor CA it causes a solid hold on the kettle handle, notwithstanding later orientations of the kettle itself.

As discussed with CA 06 MSHWA (Motor Stride to Hot Water Area), the CA's actual behaviour will be more complicated than as suggested by the flat line in its SCAM diagram. For example, while going over the drainer, or when decelerating over the right hand sink, then the hold might change; that the SCAM is over simplified at this stage of the research is not denied, it is only a start after all.

## 24   CA: COGNITIVE – LIFT KETTLE (CLK)

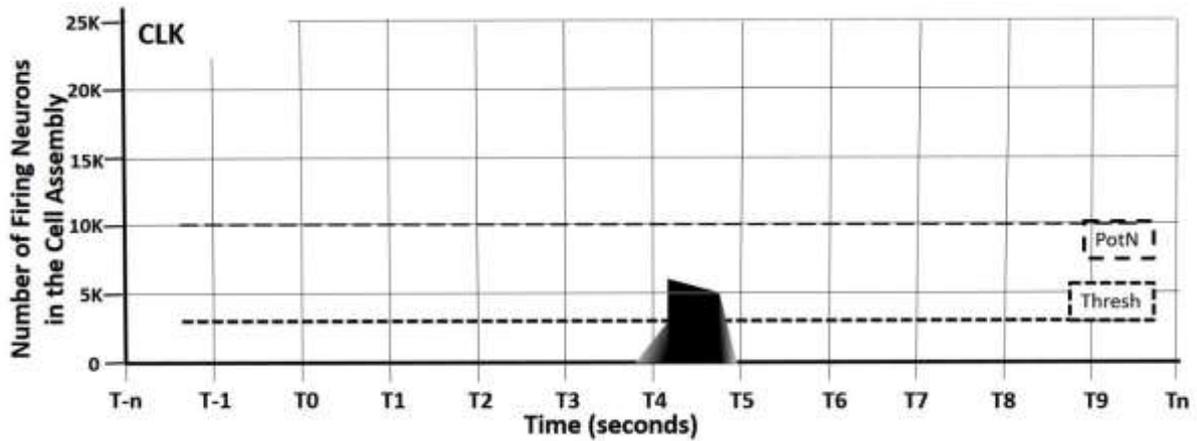

| ID | PotN | Thresh | IgMax | IgFat | P50% | IgTIg | IgTEx | D50% |
|---|---|---|---|---|---|---|---|---|
| CLK | 10 | 3 | 6 | 5 | 4.0 | 4.2 | 4.7 | 4.8 |

INPUTS:   CA: COGNITIVE – Right Hand Hold (CRHH),

CA: VISUAL – Lift Kettle (VLK)

CA: KINAESTHETIC – Kettle Weight (KKW).

OUTPUTS:   CA: VISUAL - Lift Kettle (VLK)

CA: MOTOR – Lift Kettle (MLK),

CA: KINAESTHETIC – Kettle Weight (KKW),

CA: COGNITIVE – Drainer (CD),

CA: COGNITIVE – Move Kettle to Sink (CMKS).

The ergonomics and CA perspective agree that a new subtask starts here, but within the SCAM model the line is blurred in that some CAs are already ignited and will persist beyond the duration of this analysis (CRHH and MRHH).

Empty, the kettle weighs 1.7Kg and if previously boiled water remains in it, it may weigh a third more (e.g. with about a pint/half litre: 2.3Kg/1.7Kg = 1.35). The initial vertical lift of the kettle from its base (it must be vertical because the base has a central, circular hub that the kettle locates on) critically signals the kettle's weight via kinaesthetic feedback (KKW). There is visual tracking of the kettle (VLK). The CA is well primed (P50% - IgTIg = 0.2 seconds) and it persists for longer than the motor behaviour (MLK) because it must ignite both cognitive CAs for the sub-task's continuation (CD and CMKS).

## 25   CA: MOTOR – Lift Kettle (MLK)

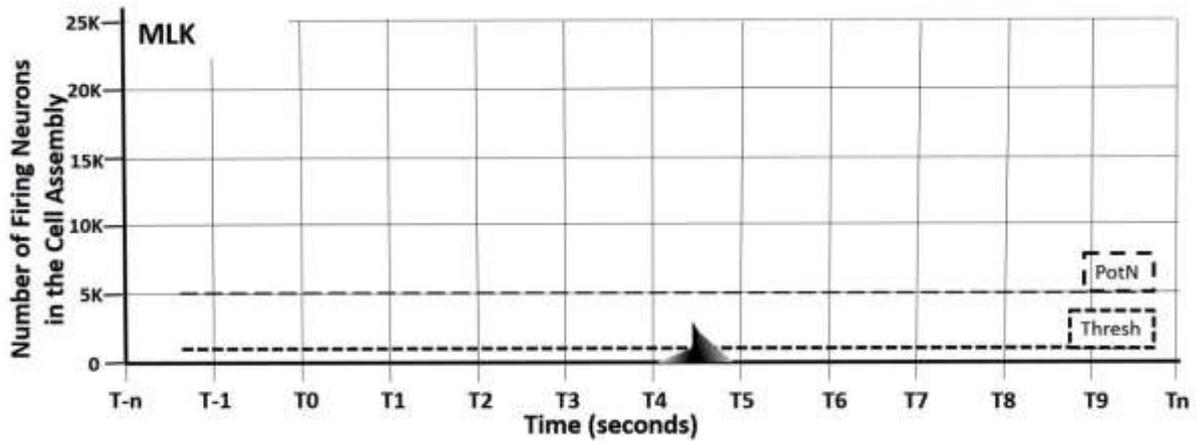

| ID | PotN | Thresh | IgMax | IgFat | P50% | IgTIg | IgTEx | D50% |
|---|---|---|---|---|---|---|---|---|
| MLK | 5 | 1 | 3 | 2 | 4.1 | 4.3 | 4.4 | 4.5 |

INPUTS: CA: COGNITIVE – Lift Kettle (CLK).

OUTPUTS: *motor behaviour ...*

Well primed (P50% - IgTIg = 0.2 seconds) because this is a highly practiced task, and with a low threshold (PotN 5K, Threshold 1k), there is an initial ballistic lift which then comes under kinaesthetic negative feedback control from KKW, which adjusts the rate of the upwards lift, and then close behind this under visual negative feedback control (VLK) via CLK, which starts to orientate the kettle by turning the right wrist clockwise.

The CA is not explicitly extinguished because it segues into the next motor operation, moving the kettle to the sink (MMKS), without a pause, but with a deceleration in the kettle's post-lift trajectory, presumably so that the kettle's path over the drainer can be determined.

## 26   CA: KINAESTHETIC – Kettle Weight (KKW)

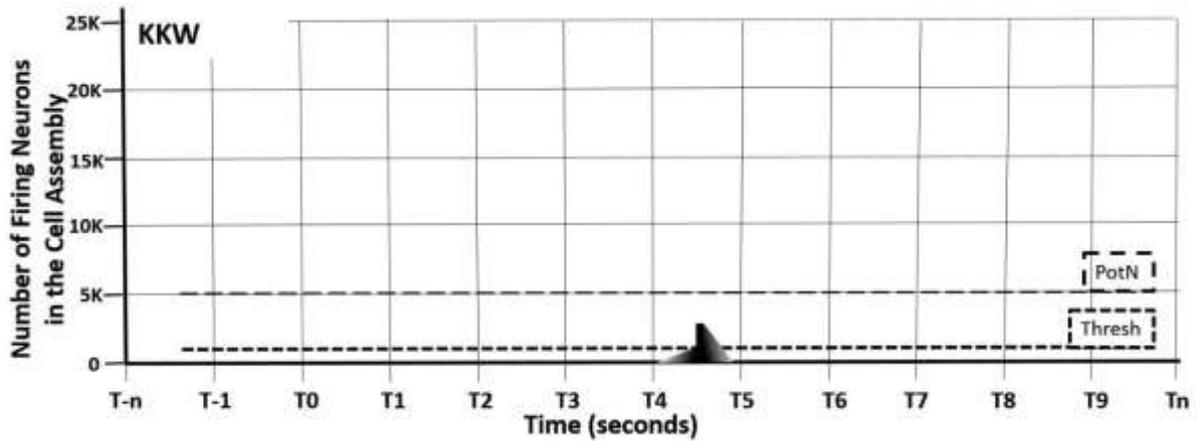

| ID | PotN | Thresh | IgMax | IgFat | P50% | IgTIg | IgTEx | D50% |
|---|---|---|---|---|---|---|---|---|
| KKW | 5 | 1 | 3 | 3 | 4.2 | 4.4 | 4.5 | 4.6 |

INPUTS:     CA: COGNITIVE – Lift Kettle (CLK).

OUTPUTS:   CA: COGNITIVE – Lift Kettle (CLK)

People have an expectation about the weight of objects before they touch them and this is easily demonstrated by the under- or over-lift people produce when such expectations are violated. While this kinaesthetic CA is undoubtedly used whenever objects are lifted, it is particularly germane here as the kettle gives no indication of how much water remains in it until it is lifted. The CA rarely has a conscious representation unless the kettle is unusually full, when, against general house policy, this signals poor energy conservation.

The CA is small (PotN 5K) and easily ignited (Threshold 1K). In this model the CA does not persist, i.e. the kettle's weight is represented in CLK.

## 27  CA: VISUAL – Lift Kettle (VLK)

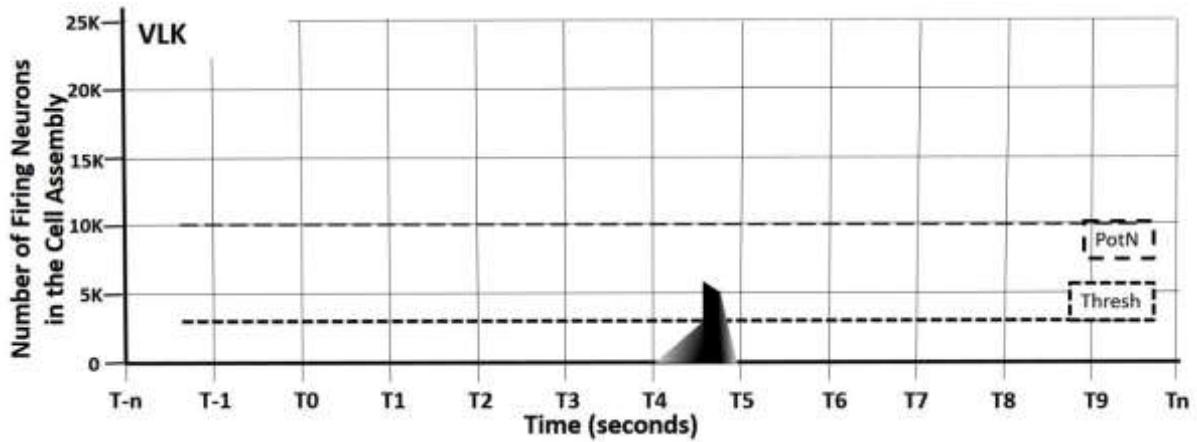

| ID  | PotN | Thresh | IgMax | IgFat | P50% | IgTIg | IgTEx | D50% |
|-----|------|--------|-------|-------|------|-------|-------|------|
| VLK | 10   | 3      | 6     | 5     | 4.3  | 4.5   | 4.6   | 4.7  |

INPUTS:    CA: COGNITIVE – Lift Kettle (CLK).

OUTPUTS:   CA: COGNITIVE – Lift Kettle (CLK).

The kettle comes more into view when it is lifted above the cluttered hot water area (it is initially also obscured by the right hand and forearm). The CA takes over from KKW providing negative feedback to CLK and starts the control of angling the kettle to the right. It is small for a visual CA (PotN 10K) as it involves object tracking and, under movement, a poor percept of the kettle itself.

## 28  CA: COGNITIVE – Drainer (CD)

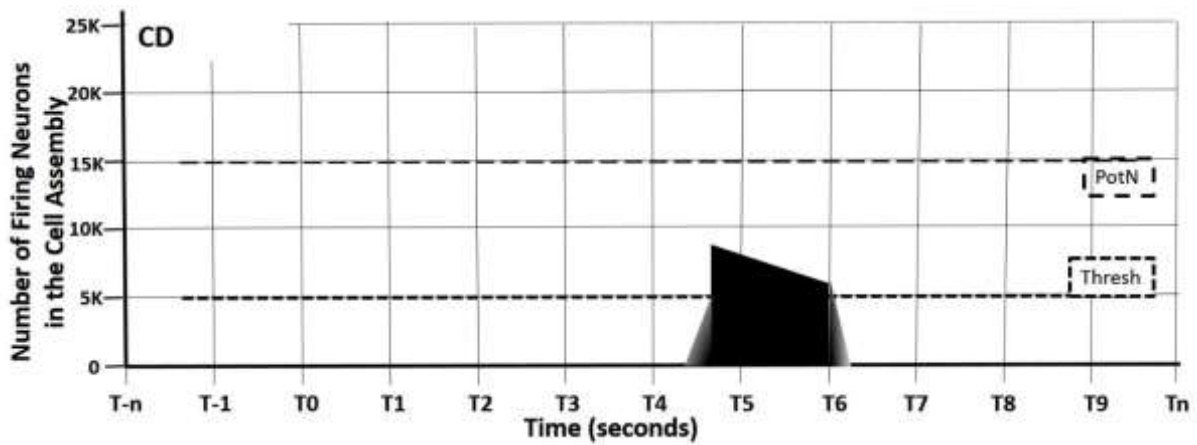

| ID | PotN | Thresh | IgMax | IgFat | P50% | IgTIg | IgTEx | D50% |
|----|------|--------|-------|-------|------|-------|-------|------|
| CD | 15   | 5      | 8     | 6     | 4.5  | 4.6   | 6.0   | 6.1  |

INPUTS:    CA: COGNITIVE – Lift Kettle (CLK),

CA: VISUAL – Drainer (VD).

OUTPUTS    CA: VISUAL – Drainer (VD),

CA: COGNITIVE – Move Kettle to Sink (CMKS).

The steel wire drainer is the most variable object associated with the task because it may be empty or it could be full of washed objects. It is 50cm in depth and 32cm along the draining board, which is the length of the kettle's path over this potential obstacle. Empty, the drainer is 10cm high but the largest pot that is regularly used has a 28cm diameter and this pot's lid, upright but at an angle in the plate rack, also has a maximum height of 28cm.

This is quite a large CA (PotN 15K) to reflect the complexity of a variable object, although the critical information extracted by CMKS is the height at particular depths over which the kettle must pass. N.B. There are other CAs concerning the drainer that are used in other tasks, such as when washing up or when putting dried objects away. The CA is ignited before CLK extinguishes and accepts the final, angled orientation of the kettle effected by CLK.

### 29    CA: VISUAL – Drainer (VD)

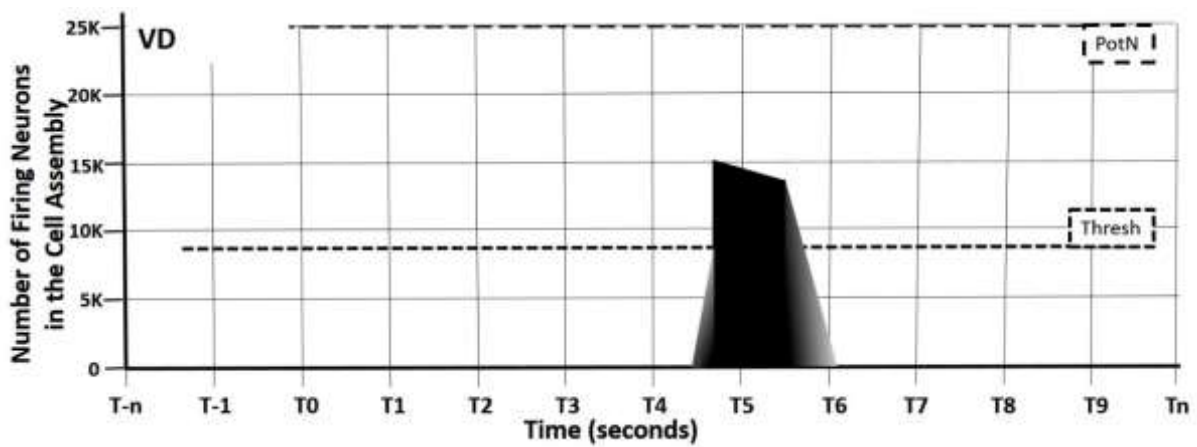

| ID | PotN | Thresh | IgMax | IgFat | P50% | IgTIg | IgTEx | D50% |
|---|---|---|---|---|---|---|---|---|
| VD | 25 | 8 | 15 | 13 | 4.6 | 4.7 | 5.5 | 5.8 |

INPUTS:    CA: COGNITIVE – Drainer (CD).

OUTPUTS    CA: COGNTIVE – Drainer (CD).

The drainer's visual CA is equivalently large (PotN 25K) to its cognitive CA (PotN 15K). As explained below (CMKS), it does not directly feed the moving the kettle to the sink CA, except via CD. It extinguishes quite early as visual attention switches to the kettle's arrival over the sinks.

## 30  CA: COGNITIVE – Move Kettle to Sink (CMKS)

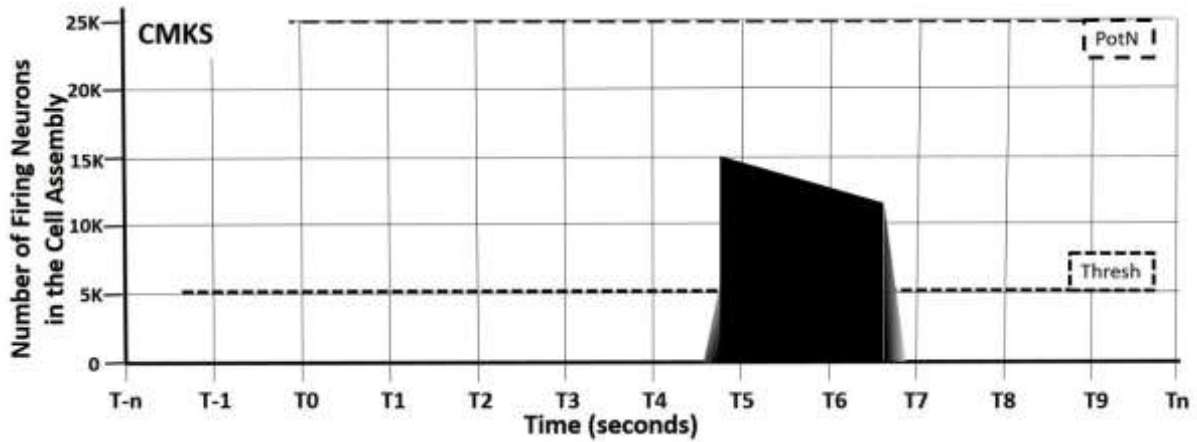

| ID | PotN | Thresh | IgMax | IgFat | P50% | IgTIg | IgTEx | D50% |
|---|---|---|---|---|---|---|---|---|
| CMKS | 25 | 5 | 15 | 12 | 4.7 | 4.8 | 6.6 | 6.7 |

INPUTS:  CA: COGNITIVE – Lift Kettle (CLK)

CA: COGNITIVE – Drainer (CD),

CA: VISUAL – Move Kettle to Sink (VMKS).

OUTPUTS:  CA: VISUAL – Move Kettle to Sink (VMKS),

CA: MOTOR – Move Kettle to Sink (MMKS)

CA: MOTOR – Left Hand Track Kettle Lid (MLHTKL)

CA: KINAESTHETIC – Left Hand track Kettle Lid (KLHTKL),

CA: MOTOR – Shuffle Body to Sink (MSBS),

CA: COGNITIVE – Left Hand Remove Kettle Lid (CLHRKL).

CA: COGNITIVE – Sink (CS).

If the kettle were an aircraft, then it would be one with terrain following radar so as to maintain height-above-ground. The kettle is flown over the drainer in a smooth path that varies in height, and to a lesser extent depth, depending on what, if anything, is in the drainer. What does not happen is that the kettle is flown around the drainer and not over it as this would require a step to be taken back, away from the hot water area, whereas CMKS involves a shuffle to the right so that the body is closer to the sink.

How much the kettle's flight path is "calculated" in advance and how much is under visual negative feedback control is moot. Performance is fast and, perhaps surprisingly, error free, i.e. objects on the drainer are never hit by the kettle even though it may be only a few centimetres above draining objects. The subjective impression following detailed observation for this research is that perhaps one course correction is made mid-flight over the drainer and a second, once that is cleared, to bring the kettle above the main, right most sink.

The CA is large for a cognitive one (PotN 25K) and with a low threshold (5K) because its ignition continues the initial kettle lift (CLK). The kettle's weight information is transferred from CLK to CMKS. At a lower level of analysis this CA might be described by a number of interacting CAs, e.g. concerning open versus negative feedback control and the varying, three dimensional accelerations applied.

## 31  CA: VISUAL – Move Kettle to Sink (VMKS)

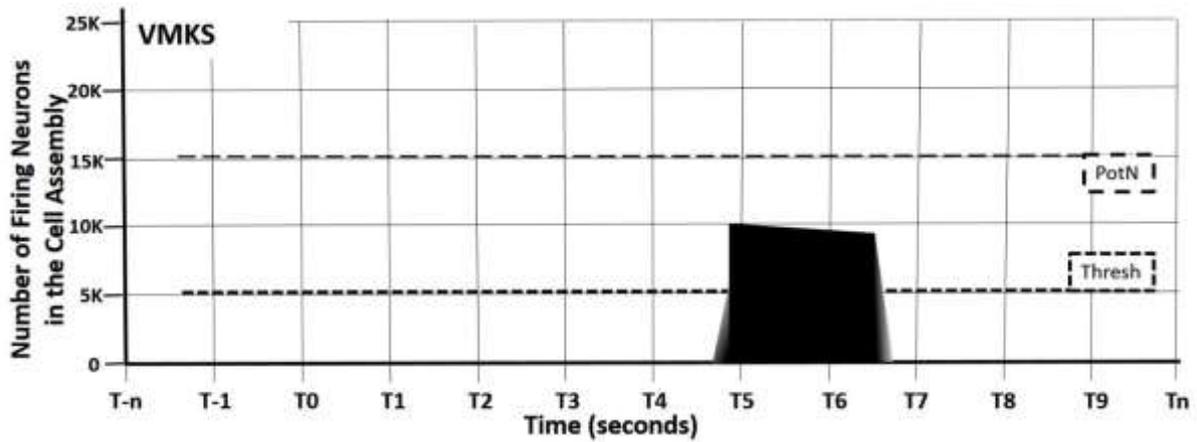

| ID | PotN | Thresh | IgMax | IgFat | P50% | IgTIg | IgTEx | D50% |
|---|---|---|---|---|---|---|---|---|
| VMKS | 15 | 5 | 10 | 9 | 4.8 | 4.9 | 6.5 | 6.6 |

INPUTS:   CA: COGNITIVE – Move Kettle to Sink (CMKS).

OUTPUTS:  CA: COGNITIVE – Move Kettle to Sink (CMKS).

Like the visual CA for lifting the kettle (VLK), this CA is quite small as it really only signals the base of the kettle over the drainer (PotN 15K) and then the general location of the kettle over the sink.

As with CMKS, at a lower level of analysis this CA might be described by several, interacting ones.

## 32 CA: MOTOR – Move Kettle to Sink (MMKS)

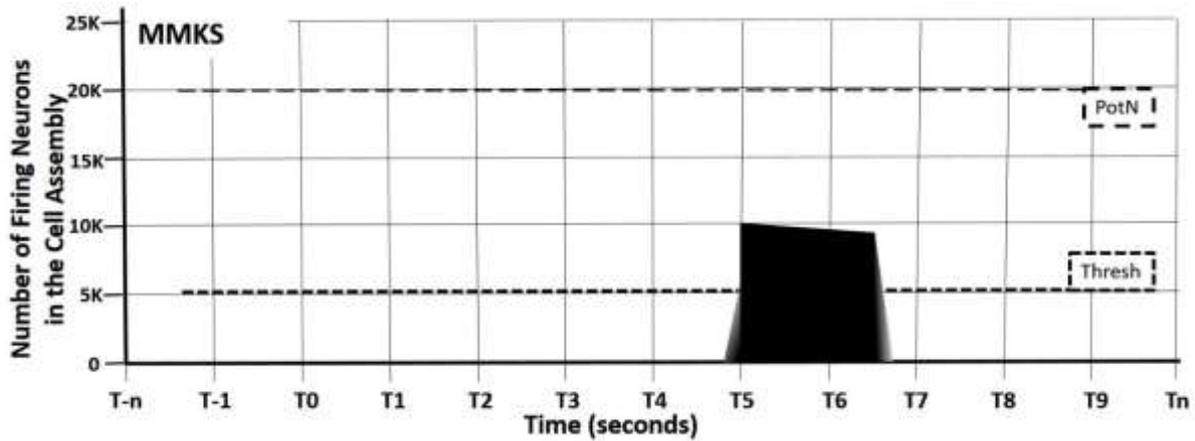

| ID | PotN | Thresh | IgMax | IgFat | P50% | IgTIg | IgTEx | D50% |
|---|---|---|---|---|---|---|---|---|
| MMKS | 20 | 5 | 10 | 9 | 4.9 | 5.0 | 6.5 | 6.6 |

INPUTS: CA: COGNITIVE – Move Kettle to Sink (CMKS).

OUTPUTS: *motor behaviour ...*

The complex motor behaviour is probably carried out by a single CA and illustrates the advantage of using a CA-based model rather a symbolic computational one, that CAs are capable of flexible learning. The CA is modelled as having several general flight paths, e.g. for when the drainer is empty, has a few low height objects, or some big ones draining, and then adapts to specific conditions to quickly and safely fly over the draining board using visual negative feedback via CMKS.

The CA is large for a motor one (PotN 20K) and perhaps only half these neurons will be involved in any particular ignition (IgMax 10K). The CA is explicitly suppressed by CMKS when the kettle is over the main sink; the actual location need not be very precise as the sink is a large target relative to the kettle.

## 33  CA: MOTOR – Left Hand Track Kettle Lid (MLHTKL)

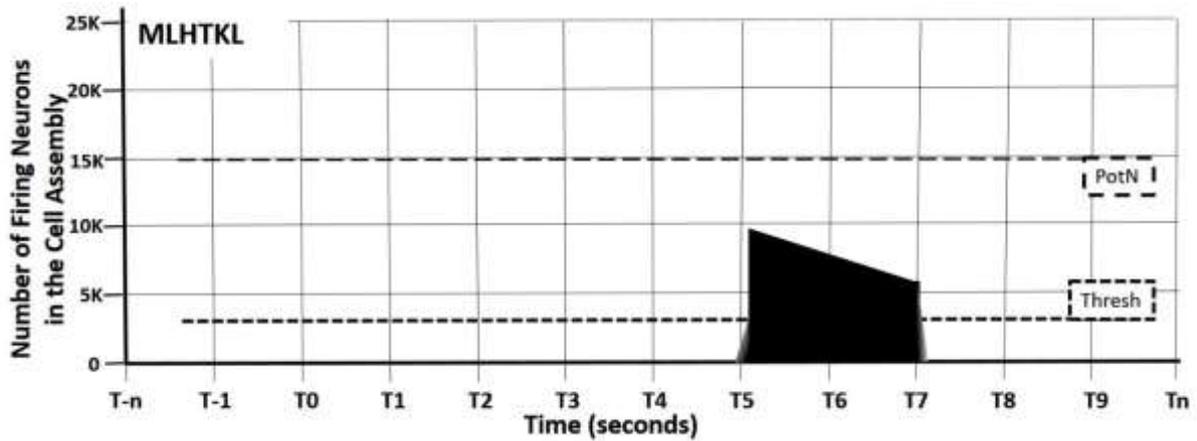

| ID | PotN | Thresh | IgMax | IgFat | P50% | IgTIg | IgTEx | D50% |
|---|---|---|---|---|---|---|---|---|
| MLHTKL | 15 | 3 | 9 | 6 | 5.0 | 5.1 | 7.0 | 7.0 |

INPUTS:   CA: COGNITIVE – Move Kettle to Sink (CMKS),

CA KINAESTHETIC – Left Hand Track Kettle Lid (KLHTKL).

CA: VISUAL – Visual Left Hand (VLH).

CA: COGNITIVE – Left Hand Remove Kettle Lid (CLHRKL).

OUTPUTS:   CA: KINAESTHETIC – Left Hand Track Kettle Lid (KLHTKL).

The left arm/hand has not so far featured in this task, being used for general balance. Out of sight, the left hand is accelerated towards, and then tracks, the kettle's lid so that the left hand is close to it when it appears (VLH). The left hand/wrist will commence to orientate to meet the kettle lid.

The CA is quite large for a motor one (PotN 15K), although we model it as a single CA because the behaviour is continuous. It has additional input once the left hand appears (VLH) and so there is then both kinaesthetic and visual negative feedback to control the final fractions of a second before the kettle lid handle is gripped, at which point this tracking CA is suppressed by CLHRKL. Ignition and visual input comes from CMKS which probably also provides kinaesthetic input about the right hand's location and movement as it grips the kettle.

## 34  CA: KINAESTHETIC – Left Hand Track Kettle Lid (KLHTKL)

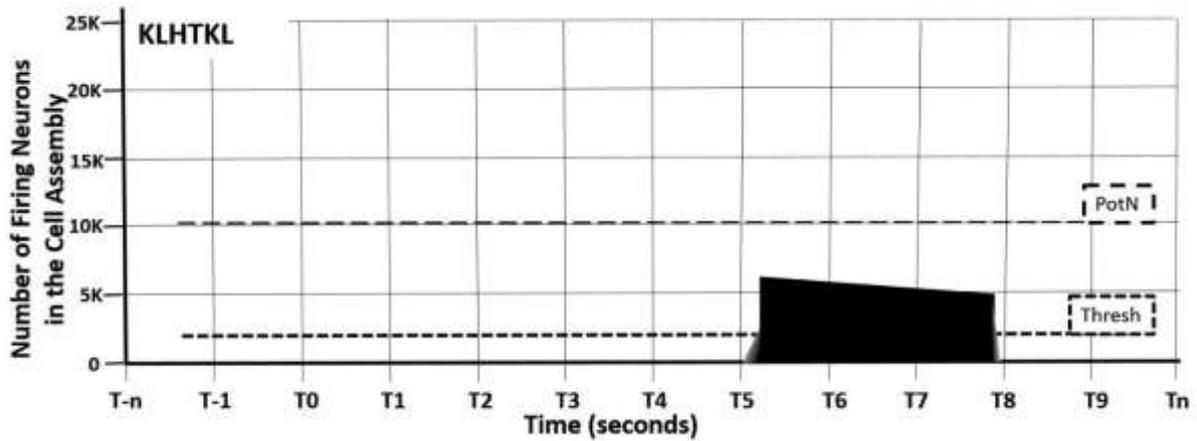

| ID | PotN | Thresh | IgMax | IgFat | P50% | IgTIg | IgTEx | D50% |
|---|---|---|---|---|---|---|---|---|
| KLHTKL | 10 | 2 | 6 | 5 | 5.1 | 5.2 | 7.8 | 7.8 |

INPUTS:   CA: COGNITIVE – Move Kettle to Sink (CMKS),

CA: MOTOR – Left Hand Track Kettle Lid (MLHTKL),

CA: COGNITIVE – Left Hand Remove Kettle Lid (CLHRKL),

CA: MOTOR – Replace Kettle Lid Left Hand (MRKLLH).

OUTPUTS:   CA: MOTOR – Left Hand Track Kettle Lid (MLHTKL),

CA: COGNITIVE – Left Hand remove Kettle Lid (CLHRKL),

CA: MOTOR – Replace Kettle Lid Left Hand (MRKLLH).

There must be all sorts of kinaesthetic feedback involved in the left hand tracking the kettle lid, then touching and gripping it, before the lid is replaced (CRKLLH).  The CA is ignited by CMKS and provides negative feedback cycles to MLHTKL and other motor CAs: CLHRKL before CRKLLH.  Unlike MLHTKL, it is not supressed but decays away once motor inputs terminate.

Kinaesthetic CAS are generally on the small side because of the quality of their output, but this one is quite large (PotN 10K) and with a low threshold (2K) and little decay (IgMax 6K, IgFat 5K) because, persisting for over two seconds, fatiguing neurons will be replaced from those so far not used within PotN.

## 35   CA: MOTOR – Shuffle Body to Sink (MSBS)

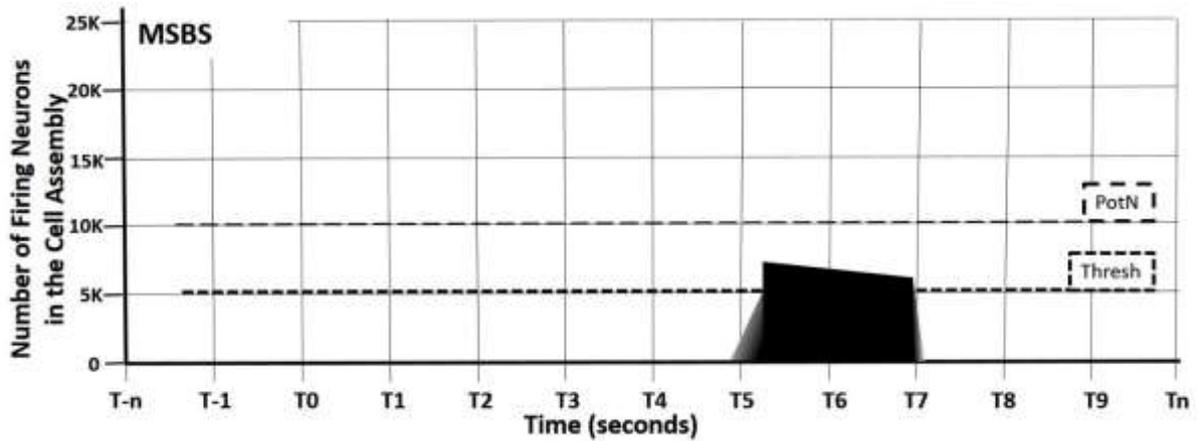

| ID   | PotN | Thresh | IgMax | IgFat | P50% | IgTIg | IgTEx | D50% |
|------|------|--------|-------|-------|------|-------|-------|------|
| MSBS | 10   | 5      | 7     | 6     | 5.1  | 5.3   | 6.9   | 7.0  |

INPUTS:    CA: COGNITIVE – Move Kettle to Sink (CMKS).

OUTPUTS:   *motor behaviour …*

This may be a super-practiced task but the movement of the body from the hot water corner to the sink is ungainly and variable, and although irrelevant, the subject isn't normally conscious of this behaviour.  The knees are close to the under sink cabinets so the move to the sink involves the hips and a sideways stretch of first the right and then the left foot and then some small foot corrections, although occasionally the final position is one where most of the body weight is on the right foot.  The shuffle may continue for some time after the kettle has reached the sink, i.e. in parallel with the next sub-task of removing the kettle's lid.

## 36   CA: COGNITIVE – Sink (CS)

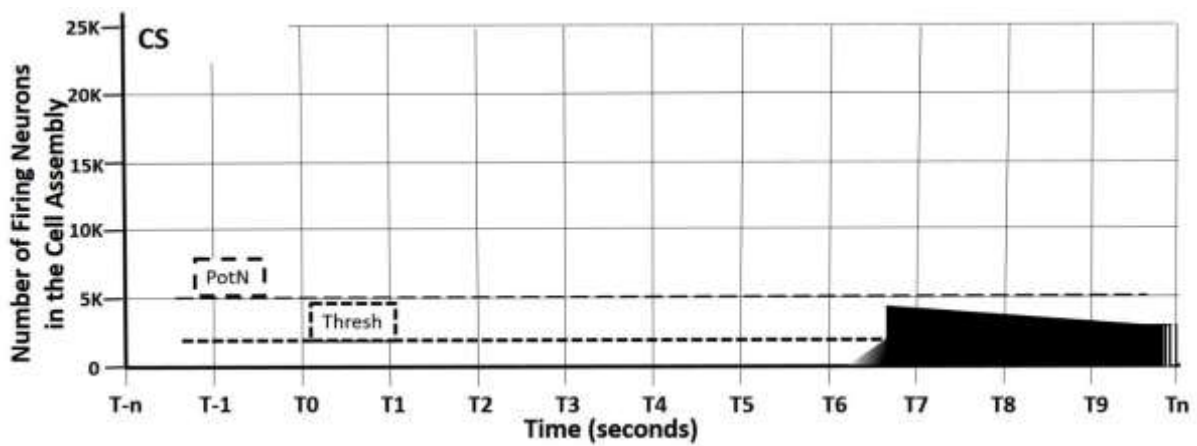

| ID | PotN | Thresh | IgMax | IgFat | P50% | IgTIg | IgTEx | D50% |
|----|------|--------|-------|-------|------|-------|-------|------|
| CS | 5    | 2      | 4     | 3     | 6.5  | 6.7   | -     | -    |

INPUTS:     CA: COGNITIVE – Move Kettle to Sink (CMKS),

CA: VISUAL – Sink (VS).

OUTPUTS:    CA: VISUAL – Sink (VS),

CA: COGNITIVE – Left Hand Remove Kettle Lid (CLHRKL).

The sink here is the larger, rightmost of the pair and it is usually empty; if it is not empty then, like CKEC at the start of this analysis, other CAs would be ignited to assess the sink's state and decide how to orientate the kettle so it can still be filled.

Empty, the sink's cognitive representation here need not be large (PotN 5K) as it is a large target relative to the kettle, which only needs to be centred above the sink so that it can be emptied.

This CA and its associated CA (VS) are assumed to persist beyond the analysis as they provide, albeit perhaps weak, context information to the following CAs (not shown on the CAAR diagram, Figure 9).

## 37    CA: VISUAL – Sink (VS)

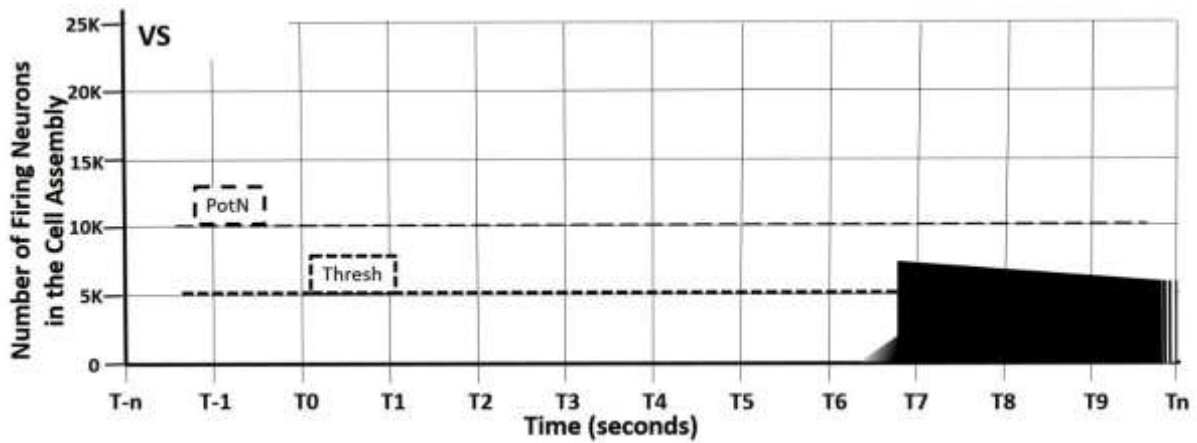

| ID | PotN | Thresh | IgMax | IgFat | P50% | IgTIg | IgTEx | D50% |
|---|---|---|---|---|---|---|---|---|
| VS | 10 | 5 | 7 | 6 | 6.6 | 6.8 | - | - |

INPUTS:     CA: COGNITIVE –Sink (CS).

OUTPUTS:    CA: COGNITIVE – Sink (CS).

Made of brushed steel, the visual representation of the sink is fairly simple (PotN 10K) as it is relatively featureless and colourless (N.B. In the human visual system there would be many low spatial frequency components; and in computational terms standard compression algorithms of a photograph would be particularly effective).

## 38   CA: COGNITIVE – Left Hand Remove Kettle Lid (CLHRKL)

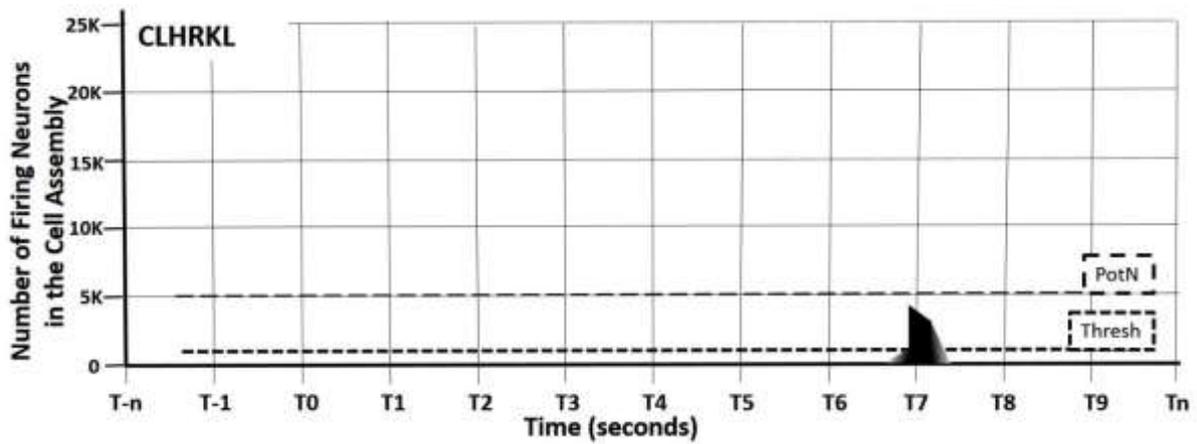

| ID | PotN | Thresh | IgMax | IgFat | P50% | IgTIg | IgTEx | D50% |
|---|---|---|---|---|---|---|---|---|
| CLHRKL | 5 | 1 | 4 | 3 | 6.8 | 6.9 | 7.2 | 7.3 |

INPUTS:   CA: COGNITIVE – Move Kettle to Sink (CMKS)

CA: COGNITIVE – Sink (CS),

CA: KINAESTHETIC – Left Hand Track Kettle Lid (KLHTKL),

CA: VISUAL – Kettle Lid (VKL)

CA: VISUAL – Left Hand (VLH),

CA: VISUAL – Kettle Without Lid (VKWL).

OUTPUTS:   CA: KINAESTHETIC – Left Hand Track Kettle Lid (KLHTKL),

CA: VISUAL – Kettle Lid (VKL)

CA: VISUAL – Left Hand (VLH),

CA: MOTOR – Left Hand Remove Kettle Lid (MLHRKL),

CA: VISUAL – Kettle Without Lid (VKWL),

CA: COGNITIVE – Empty Kettle (CEK).

CA: MOTOR – Left Hand Track Kettle Lid (MLHTKL).

On the adage that the act of doing a TA improves, by method iteration (Section 1), even the earliest analysis stages, then this CA provides a good example. Initially the subtask seemed remarkable for its speed (say a third of a second) and accuracy (it virtually never fails on the first attempt); it took careful, further observation for this research to be able to model it. The initial problem was that the first analysis only included the left arm/hand once it came into operation to remove the kettle's lid. Further observation showed that the left hand was tracking the kettle's lid soon after the kettle starts moving towards the sink (CMKS) and that the lid is closely tracked by the left hand (MLHTKL and KLHTKL) during its flight over the drainer.

Like the right hand approaching the kettle (CRHA), it must start with a ballistic movement as the left hand is not in view and then it must come under visual negative feedback control for the fingers to grip the kettle lid's handle, which can be at any angle on top of the kettle, but this is far less variability than exists in the hot water area.

This CA and its associates could be analysed in much greater detail than is provided at the level of analysis we've chosen. In the analysis offered the CA is small (PotN 5K), with a low threshold (1K), and, being highly specialised, IgMax is proportionally high (4K). In an alternative CAA this CA could be larger or, as we suspect, there are many component CAs that we have not modelled in our analysis.

## 39   CA: VISUAL – Kettle Lid (VKL)

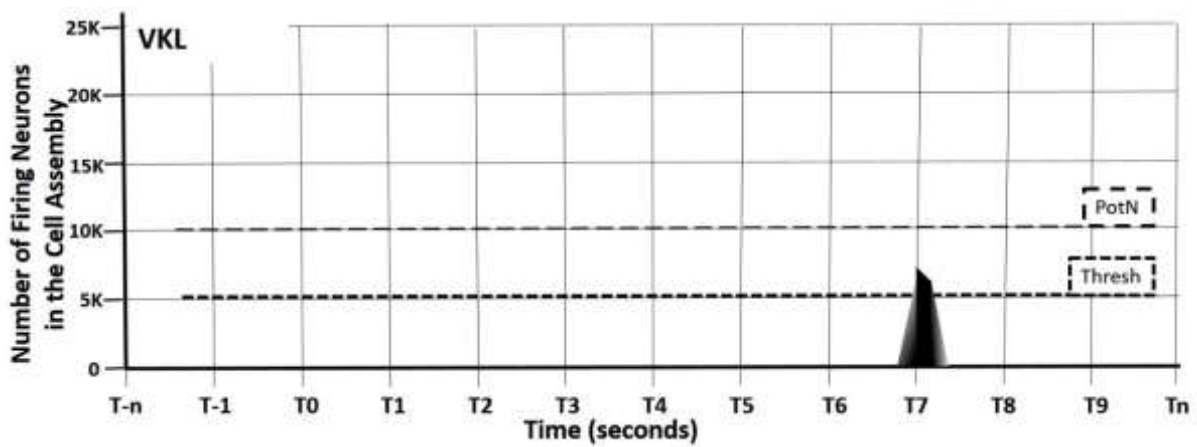

| ID  | PotN | Thresh | IgMax | IgFat | P50% | IgTIg | IgTEx | D50% |
|-----|------|--------|-------|-------|------|-------|-------|------|
| VKL | 10   | 5      | 7     | 6     | 6.9  | 7.0   | 7.1   | 7.2  |

INPUTS:     CA: COGNITIVE – Left Hand Remove Kettle Lid (CLHRKL).

OUTPUTS:    CA: COGNITIVE – Left Hand Remove Kettle Lid (CLHRKL).

The kettle lid is a black/dark grey plastic with an inverted dished top and a simple sold bar across this to act as the handle. What the CA needs to represent is the angle of the handle and the three dimensional location of the lid on the top of the kettle; the latter is no doubt determined by binocular parallax (the different images in the eyes caused by the eyes' horizontal separation). It doesn't need to be large for a visual CA (PotN 10K).

## 40  CA: VISUAL – Left Hand (VLH)

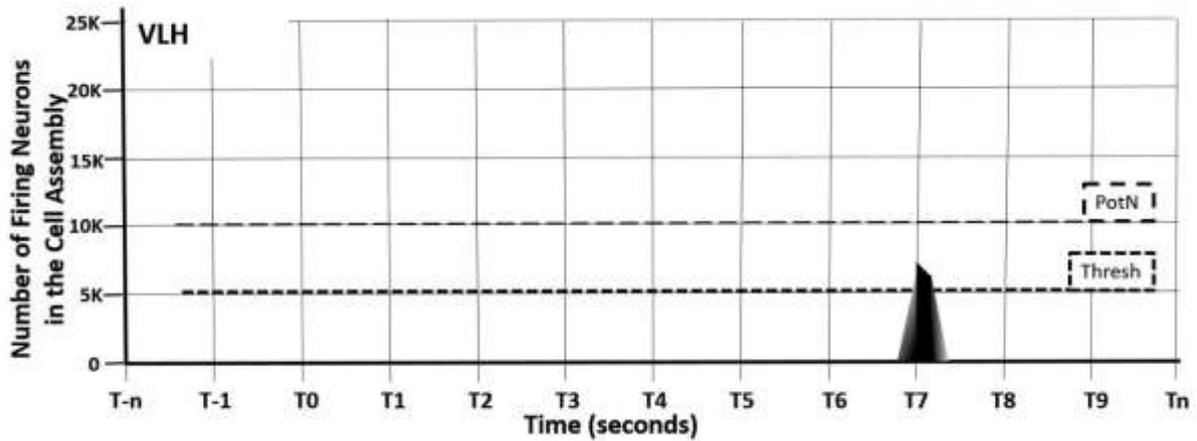

| ID | PotN | Thresh | IgMax | IgFat | P50% | IgTIg | IgTEx | D50% |
|---|---|---|---|---|---|---|---|---|
| VLH | 10 | 5 | 7 | 6 | 6.9 | 7.0 | 7.1 | 7.2 |

INPUTS:     CA: COGNITIVE – Left Hand Remove Kettle Lid (CLHRKL).

OUTPUTS:    CA: COGNITIVE – Left Hand Remove Kettle Lid (CLHRKL).

This doesn't need to be a big visual CA (PotN 10K) as its purpose is only for control of the left hand's final approach to the kettle's lid.  It is also assumed, because of the speed and accuracy (see CLHRKL), that the CA is not that large.

## 41  CA: MOTOR – Left Hand Remove Kettle Lid (MLHRKL)

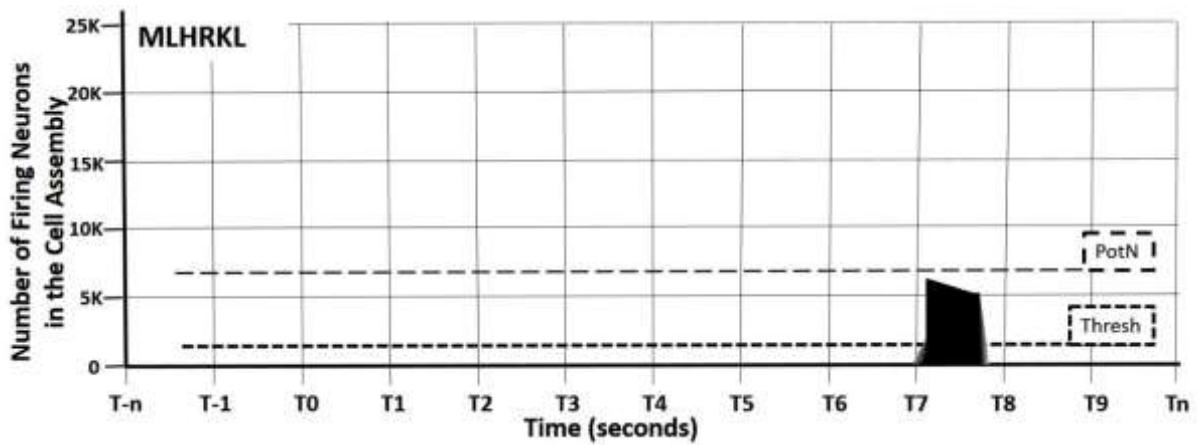

| ID | PotN | Thresh | IgMax | IgFat | P50% | IgTIg | IgTEx | D50% |
|---|---|---|---|---|---|---|---|---|
| MLHRKL | 7 | 2 | 6 | 5 | 7.0 | 7.1 | 7.7 | 7.7 |

INPUTS:     CA: COGNITIVE – Left Hand Remove Kettle Lid (MLHRKL).

OUTPUTS:    *motor behaviour …*

This is a snatch, hold and move away action which in this analysis is modelled by a single CA (PotN 7K) because, again, of the speed of the initial behaviour, although at a more detailed level it might be treated as compound behaviour involving several CAs. On the other hand, a single CA, as here, seems equally plausible, with it smoothly combining the component behaviours. The CA remains ignited, holding the lid away from the kettle, until the lid is replaced about half a second later (CRKLLH and MRKLLH).

## 42   CA: VISUAL – Kettle Without Lid (VKWL)

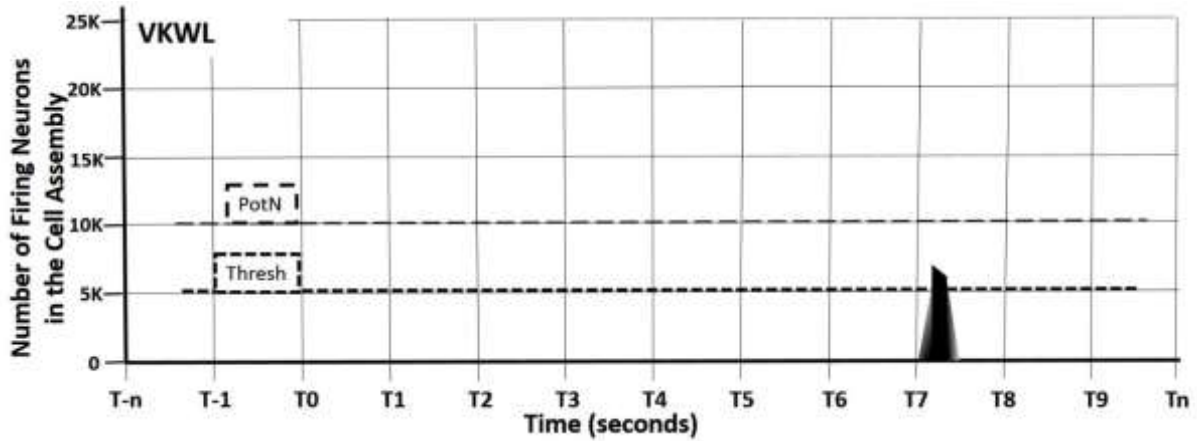

| ID | PotN | Thresh | IgMax | IgFat | P50% | IgTIg | IgTEx | D50% |
|---|---|---|---|---|---|---|---|---|
| VKWL | 10 | 5 | 7 | 6 | 7.1 | 7.2 | 7.3 | 7.4 |

INPUTS:   CA: COGNITIVE – Left Hand Remove Kettle Lid (CLHRKL).

OUTPUTS:   CA: COGNITIVE – Left Hand Remove Kettle Lid (CLHRKL).

This is a small (PotN 10K) visual CA that confirms the kettle's lid is off. It provides feedback to CLHRLK which then allows the ignition of CEK to empty the kettle.

## 43    CA: COGNITIVE – Empty Kettle (CEK)

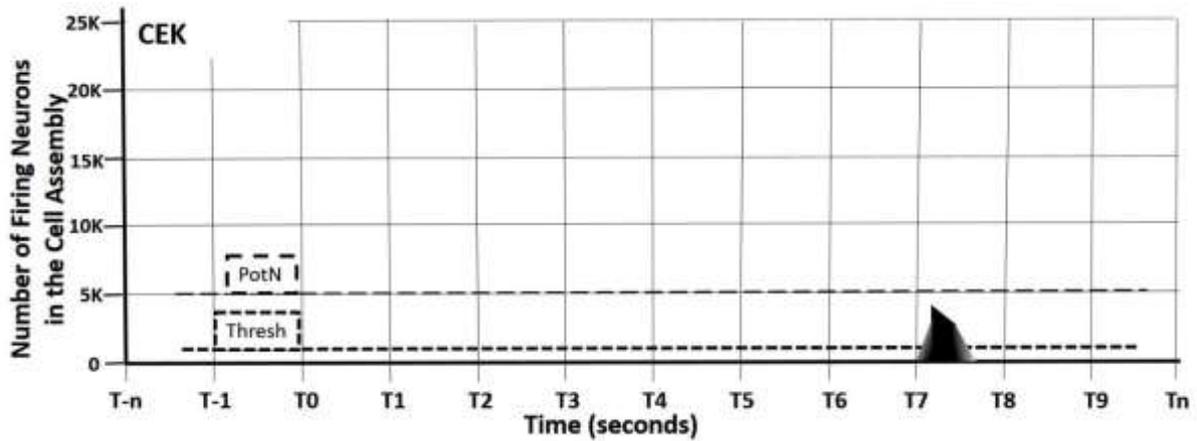

| ID | PotN | Thresh | IgMax | IgFat | P50% | IgTIg | IgTEx | D50% |
|---|---|---|---|---|---|---|---|---|
| CEK | 5 | 1 | 4 | 3 | 7.1 | 7.2 | 7.4 | 7.5 |

INPUTS:    CA: COGNITIVE – Left Hand Remove Kettle Lid (CLHRKL),

CA: VISUAL – Kettle Empty (VKE).

OUTPUTS:    CA: MOTOR – Right Hand Invert Kettle (MRHIK),

CA: VISUAL – Kettle Empty (VKE)

CA: COGNITIVE – Kettle Empty (CKE).

Residue water in the kettle is never re-boiled. The kettle is emptied very rapidly by turning the kettle upside down; there is a brief physical delay as the water falls out. Inverting the kettle, however, involves a single, fast right wrist rotation to the left. The CA is small (PotN 5K) with a low threshold (1K).

## 44  CA: MOTOR – Right Hand Invert Kettle (MRHIK)

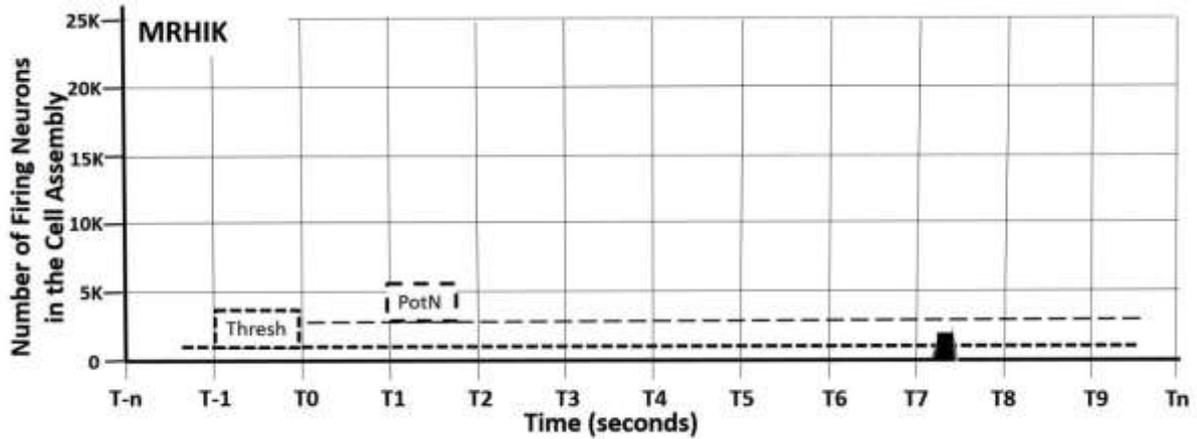

| ID | PotN | Thresh | IgMax | IgFat | P50% | IgTIg | IgTEx | D50% |
|---|---|---|---|---|---|---|---|---|
| MRHIK | 3 | 1 | 2 | 2 | 7.2 | 7.3 | 7.4 | 7.4 |

INPUTS:  CA: COGNITIVE – Empty Kettle (CEK).

OUTPUTS:  *motor behaviour …*

This is a really small CA (PotN 3K) as the right wrist is rotated to the left (anticlockwise) to its maximum extent. Mostly it is open loop control, although there is probably kinaesthetic negative feedback control, which isn't modelled in this analysis, and might involve the right elbow and shoulder which, starting to lift as the wrist rotation approaches its maximum, may contribute to the CA extinguishing.

## 45  CA: VISUAL – Kettle Empty (VKE)

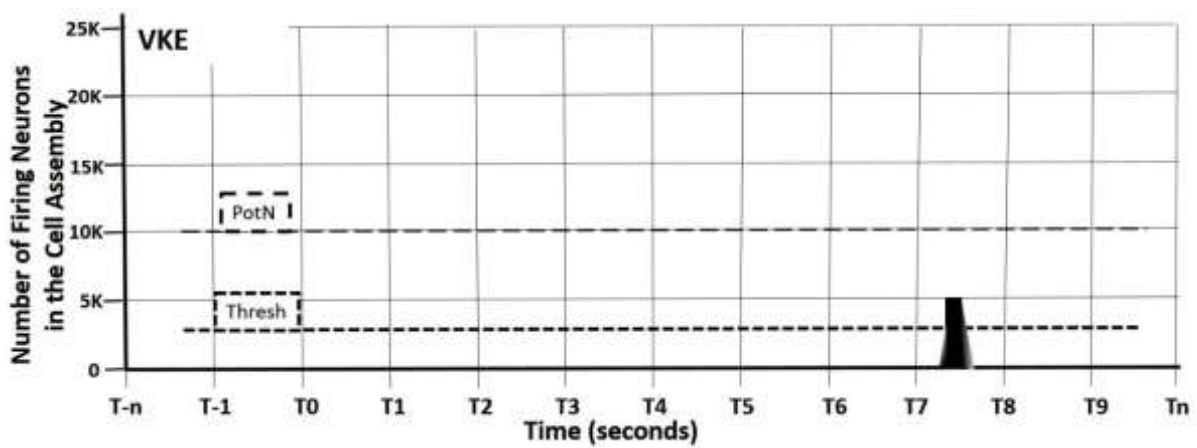

| ID | PotN | Thresh | IgMax | IgFat | P50% | IgTIg | IgTEx | D50% |
|---|---|---|---|---|---|---|---|---|
| VKE | 10 | 3 | 5 | 5 | 7.3 | 7.4 | 7.5 | 7.6 |

INPUTS:  CA: COGNITIVE – Empty Kettle (CEK).

OUTPUTS: CA: COGNITIVE –Empty Kettle (CEK).

The water falls out of the kettle in a lump; the splash remains within the sink; the critical thing for the CA is that the event has ended. The CA ignites CKE via CEK.

## 46    CA: COGNITIVE – Kettle Empty (CKE)

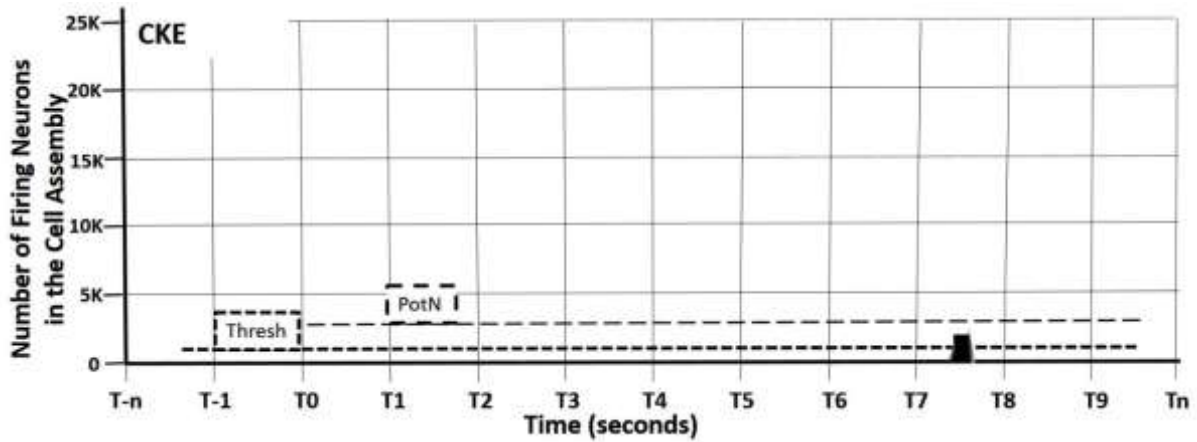

| ID  | PotN | Thresh | IgMax | IgFat | P50% | IgTIg | IgTEx | D50% |
|-----|------|--------|-------|-------|------|-------|-------|------|
| CKE | 3    | 1      | 2     | 2     | **7.4** | 7.5 | 7.6 | 7.6 |

INPUTS: CA: COGNITIVE –Empty Kettle (CEK).

OUTPUTS: CA: COGNITIVE – Right Hand Orientate Kettle (CRHOK).

Primarily concerned with signally that the kettle is empty, rationally this CA ought to exist, but in the CAA described it really only functions as a place marker that ignites CRHOK. An alternative CAA could equally plausible have CRHOK ignited by VKE. It is modelled as a very small CA (PotN 3K) and transient, lasting 100ms or less.

## 47   CA: COGNITIVE – Right Hand Orientate Kettle (CRHOK)

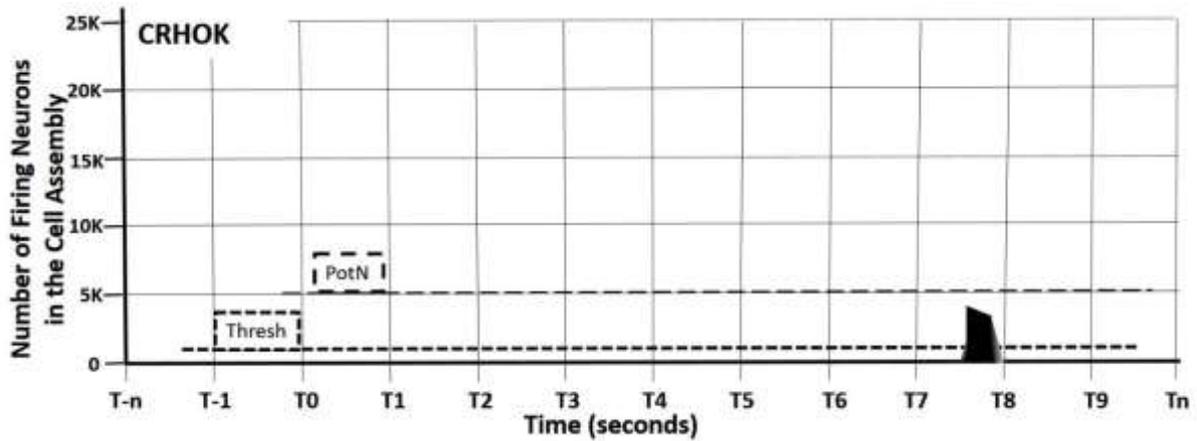

| ID | PotN | Thresh | IgMax | IgFat | P50% | IgTIg | IgTEx | D50% |
|---|---|---|---|---|---|---|---|---|
| CRHOK | 5 | 1 | 4 | 3 | 7.5 | 7.6 | 7.8 | 7.9 |

INPUTS:  CA: COGNITIVE – Kettle Empty (CKE),

CA: VISUAL – Right Hand Orientate Kettle (VRHOK).

OUTPUTS:  CA: VISUAL – Right Hand Orientate Kettle (VRHOK)

CA: MOTOR – Right Hand Orientate Kettle (MRHOK),

CA: COGNITIVE – Replace Kettle Lid with Left Hand (CRKLLH).

This is the opposite of inverting the kettle (CRHIK) and involves a right wrist rotation of about 100 degrees so that the kettle is returned to being upright and roughly angled towards the filtered water tap.  It is a small CA (PotN 5K) and visual negative feedback control (VRHOK) primarily concerns the end of the rotation and halting its motor CA (MRHOK)

### 48  CA: VISUAL – Right Hand Orientate Kettle (VRHOK)

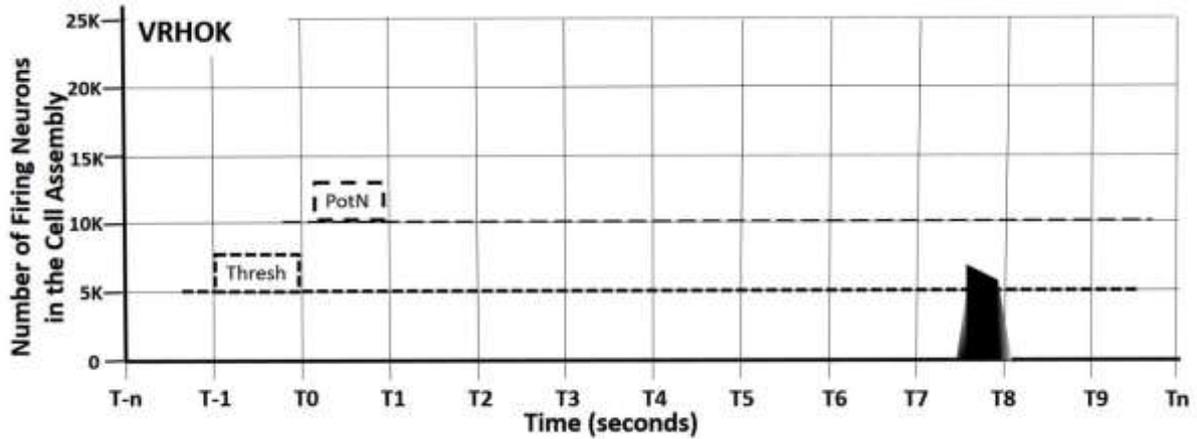

| ID | PotN | Thresh | IgMax | IgFat | P50% | IgTIg | IgTEx | D50% |
|---|---|---|---|---|---|---|---|---|
| VRHOK | 10 | 5 | 7 | 6 | 7.5 | 7.6 | 7.9 | 8.0 |

INPUTS:   CA: COGNITIVE – Right Hand Orientate Kettle (CRHOK).

OUTPUTS:   CA: COGNITIVE – Right Hand Orientate Kettle (CRHOK).

Since the previous visual target was the emptied, inverted kettle, then visual attention is already directed to the kettle.  The initial wrist rotation is fast but as it decelerates to a halt then this CA provides the final control that orientates the kettle and then leads to the suppression of MRHOK via CRHOK.

### 49  CA: MOTOR – Right Hand Orientate Kettle (MRHOK)

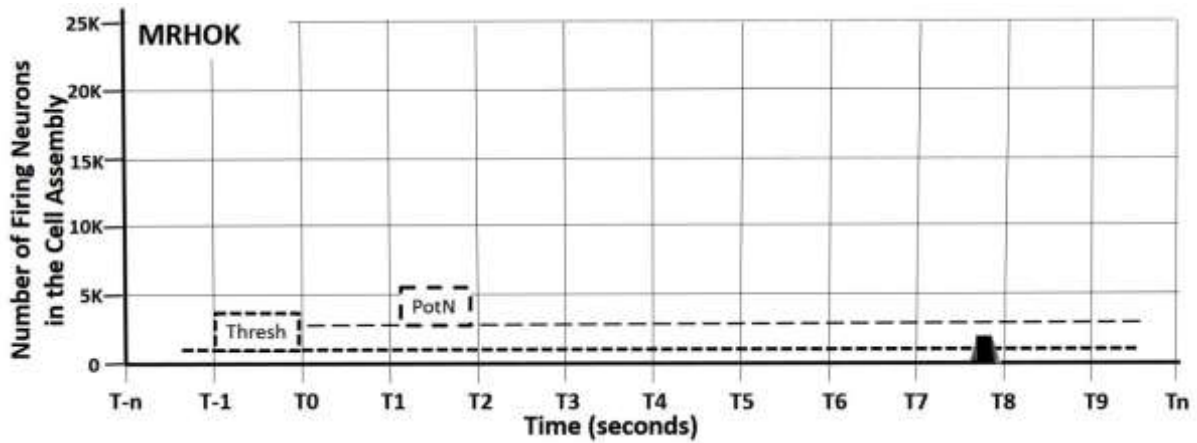

| ID | PotN | Thresh | IgMax | IgFat | P50% | IgTIg | IgTEx | D50% |
|---|---|---|---|---|---|---|---|---|
| MRHOK | 3 | 1 | 2 | 2 | 7.6 | 7.7 | 7.8 | 7.9 |

INPUTS:   CA: COGNITIVE – Right Hand Orientate Kettle (CRHOK).

OUTPUTS:   *motor behaviour ...*

A small motor CA (PotN 5K), it is ignited and then supressed by CRHOK. Like MRHIK, there is probably kinaesthetic feedback which is not modelled here.

## 50   CA: COGNITIVE – Replace Kettle Lid with Left Hand (CRKLLH)

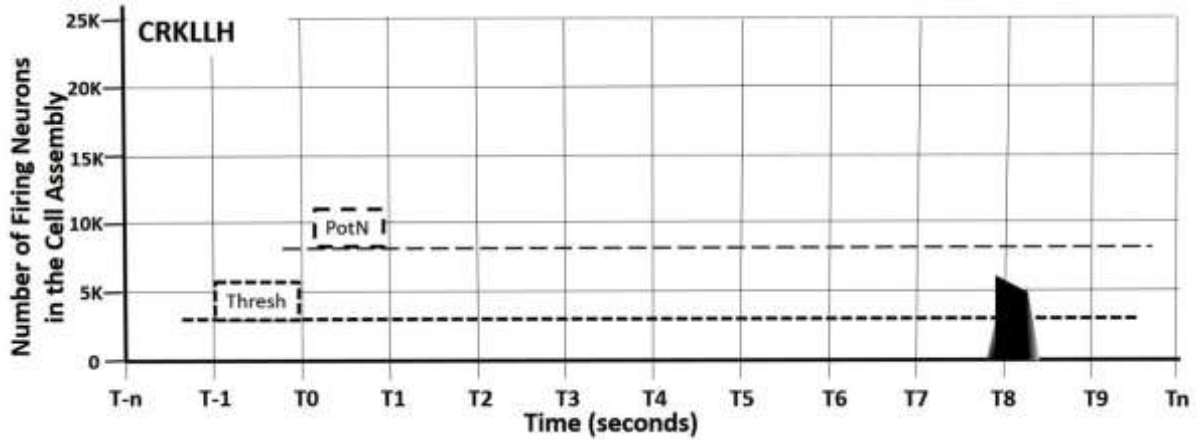

| ID | PotN | Thresh | IgMax | IgFat | P50% | IgTIg | IgTEx | D50% |
|---|---|---|---|---|---|---|---|---|
| CRKLLH | 8 | 3 | 6 | 5 | 7.8 | 7.9 | 8.2 | 8.3 |

INPUTS:    CA: COGNITIVE – Right Hand Orientate Kettle (CRHOK),

CA: VISUAL – Replace Kettle Lid with Left Hand (VRKLLH).

OUTPUTS:   CA: VISUAL – Replace Kettle Lid with Left Hand (VRKLLH)

CA: MOTOR – Replace Kettle Lid with Left Hand (MRKLLH),

CA: COGNITIVE – Move Kettle to Tap (CMKT).

The kettle is filled through its spout so its lid is replaced by the left hand before filling. The lid has been held in roughly the correct position, slightly above the top of the kettle. The lid is nearly always accurately inserted into the kettle in a single motion under visual negative feedback control.

A slight wobble from the left wrists ensures the lid is correctly located in the final few tens of milliseconds. Whatever kinaesthetic feedback from left, and right, hands is not modelled being too fast and at too low a level of detail. In any case, bringing the hands together, with or without an intervening object, are expert skills everyone learns very early in life. Similarly, we have not modelled audio inputs, but there is a "click" when the lid locates, although this is only noticeable in its absence, e.g. when the washing machine is making so much noise that such quiet noises cannot be detected.

The CA ignites its motor component (MRKLLH) and then explicitly supresses it on confirmation that the lid is correctly in place.

### 51    CA: VISUAL – Replace Kettle Lid with Left Hand (VRKLLH)

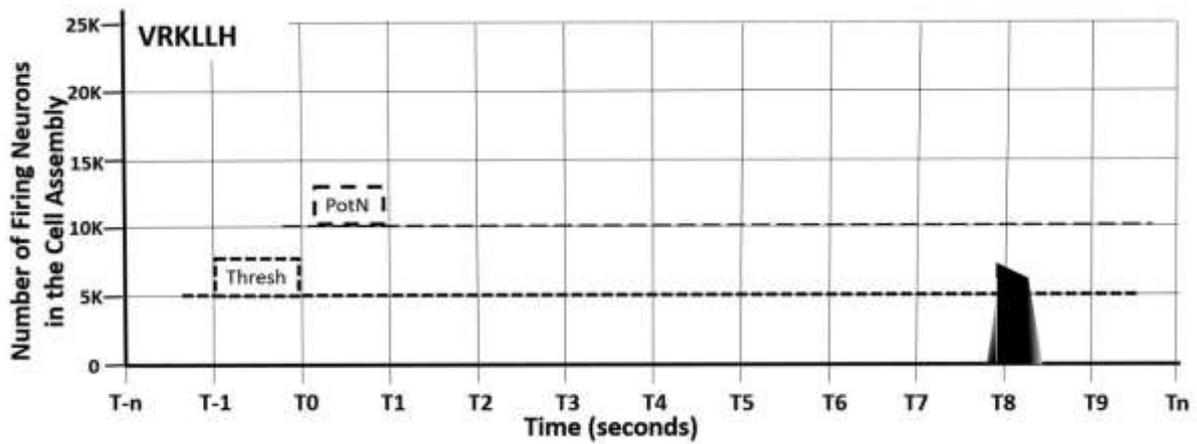

| ID | PotN | Thresh | IgMax | IgFat | P50% | IgTIg | IgTEx | D50% |
|---|---|---|---|---|---|---|---|---|
| VRKLLH | 10 | 5 | 7 | 6 | 7.8 | 7.9 | 8.2 | 8.3 |

INPUTS:    CA: COGNITIVE – Replace Kettle Lid with Left Hand (CRKLLH).

OUTPUTS:   CA: COGNITIVE – Replace Kettle Lid with Left Hand (CRKLLH).

This is a straightforward, short distance, tracking task for the visual system. It's part of the vast suite of potential CAs involved with manipulating objects with our hands.

### 52    CA: MOTOR – Replace Kettle Lid with Left Hand (MRKLLH)

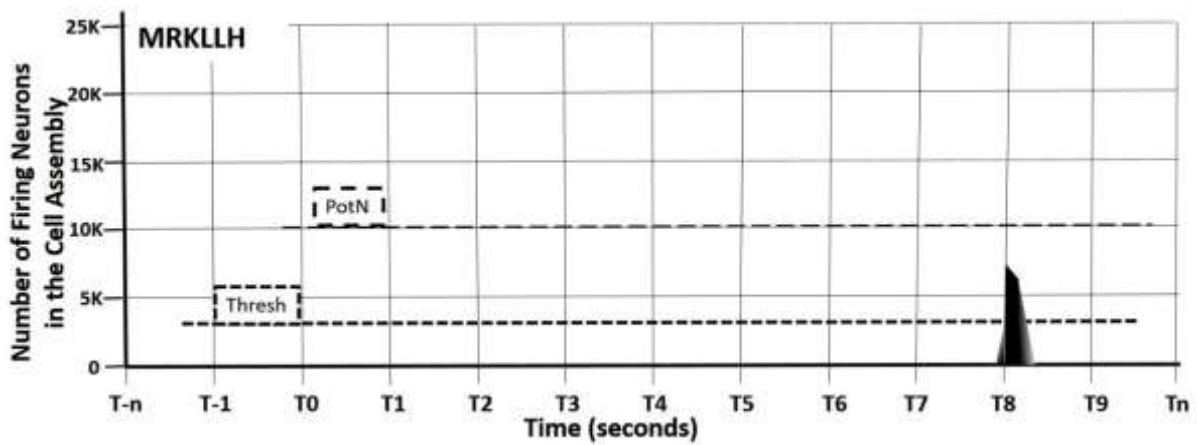

| ID | PotN | Thresh | IgMax | IgFat | P50% | IgTIg | IgTEx | D50% |
|---|---|---|---|---|---|---|---|---|
| MRKLLH | 10 | 3 | 7 | 6 | 7.9 | 8.0 | 8.1 | 8.2 |

INPUTS:    CA: COGNITIVE – Replace Kettle Lid with Left Hand (CRKLLH),

           CA: KINAESTHETIC – Left Hand Track Kettle Lid (KLHTKL).

OUTPUTS:   CA: KINAESTHETIC – Left Hand Track Kettle Lid (KLHTKL).

Ignited by CRKLLH it is then supressed by it once the kettle's lid is located. Once ignited it establishes a negative feedback loop with KLHTKL. The CA has three motor components: the movement to the kettle; a wobble to locate the lid securely; and the final operation is to move the left hand away and leave it hovering before the next behaviour, moving the left hand to operate the filter water tap's switch. Thus, the CA is relatively large (PotN 10K).

## 53   CA: COGNITIVE – Move Kettle to Tap (CMKT)

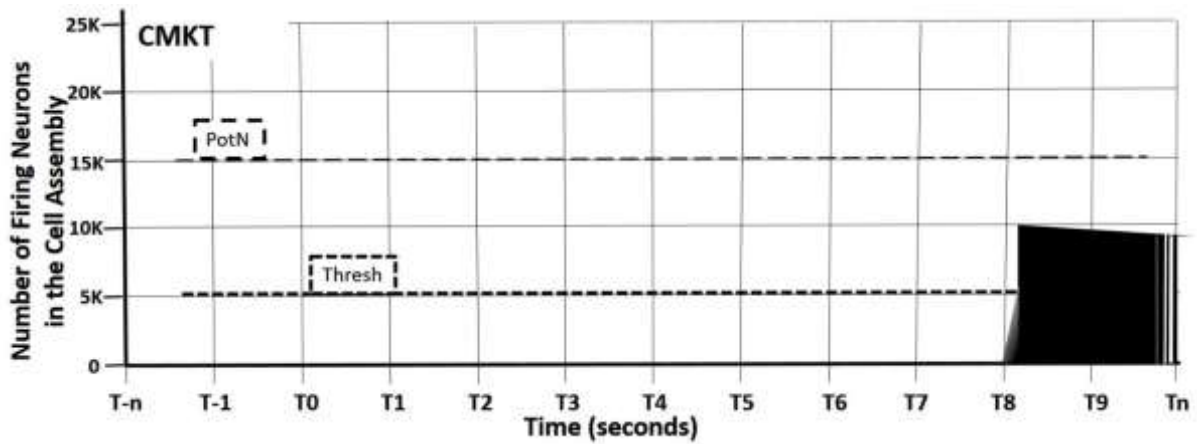

| ID   | PotN | Thresh | IgMax | IgFat | P50% | IgTIg | IgTEx | D50% |
|------|------|--------|-------|-------|------|-------|-------|------|
| CMKT | 15   | 5      | 10    | 9     | 8.1  | 8.2   | -     | -    |

INPUTS:     CA: COGNITIVE – Replace Kettle Lid with Left Hand (CRKLLH),

                 CA: VISUAL – Tap (VT)

                 CA: VISUAL – Kettle (VK).

OUTPUTS:    CA: VISUAL – Tap (VT)

                 CA: VISUAL – Kettle (VK),

                 CA: MOTOR – Move Kettle to Tap (MMKT).

                 CA: MOTOR – Hold Kettle to Tap (MHKT),

                 CA: COGNITIVE – Move Left Hand to Tap Switch (CMLHTS)

While similar to moving the kettle to the sink (CMKS), the flight path here is only short, say 15cm, and unobstructed given the sink's usual, empty state. It is a complex behaviour in that the kettle needs some small, careful rotations under visual negative feedback control so that the kettle's spout is accurately located directly under the water filter's spout, preferably without the two touching. Hence a PotN of 15K.

The CAs final operation is to supress the movement of the kettle (MMKT) and to hold the kettle still (MHKT) while the kettle is being filled. It ignites CMLHTS to move the left hand to the filtered water's tap switch.

## 54   CA: VISUAL – Tap (VT)

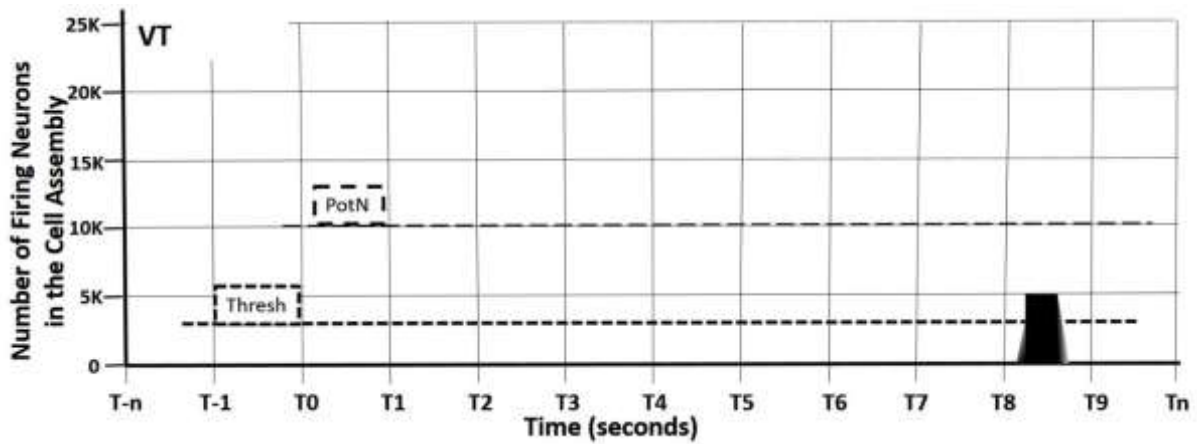

| ID | PotN | Thresh | IgMax | IgFat | P50% | IgTIg | IgTEx | D50% |
|----|------|--------|-------|-------|------|-------|-------|------|
| VT | 10   | 3      | 5     | 5     | 8.2  | 8.3   | 8.6   | 8.7  |

INPUTS:      CA: COGNITIVE – Move Kettle to Tap (CMKT).

OUTPUTS:   CA: COGNITIVE – Move Kettle to Tap (CMKT).

The filtered water tap consists of a tubular spout that rises beside the sink and turns 180 degrees vertically so that water flows down into the far right corner of the sink; the tap and spout are in a fixed position that does not change.  Such invariance, unlike even in the hot water area, means that a large visual CA is not necessary (PotN 10K).

## 55   CA: VISUAL – Kettle (VK)

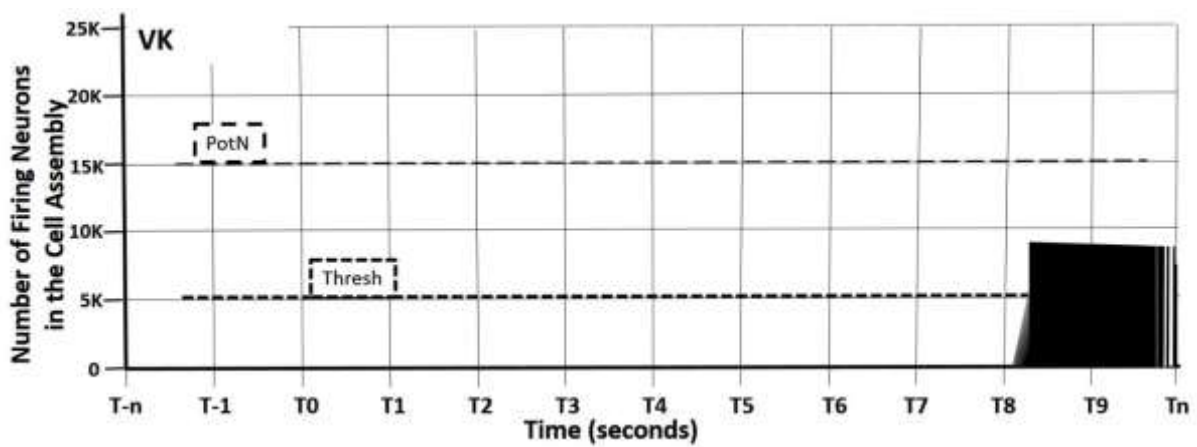

| ID | PotN | Thresh | IgMax | IgFat | P50% | IgTIg | IgTEx | D50% |
|----|------|--------|-------|-------|------|-------|-------|------|
| VK | 15   | 5      | 8     | 7     | 8.2  | 8.3   | -     | -    |

INPUTS:      CA: COGNITIVE – Move Kettle to Tap (CMKT).

OUTPUTS:   CA: COGNITIVE – Move Kettle to Tap (CMKT).

Visual attention is primarily on the kettle's spout and its three dimensional location with respect to the fixed location of the filtered water's spout, which is a small silver coloured target against a similarly coloured background, the sink. Visually fiddly but highly practiced, it has a PotN of 15K.

## 56   CA: MOTOR – Move Kettle to Tap (MMKT)

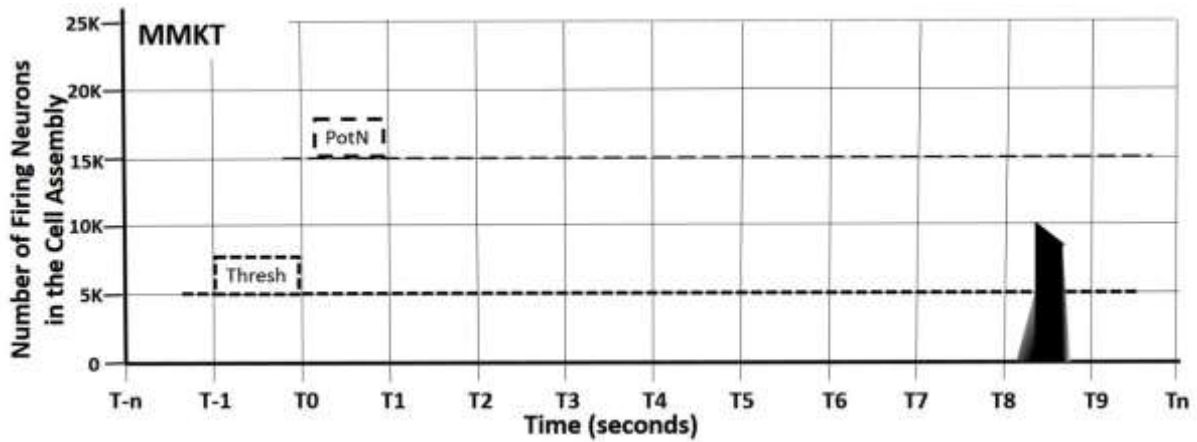

| ID | PotN | Thresh | IgMax | IgFat | P50% | IgTIg | IgTEx | D50% |
|---|---|---|---|---|---|---|---|---|
| MMKT | 15 | 5 | 10 | 8 | 8.3 | 8.4 | 8.6 | 8.6 |

INPUTS:   CA: COGNITIVE – Move Kettle to Tap (CMKT).

OUTPUTS:   *motor behaviours ...*

Involving hand, wrist and arm movement is relatively complex and requires accuracy if the kettle and tap spouts are not to make contact (PotN 15K). It is ignited by CMKT and then extinguished by CMKT so that the kettle can be held in its final, filling location (MKHT).

## 57   CA: MOTOR – Hold Kettle to Tap (MHKT)

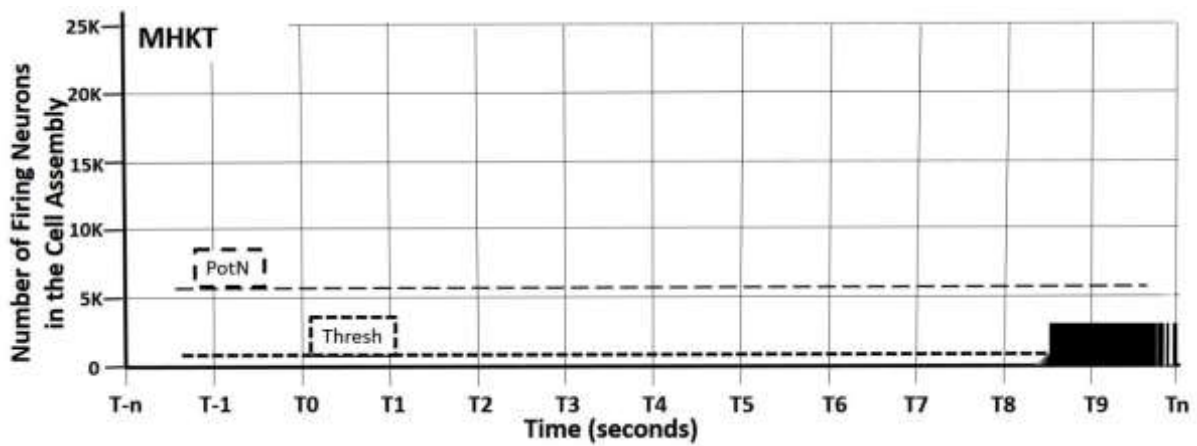

| ID | PotN | Thresh | IgMax | IgFat | P50% | IgTIg | IgTEx | D50% |
|---|---|---|---|---|---|---|---|---|
| MHKT | 6 | 1 | 3 | 3 | 8.4 | 8.5 | - | - |

INPUTS:   CA: COGNITIVE – Move Kettle to Tap (CMKT).

OUTPUTS:   *motor behaviours ...*

This is a small CA (PotN 6K) that is of a stationary class involving maintaining the position of an object. It is easily ignited (Threshold 1K) and can maintain itself indefinitely, i.e. neurons firing as other fatigue, indeed, the muscle fibres similarly fatigue and rotate contraction amongst themselves.

## 58   CA: COGNITIVE – Move Left Hand to Tap Switch (CMLHTS)

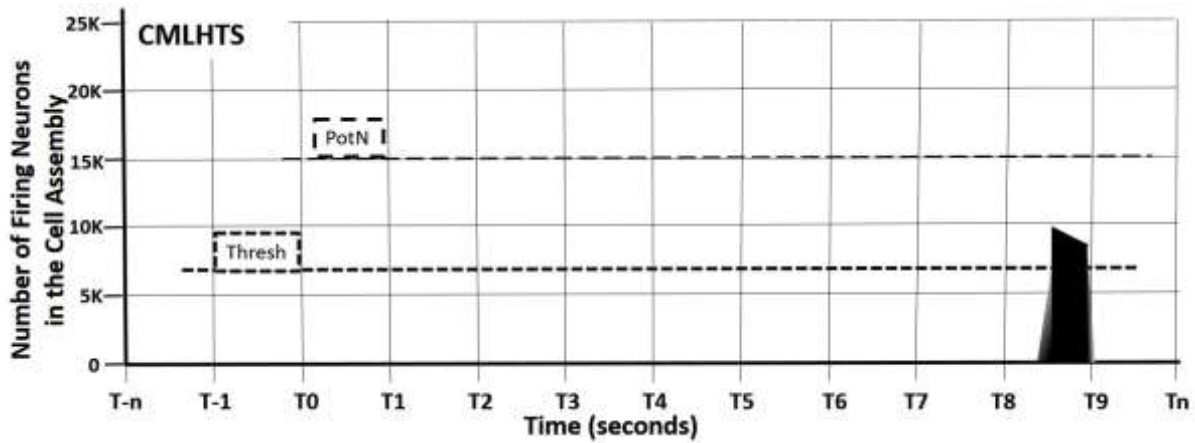

| ID | PotN | Thresh | IgMax | IgFat | P50% | IgTIg | IgTEx | D50% |
|---|---|---|---|---|---|---|---|---|
| CMLHTS | 15 | 7 | 10 | 8 | 8.3 | 8.5 | 8.9 | 9.0 |

INPUTS:   CA: COGNITIVE – Move Kettle to Tap (CMKT),

CA: VISUAL – Left Hand to Tap Switch (VLHTS)

CA: VISUAL – Tap Switch (VTS),

CA: TOUCH – Left Hand on Tap Switch (TLHTS).

OUTPUTS:   CA: VISUAL – Left Hand to Tap Switch (VLHTS)

CA: VISUAL – Tap Switch (VTS),

CA: MOTOR – Move Left Hand to Tap Switch (MMLHTS)

CA: TOUCH – Left Hand on Tap Switch (TLHTS),

CA: COGNITIVE: Fill Kettle (CFK).

The left hand has been hovering, waiting for the kettle to start to move towards the filtered water tap (CMKT). The hand loosely follows behind the top of the kettle and then the elbow and wrist have to make adjustments for the left hand's awkward reach behind the tap to the tap's switch. It's quite a large cognitive CA (PotN 15K) to reflect the movement's complexity and has a high threshold (7K) to reflect the variability of when the CA ignites and the reaching behaviour starts (IgTIg – P50% = 0.2 seconds, i.e. perhaps nearly half a second of priming).

With detailed observation it seems about 10% of the time the left hand takes an alternative route, in between the tap and the kettle, rather than behind both, and this seems to be determined by how close are objects behind the tap switch (an area used to stack things waiting to be washed-up) that might interfere with the left fingers. Thus, a decision is made early on by this CA as to which path the left hand will follow.

## 59   CA: VISUAL – Left Hand to Tap Switch (VLHTS)

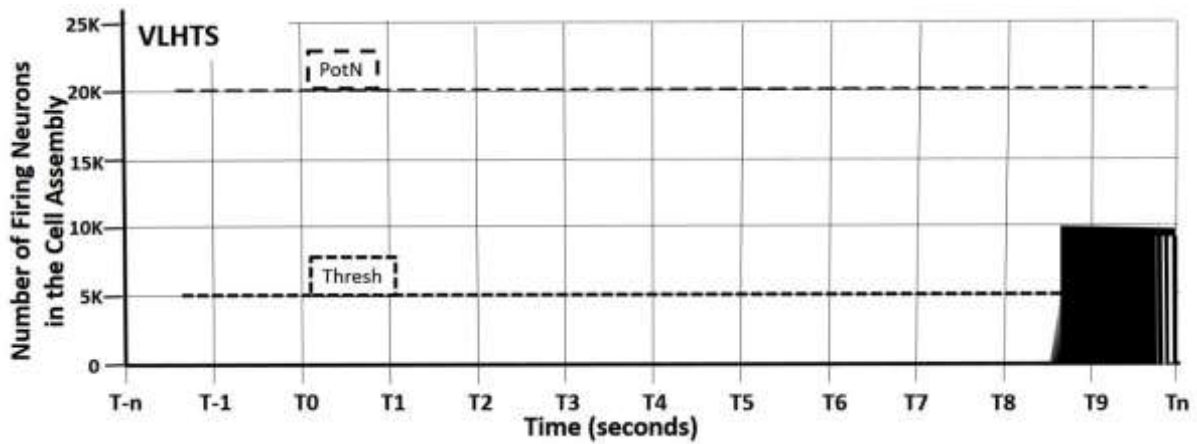

| ID | PotN | Thresh | IgMax | IgFat | P50% | IgTIg | IgTEx | D50% |
|---|---|---|---|---|---|---|---|---|
| VLHTS | 20 | 5 | 10 | 7 | 8.5 | 8.6 | - | - |

INPUTS:   COGNITIVE – Move Left Hand to Tap Switch (CMLHTS).

OUTPUTS:  COGNITIVE – Move Left Hand to Tap Switch (CMLHTS).

The CA is larger than one might initially expect (PotN 20K) because of the awkwardness of the movement, first tracking the kettle top and then providing feedback to control adjusting the hand to reach behind the tap to the switch.

## 60    CA: VISUAL – Tap Switch (VTS)

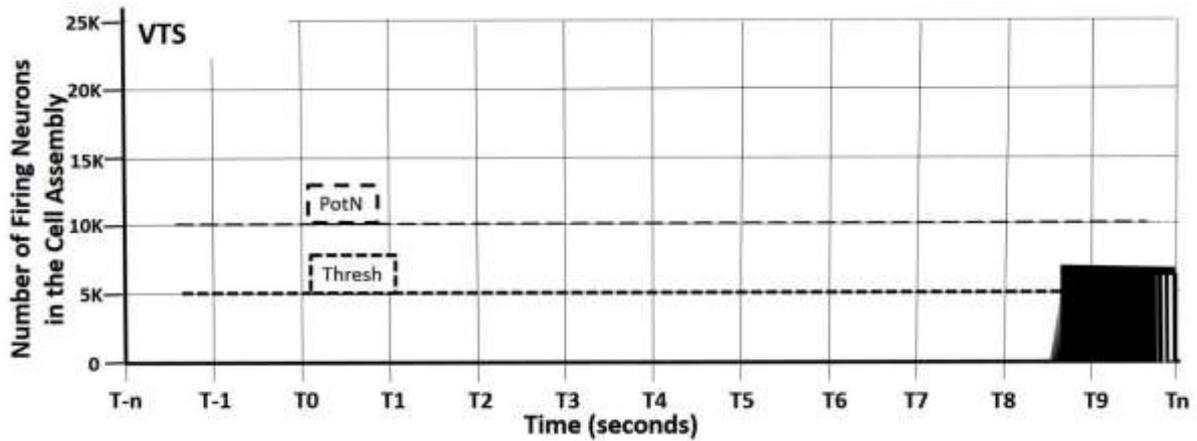

| ID | PotN | Thresh | IgMax | IgFat | P50% | IgTIg | IgTEx | D50% |
|---|---|---|---|---|---|---|---|---|
| VTS | 10 | 5 | 7 | 6 | 8.6 | 8.7 | - | - |

INPUTS:    COGNITIVE – Move Left Hand to Tap Switch (CMLHTS).

OUTPUTS:   COGNITIVE – Move left Hand to Tap switch (CMLHTS).

The invariance of the filtered water tap's location means that this is small for a visual CA (PotN 10K). Indeed, while this CA rationally needs to exist, one might suggest that its effect on behaviour is limited and perhaps, if the tap switch were absent (broken off), then CMLHTS and MMLTS would still ignite and the hand reach the switch before its absence was discovered, perhaps even discovered kinaesthetically (TLHTS).

## 61    CA: MOTOR – Move Left Hand to Tap Switch (MMLHTS)

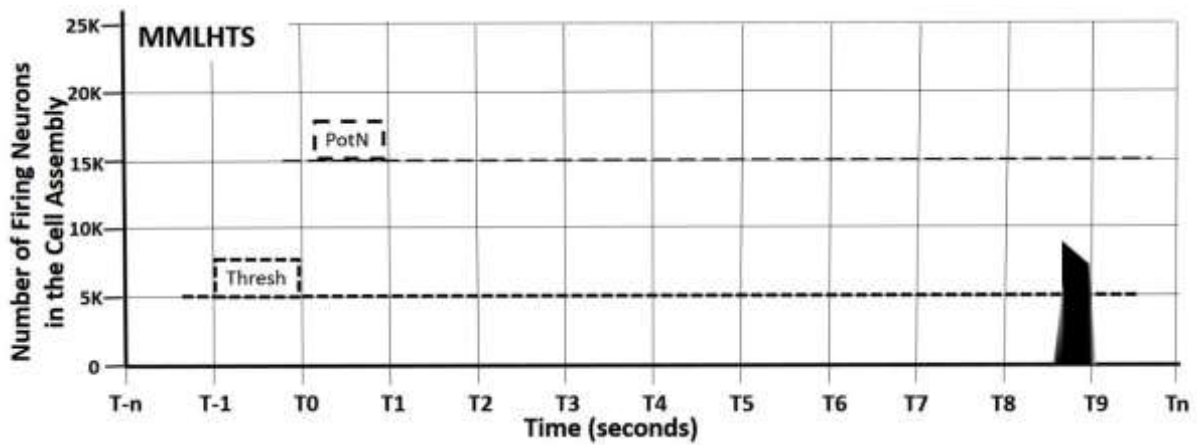

| ID | PotN | Thresh | IgMax | IgFat | P50% | IgTIg | IgTEx | D50% |
|---|---|---|---|---|---|---|---|---|
| MMLHTS | 15 | 5 | 8 | 7 | 8.7 | 8.7 | 8.9 | 9.0 |

INPUTS:    COGNITIVE – Move Left Hand to tap Switch (CMLHTS).

OUTPUTS: *motor behaviour ...*

With a PotN of 15K, this is a relatively large motor CA to reflect the compound nature of the behaviour. At a lower level of analysis this CA might be broken into several tightly bound ones, although some CAA models might still prefer a single CA as used here. It is ignited and then extinguished by CMLHTS. The latter based on touch feedback (TLHTS).

## 62  CA: TOUCH –Left Hand on Tap Switch (TLHTS)

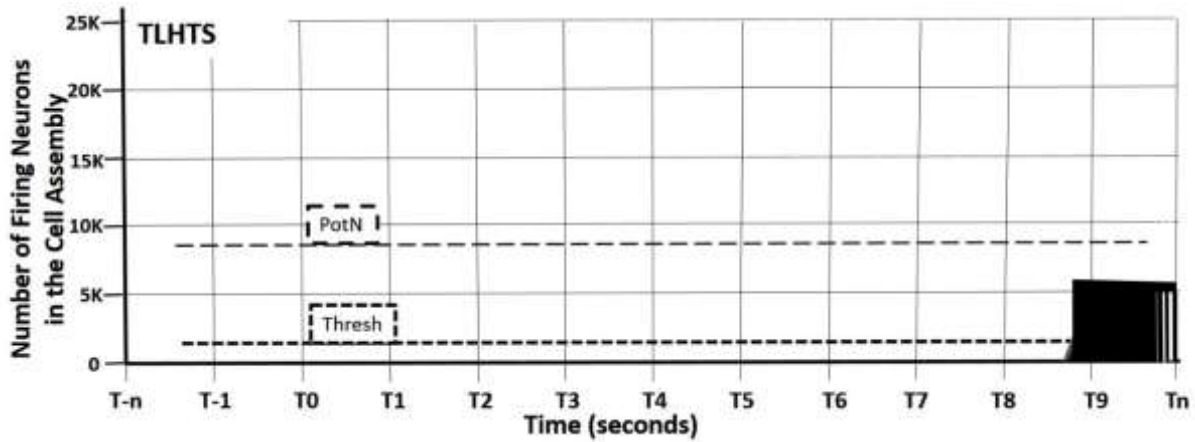

| ID | PotN | Thresh | IgMax | IgFat | P50% | IgTIg | IgTEx | D50% |
|---|---|---|---|---|---|---|---|---|
| TLHTS | 8 | 2 | 6 | 5 | 8.7 | 8.8 | - | - |

INPUTS: COGNITIVE – Move Left Hand to Tap Switch (CMLHTS).

OUTPUTS: COGNITIVE – Move left Hand to Tap switch (CMLHTS).

The left hand's final approach behind the tap switch target might be described as a fumble; whether it's the first two or the middle pair of fingers which come to rest under the switch appears to vary across task performances. An alternative description is that this CA is one of a common class of small ones (here PotN = 8K) that are used in fixed environments, are highly practiced, and use only limited visual feedback for approximate control, instead relying on a final fumble and negative feedback from touching the target object (putting one's coffee mug down on one's desk while still looking at the computer screen might be a particularly common example).

## 63   CA: COGNITIVE – Fill Kettle (CFK)

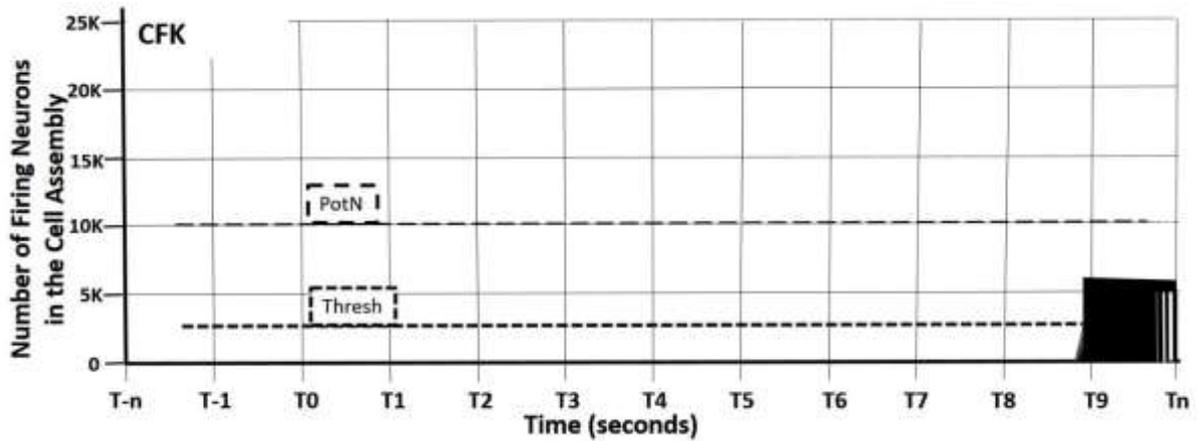

| ID  | PotN | Thresh | IgMax | IgFat | P50% | IgTIg | IgTEx | D50% |
|-----|------|--------|-------|-------|------|-------|-------|------|
| CFK | 10   | 3      | 7     | 6     | 8.8  | 8.9   | -     | -    |

INPUTS:    CA: COGNITIVE – Move Left Hand to Tap Switch (CMLHTS).

OUTPUTS:   CA: MOTOR – Pull Tap Switch Up (MPTSU),

CA: COGNITIVE – Make Coffee (CMC).

Ignition of this CA starts the first of three long pauses in the making a mug of coffee task; the other, longer two, are (1) boiling the kettle; and (2) waiting for the coffee to filter into the mug. It doesn't have to be a large CA (PotN 10K) as it needs only to ignite MPTSU to start the kettle filling process and then to reignite CMC.

## 64   CA: MOTOR – Pull Tap Switch Up (MPTSU)

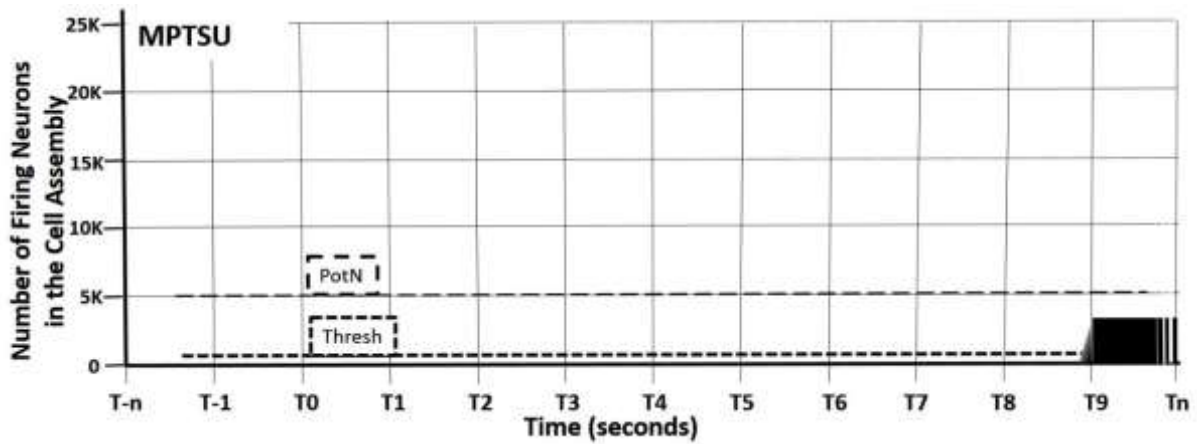

| ID    | PotN | Thresh | IgMax | IgFat | P50% | IgTIg | IgTEx | D50% |
|-------|------|--------|-------|-------|------|-------|-------|------|
| MPTSU | 5    | 1      | 3     | 3     | 8.9  | 9.0   | -     | -    |

INPUTS: CA: COGNITIVE – Fill Kettle (CFK)

This involves a simple flick of the fingers upwards under open loop control (PotN 5K). After this flick the left hand may or may not be removed from the tap switch, immediately or at some later time during the filling process; this depends on other mental activities during the filling time.

**03    CA: Cognitive – Make Coffee (CMC)** …

This is the end of the 'First Steps to Making Coffee' analysis. At this point there remains ten ignited CAs:

| | |
|---|---|
| CRHH & MRHH | – The right hand is holding the kettle by its handle. |
| CS & VS | – The sink is still a major feature of the task environment. |
| CMKT & MHKT | – The kettle is held to the tap as the kettle fills. |
| VK | – The kettle remains a major object in the task environment. |

TLHTS, CFK & MPTSU – The left hand may make a variety of movements, including none, after the end of the analysis period.

**Acronym Glossary**

**CAs** (*in order of analysis appearance*)

| | | |
|---|---|---|
| 01 CKEC | COGNITIVE – Kitchen Entrance Check (CKEC) |
| 02 VKEG | VISUAL – Kitchen Entrance General (VKEG) |
| 03 CMC | COGNITIVE – Make Coffee (CMC) |
| 04 CAHWA | COGNITIVE – Approaching Hot Water Area (CAHWA) |
| 05 VAHWA | VISUAL – Approaching Hot Water Area (VAHWA) |
| 06 MSHWA | MOTOR – Stride To Hot Water Area (MSHWA) |
| 07 CKHWA | COGNITIVE – Kettle In Hot Water Area (CKHWA) |
| 08 VKHWA | VISUAL – Kettle In Hot Water Area (VKHWA) |
| 09 CKH | COGNITIVE – Kettle Handle (CKH) |
| 10 VKH | VISUAL – Kettle Handle (VKH) |
| 11 MRAB | MOTOR – Right Arm Ballistic (MRAB) |
| 12 VRH | VISUAL – Right hand (VRH) |
| 13 CRH | COGNITIVE – Right hand (CRH) |
| 14 CHWA | COGNITIVE – Hot water Area (CHWA) |
| 15 VHWA | VISUAL – Hot Water Area |
| 16 CRHA | COGNITIVE – Right Hand Approach (CRHA) |
| 17 VRHA | VISUAL – Right Hand Approach (VRHA) |
| 18 MRHA | MOTOR – Right Hand Approach (MRHA) |
| 19 TRHKH | TOUCH – Right Hand to Kettle Handle (TRHKH) |
| 20 CRHG | COGNITIVE – Right Hand Grip (CRHG) |
| 21 MRHG | MOTOR – Right Hand Grip (MRHG) |
| 22 TRHG | TOUCH – Right Hand Grip (TRHG) |
| 23 CRHH | COGNITIVE – Right Hand Hold (CRHH) |
| 24 MRHH | MOTOR – Right Hand Hold (MRHH) |
| 25 CLK | COGNITIVE – Lift Kettle (CLK) |
| 26 MLK | MOTOR – Lift Kettle (MLK) |
| 27 KKW | KINAESTHETIC –Kettle Weight (KKW) |
| 28 VLK | VISUAL – Lift Kettle (VLK) |

29 CD COGNITIVE – Drainer (CD)

30 VD VISUAL – Drainer (VD)

31 CMKS COGNITIVE – Move Kettle to Sink (CMKS)

32 VMKS VISUAL – Move Kettle to Sink (VMKS)

33 MMKS MOTOR – Move Kettle to Sink (MMKS)

34 MLHTKL MOTOR – Left Hand Track Kettle Lid (KLHTKL)

35 KLHTKL KINAESTHETIC – Left Hand Track Kettle Lid (KLHTKL)

36 MSBS MOTOR – Shuffle Body to Sink (MSBS)

37 CS COGNITIVE – Sink (CS)

38 VS VISUAL – Sink (VS)

39 CLHRKL COGNITIVE – Left Hand Remove Kettle Lid (CLHRKL)

40 VKL VISUAL – Kettle Lid (VKL)

41 VLH VISUAL – Left Hand (VLH)

42 MLHRKL MOTOR – Left Hand Remove Kettle Lid (MLHRKL)

43 VKWL VISUAL – Kettle Without Lid (VKWL)

44 CEK COGNITIVE – Empty Kettle (CEK)

45 MRHIK MOTOR – Right Hand Invert Kettle (MRHIK)

46 VKE VISUAL – Kettle Empty (VKE)

47 CKE COGNITIVE – Kettle Empty (CKE)

48 CRHOK COGNITIVE – Right Hand Orientate Kettle (CRHOK)

49 VRHOK VISUAL – Right Hand Orientate Kettle (VRHOK)

50 MRHOK MOTOR – Right Hand Orientate Kettle (MRHOK)

51 CRKLLH COGNITIVE – Replace Kettle Lid with Left Hand (CRKLLH)

52 VRKLLH VISUAL – Replace Kettle Lid with Left Hand (VRKLLH)

53 MRKLLH MOTOR – Replace Kettle Lid with Left Hand (MRKLLH)

54 CMKT COGNITIVE – Move Kettle to Tap (CMKT)

55 VT VISUAL – Tap (VT)

56 VK VISUAL – Kettle (VK)

57 MMKT MOTOR – Move Kettle to Tap (MMKT)

58 MHKT MOTOR – Hold Kettle to Tap (MHKT)

| | |
|---|---|
| 59 CMLHTS | COGNITIVE – Move Left Hand to Tap Switch (CMLHTS) |
| 60 VLHTS | VISUAL – Left Hand to Tap Switch (VLHTS) |
| 61 VTS | VISUAL – Tap Switch (VTS) |
| 62 MMLHTS | MOTOR – Move Left Hand to Tap Switch (MMLHTS) |
| 63 TLHTS | TOUCH – Left Hand on Tap Switch (TLHTS) |
| 64 CFK | COGNITIVE – Fill Kettle (CFK) |
| 65 MPTSU | MOTOR – Pull Tap Switch Up (MPTSU) |
| 03 CMC… | COGNTIVE – Make Coffee (CMC) |

**Other**

| | |
|---|---|
| ACT-R | Active Control of Thought – R version |
| AI | Artificial Intelligence |
| AL | Activity List |
| ANN | Artificial Neural Network |
| CA | Cell Assembly |
| CAA | Cell Assembly Architecture |
| CAAR | Cell Assembly Architecture Relationship |
| CABot | Cell Assembly roBot |
| EPIC | Executive-Process/Interactive Control |
| FLIF | Fatiguing Leaky Integrate and Fire |
| GOMS | Goals, Operators, Methods and Selection-rules |
| HTA | Hierarchical Task Analysis |
| LTM | Long Term Memory |
| NL | Natural Language |
| QPID | Quiescent, Priming, Ignition and Decay |
| SCAM | Simplified Cell Assembly Model |
| STM | Short Term Memory |
| TA | Task Analysis |
| TACAP | Task Analysis Cell Assembly Perspective |